\documentclass[lettersize,journal]{IEEEtran}
\usepackage{cite}
\usepackage{amsmath,amssymb,amsfonts,amsthm}
\usepackage{algorithm}
\usepackage{algpseudocode}
\usepackage{graphicx}
\usepackage{epstopdf}
\usepackage{caption}
\usepackage{subcaption}
\usepackage{float}
\usepackage{textcomp}
\usepackage{color,soul}
\usepackage[abs]{overpic}
\usepackage{mathtools}
\usepackage{stfloats}
\usepackage{adjustbox}
\usepackage{booktabs}
\usepackage{array}
\usepackage{makecell}
\usepackage{placeins}

\usepackage{xcolor,varwidth}

\newcommand{\myproof}{\textbf{Proof:~}}
\newtheorem{remark}{Remark}
\newtheorem{lem}{Lemma}
\newtheorem{thm}{Theorem}

\def\BibTeX{{\rm B\kern-.05em{\sc i\kern-.025em b}\kern-.08em
		T\kern-.1667em\lower.7ex\hbox{E}\kern-.125emX}}

\begin{document}

\title{Real Time Adaptive Estimation of Li-ion Battery Bank Parameters}

\author{Hafiz M. Usman$^{\dagger *}$,
	Shayok Mukhopadhyay$^{*}$, and Habibur Rehman$^{*}$\\
$^{*}$American University of Sharjah, Dept. of Electrical Engineering, P.O. Box 26666, Sharjah, United Arab Emirates\\
$^{\dagger}$University of Waterloo, Department of Electrical and Computer Engineering, Waterloo, ON, Canada
\thanks{This work was supported in part by the Office of Research and Graduate Studies at the American University of Sharjah through the Faculty Research Grant FRG17-R-34.}}



\maketitle

\begin{abstract}
This paper proposes an accurate and efficient Universal Adaptive Stabilizer (UAS) based online parameters estimation technique for a 400 V Li-ion battery bank. The battery open circuit voltage, parameters modeling the transient response, and series resistance are all estimated in a single real-time test. In contrast to earlier UAS based work on individual battery packs, this work does not require prior offline experimentation or any post-processing. Real time fast convergence of parameters' estimates with minimal experimental effort enables self-update of battery parameters in run-time. The proposed strategy is mathematically validated and its performance is demonstrated on a 400 V, 6.6 Ah Li-ion battery bank powering the  induction motor driven prototype electric vehicle (EV) traction system.
\end{abstract}


\section{Introduction}
\label{sec1}
\IEEEPARstart{H}{igh} energy density and low self-discharge rate have made Li-ion batteries a premium candidate for electric vehicle (EV) applications. Accurate estimation of open circuit voltage (OCV), series resistance, and State-of-Charge (SoC) are indispensable for an effective battery management system. Precise estimates of internal states of a Li-ion battery like SoC, State-of-Health (SoH) also rely on an accurate battery model. The Chen and Mora equivalent circuit model \cite{CM} has been widely adopted in the literature for Li-ion battery modeling. The salient features of this model which make it attractive for the proposed work are: it can model real time voltage and current dynamics; can capture temperature effects and number of charge-discharge cycles; it is simple to implement for a run-time battery management system; has low computational effort, and it includes SoC dependent equivalent circuit elements without requiring to solve partial differential equations (PDEs) common in electrochemical Li-ion battery models. Therefore, Chen and Mora's battery model \cite{CM} has been utilized for this and our previous work \cite{Daniyal,ISMA,Access}. Different strategies are available in the literature for extracting Li-ion battery model parameters \cite{R1,R2,R3,R4,R5,R6,new1,new2,new3,new4,new5,new6}.

Not so long ago, dual unscented Kalman filter \cite{R1} and $H_\infty$ Kalman filter \cite{R2} based approaches were proposed to overcome the limitations of Kalman Filters (KFs) and Extended Kalman Filters (EKFs) for accurate battery SoC estimation. Usually, model-based KF and EKF methods require prior knowledge of battery parameters via some offline method, which is normally time-consuming and could be prone to error. However, the strategies presented in \cite{R1}, and \cite{R2} simultaneously identify both the battery model circuit elements and SoC. A fractional calculus theory-based intuitive and highly accurate fractional-order equivalent circuit model of Li-ion battery is presented in \cite{R3}. The fractional-order circuit is capable of modeling many electrochemical aspects of a Li-ion battery, which are typically ignored by integer-order RC equivalent circuit models. The authors in \cite{R3} used a modified version of Particle Swarm Optimization algorithm for accurate estimation of equivalent circuit elements, and validated their results for various operating conditions of a Li-ion battery. Yet this strategy requires a precise knowledge of open circuit voltage, and optimization based strategies can be susceptible to high computational effort. The authors in \cite{R4} proposed a moving window based least squares method for reducing the complexity and computational cost of online equivalent circuit elements' identification, along with the battery SoC estimation. The technique presented in \cite{R4} utilizes a piece-wise linear approximation of the open circuit voltage curve. Nevertheless, the length of the linear approximation window may affect the overall  accuracy of the estimated equivalent circuit elements. The authors in \cite{R5} attempted to identify the equivalent circuit elements of a Li-ion battery model by means of voltage relaxation characteristics. Although the strategy described in \cite{R5} requires several pulse charging and discharging experiments, yet it extracts the equivalent circuit elements with good accuracy. A possible drawback of this strategy includes offline identification, and similar to other techniques described earlier, it relies on accurate open circuit voltage measurement. Two extended Kalman filters (named as dual EKF) are combined in \cite{R6} for simultaneous estimation of Li-ion battery model parameters and SoC. A dead-zone is utilized in \cite{R6} to overcome the issue of dual EKF's high computational cost. The dead-zone defines the duration for which adaptive estimation of parameters and SoC is stopped, while the terminal voltage estimation error stays within the user-defined error limit. However, the accuracy of estimated parameters and open circuit voltage are not analyzed in \cite{R6}. %new from usman

\begin{table*}[!t]
	\centering
	\caption{{Comparison of important attributes for the implementation of real-time battery parameters estimation strategy.}}
	\begin{adjustbox}{width=1\linewidth}
		\begin{tabular}{c c c c c} 
			\hline\hline
			Techniques & \thead{No prior knowledge\\/pre-processing} & \thead{Determines open-circuit\\voltage} & \thead{Low computation\\time} & \thead{Ease of assuring convergence of estimate\\close to actual values}  \\
			\hline \\    
			Kalman filtering-based approaches \cite{R1,R2,R6} & $\times$ & $\times$ & $\checkmark$ & $\checkmark$ \\
			Least squares-based approaches \cite{new1,new2,new3} & $\times$ & $\checkmark$ & $\checkmark$ & $\times$ \\
			Metaheuristic optimization (PSO, GN) \cite{R3}, \cite{new4} & $\times$ & $\checkmark$ & $\times$ & $\times$ \\
			Artificial intelligence-based approaches \cite{new6} & $\times$ & $\checkmark$ & $\times$ & $\times$ \\
			\textbf{Proposed UAS-based approach} & $\checkmark$ & $\checkmark$ & $\checkmark$ & $\checkmark$ \\
			\hline
		\end{tabular}
		\label{tab_compare}
	\end{adjustbox}
\end{table*}

As for more recent methods, a variable time window-based least squares method in \cite{new1} models the hysteresis effect and effectively captures the nonlinear dynamics of a Li-ion battery. Similarly, a partial adaptive forgetting factor-based least squares method is proposed in \cite{new2} for Li-ion battery parameters estimation in electric vehicles. The method in \cite{new2} also incorporates different exogenous factors such as driver behavior, environmental conditions, and traffic congestion in problem formulation. Likewise, a trust region optimization-based least squares approach is proposed in \cite{new3}, which claims to reduce the complexity, and thus the estimation time, of a conventional least squares estimation procedure. To overcome the potential limitations of Genetic Algorithm (GN), such as higher computational efforts, and possible convergence to local minima, the authors in \cite{new4} deployed Particle Swarm Optimization (PSO) routine after GN for accurate identification of both temperature and SoC dependent Li-ion battery parameters. PSO routine not only helps to obtain a near global solution but also refines the GN results. Recently, a sequential algorithm  based on high pass filter and active current injections  is developed in \cite{new5} for accurate and quick estimation of Li-ion battery parameters. It is shown in \cite{new5} that higher frequencies in an injected current improves the performance of parameters estimation process. Various Neural Network (NN)-based data-driven strategies have also been reported in the literature for Li-ion battery parameters estimation. Different variants of NN-based methods, such as \cite{new6} learn and capture the dynamics of a Li-ion battery model. However, the major downsides of several recent state-of-the-art methods \cite{new1}-\cite{new4} include some kind of offline pre-processing for appropriate selection of initial parameters, offline open-circuit voltage determination, appropriate tuning of optimization parameters, higher computational efforts, and unsatisfactory convergence performance. Moreover, some additional constraints in the recent mainstream methods are as follows. The Hessian matrix approximation undermines the accuracy of GN algorithm in \cite{new4}, the exogenous factors in \cite{new2} are not easily accessible, and the battery current profile in [5] cannot be altered to inject the signal enriched with enough frequencies. The performance of NN-based methods \cite{new6} relies on effective training with large datasets, requiring large memory and high computations, which may be infeasible in many battery management systems (BMS) and real-time EV applications. Furthermore, the training datasets may not be enriched with rarely occurring instances in a Li-ion battery, such as short circuit, overcharging, and overdischarging.

{To highlight the advantages of the proposed UAS-based scheme compared to the mainstream methods, we present a comparative analysis of different techniques in Table} \ref{tab_compare} {below. The attributes in Table} \ref{tab_compare} {are considered important for real-time battery parameters estimation of an electric vehicle. An effective online strategy for battery parameters estimation should have the following attributes: (i) does not require any prior knowledge for parameters initialization or offline pre-processing, (ii) determines open-circuit voltage without offline experimentations, (iii) has low computation cost, and (iv) guarantees parameters convergence. Based on the experimental work presented in this paper, the proposed UAS-based scheme features the above-mentioned attributes and, thus, is best suitable for real-time battery parameters estimation of an electric vehicle.}

This work proposes a UAS-based adaptive parameters estimation scheme for a Li-ion battery that neither needs any kind of offline pre-processing. Unlike optimization and NN-based methods, the proposed method requires very less memory and low computations, and thus it is very quick and yet effective for BMS and real-time EV applications. The proposed method has been tested and verified at the battery cell, pack and bank levels for simultaneous estimation of battery parameters, and open circuit voltage. This work utilizes a high-gain universal adaptive stabilization (UAS) based observer. The switching function required by UAS \cite{Ilchmann} is realized by a Nussbaum function. A Nussbaum function has rapid oscillations and variable frequency by definition \cite{Ilchmann}. When a Nussbaum function is input to the observer, it injects enough sinusoids into the high-gain observer, satisfying the required persistence of excitation (PE) condition \cite{sastry2013}. Therefore, our previous \cite{Daniyal,usman2019,VCollapse,alkhawaja2018} and the present work are theoretically and experimentally verified without explicitly mathematically imposing the PE condition. The above mentioned properties of a Nussbaum function result in accurate parameter estimation, even without mathematically imposing PE. It is also worth noting that some other work \cite{stanislav2015} also exists in the literature which does not explicitly impose PE condition for parameters estimation.

This work extends our previous work \cite{Daniyal} to another level, by estimating Li-ion battery open circuit voltage, series resistance and other battery model parameters, all by a single experiment conducted in real-time. The proposed approach is validated at the battery cell level as well as on a prototype battery bank setup for an EV traction system. In our previous work, open circuit voltage and series resistance parameters were found by the voltage relaxation test and curve fitting, respectively, and then the remaining parameters were estimated using a UAS based strategy. %Thus motivated, a modified UAS-based APE strategy is proposed in this work which estimates all parameters of a Li-ion battery model along with open circuit voltage, and series resistance in a single adaptation run.
The previous offline adaptive parameters estimation (APE) strategy in \cite{Daniyal} required eight experiments to estimate all battery model parameters, while the proposed online APE scheme runs online requiring only one experiment for parameters estimation. Furthermore, in contrast to \cite{R1,R2,R3,R4,R5,R6}, our proposed strategy does not require any experimental effort towards acquiring prior knowledge of open circuit voltage, rather the open circuit voltage is also estimated by the strategy proposed in this paper. 

Following are the main contributions of this research work.
\begin{itemize}
	
	\item The proposed online APE scheme estimates all equivalent circuit elements, including open circuit voltage, and series resistance of a Li-ion battery model at the cell/pack/bank level in one real-time experimental run. %This mathematical proof is an extension of previous work provided in \cite{Daniyal}, where open circuit voltage and series resistance parameters were found by the voltage relaxation test and curve fitting, respectively. 
	
	\item The proposed strategy is formulated and proved mathematically. 
	
	\item The accuracy of parameters estimation is validated by the following simulations and experiments:
	
	\begin{itemize}
		
		\item The parameters estimated in simulation using the proposed online APE approach are compared against the ones experimentally obtained by Chen and Mora \cite{CM} for a  4.1 V, 270 mAh Li-ion battery.
		
		%The accuracy of the proposed modified APE approach is validated by comparing the simulation results obtained by the proposed approach and the ones acquired by Chen and Mora \cite{CM} on a  4.1 V, 270 mAh Li-ion battery.
		
		\item The parameters estimated online using experimental data are compared with the previous offline parameters estimation \cite{Daniyal} results for a 22.2 V, 6.6 Ah Li-ion battery.
		
		\item Finally, the proposed online APE strategy is implemented on a 400 V, 6.6 Ah Li-ion battery bank powering a prototype EV traction system.
		
		%The last contribution of this work includes parameters estimation of a 400 V, 6.6 Ah Li-ion battery bank via proposed modified APE strategy. The 400 V, 6.6 Ah Li-ion battery bank is developed by connecting sixteen 25 V, 6.6 Ah Li-ion batteries in series. The offline verification of estimated parameters on a 400 V battery bank allows us to implement the modified APE strategy for an induction motor driven EV traction system. The EV traction system used in this work is controlled by indirect field orientation technique and powered by a 400 V, 6.6 Ah Li-ion battery bank.
	\end{itemize}
	
\end{itemize}

The rest of the article is organized as follows. Necessary background information about the CM \cite{CM} Li-ion battery equivalent circuit model and UAS are provided in Section \ref{sec2}. Section \ref{sec3} formulates the proposed UAS based high gain adaptive observer for parameters estimation.  Section \ref{sec4} provides mathematical justification of our proposed method. Simulation and experimental results are presented in Section \ref{sec5} and \ref{sec6} respectively for validating the proposed online APE strategy. Real time implementation results for an EV traction system are shared in Section \ref{sec7}. Finally, the concluding remarks are made in Section \ref{sec8} of this article.	

\section{Background} \label{sec2}
This section provides information about the CM Li-ion battery equivalent circuit model and UAS used in this work. The battery equivalent circuit model is described in Section 2.1, while Section 2.2 presents the formulation of a Nussbaum type switching function employed in the proposed online APE algorithm.
\subsection{Li-ion Battery Equivalent Circuit Model}
\begin{figure}[!b]
	\centering
	\includegraphics[scale=0.321]{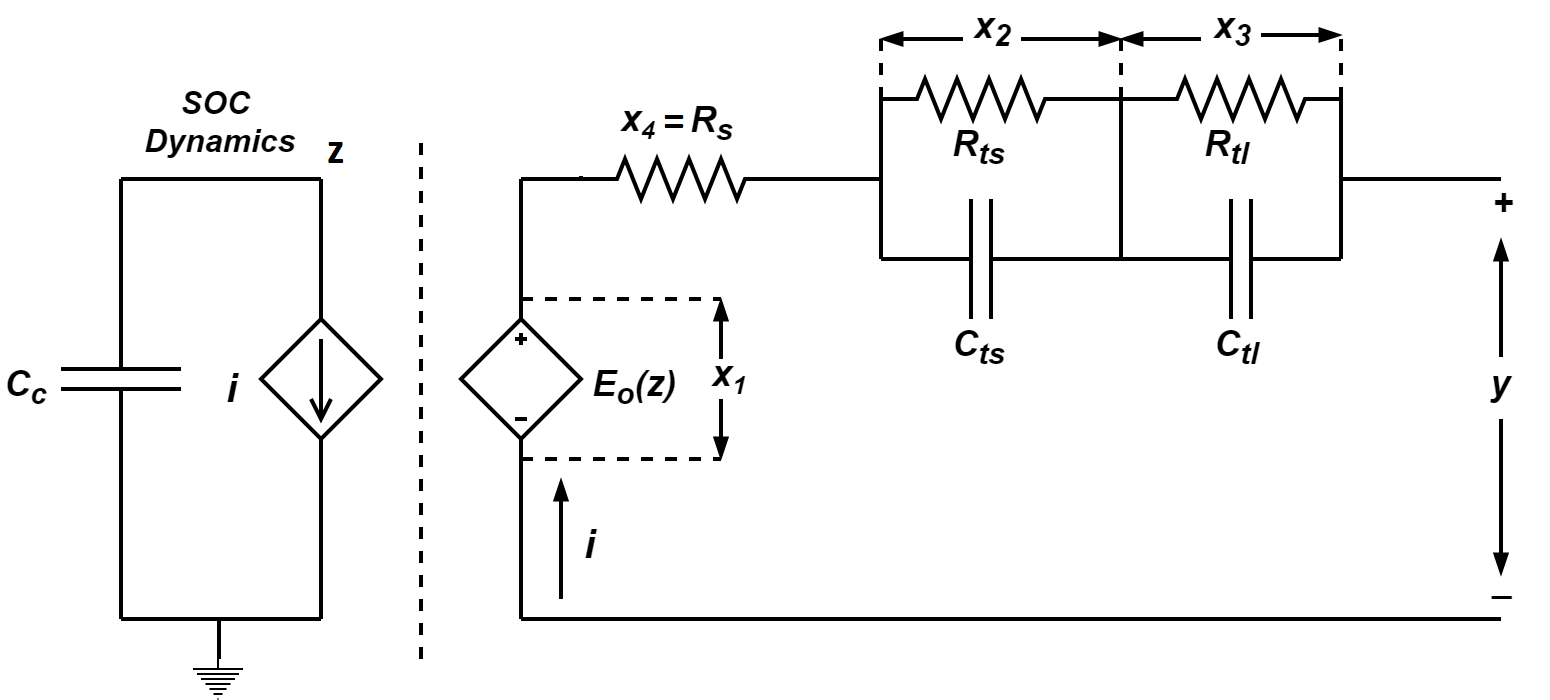}
	\caption{Li-ion battery equivalent circuit model.}
	\label{fig1}
\end{figure}

The Chen and Mora \cite{CM} equivalent circuit model of a Li-ion battery is shown in Figure \ref{fig1}. This work aims at providing an accurate and simple online adaptive parameters estimation method, for a battery at the cell/pack/bank level using the Li-ion battery model shown in Figure \ref{fig1}. The state space representation of Figure \ref{fig1} is described by (\ref{pf1})-(\ref{pf4}).

\begin{align}
\dot{z}(t)&=-\frac{1}{C_c}i(t), \,\,\,\, C_c=3600Cf_1 f_2 f_3 \label{pf1} \\
\dot{x}_1(t)& = \frac{\partial x_1(z)}{\partial z(t)}\dot{z}(t), \mbox{ therefore} \nonumber \\ \dot{x}_1(t)&=-\bigg(r_1r_2e^{-r_2z}+r_4-2r_5z+3r_6z^2\bigg)\frac{i(t)}{C_c} \label{pf5}\\
\dot{x}_2(t)&=-\frac{x_2(t)}{R_{ts}(z)C_{ts}(z)}+\frac{i(t)}{C_{ts}(z)} \label{pf2} \\
\dot{x}_3(t)&=-\frac{x_3(t)}{R_{tl}(z)C_{tl}(z)}+\frac{i(t)}{C_{tl}(z)} \label{pf3} \\
\dot{x}_4(t)& = \frac{\partial x_4(z(t))}{\partial z(t)}\dot{z}(t) = \bigg(r_{19}r_{20}e^{-r_{20}}z \bigg)\frac{i(t)}{C_c}  \label{pf10} \\
y(t)&=x_1(z)-x_2(t)-x_3(t)-i(t)x_4(t). \label{pf4}
\end{align}

Here, the battery SoC is denoted by $z \in [0,1]$. The states $x_1$, $x_2$, $x_3$, $x_4$, represent the open circuit voltage, the voltage across $R_{ts}||C_{ts}$, the voltage across $R_{tl}||C_{tl}$, and the battery series resistance $R_s$ respectively. The term $C_c$ and $y(t)$ denote the battery capacity in ampere-hour (Ah) and battery terminal voltage. The factors $f_1$, $f_2$, $f_3 \in [0,1]$ account for the effects of temperature, charge-discharge cycles, and self discharging respectively. The battery open circuit voltage $x_1$ in (\ref{pf5}), battery series resistance $x_4$ in (\ref{pf10}), and equivalent circuit elements $R_{ts},R_{tl},C_{ts},C_{tl}$ can be defined from Chen and Mora's work \cite{CM} by (\ref{pf6})-(\ref{pf9}). Note that the formulation in \eqref{pf1}-\eqref{pf10} is novel compared to \cite{Daniyal}, as the notation introduced here for the CM model specifically allows simultaneous online estimation of battery parameters, and open circuit voltage.
\begin{align}
&E_o(z) = -r_1e^{-r_2z} + r_3 + r_4z -r_5z^2 + r_6z^3 =x_1(z) \label{pf6} \\
&R_{ts}(z)=r_{7}e^{-r_8z}+r_9  \label{pf_6} \\
&R_{tl}(z)=r_{10}e^{-r_{11}z}+r_{12}  \label{pf7} \\
&C_{ts}(z)= -r_{13}e^{-r_{14}z}+r_{15}  \label{pf8} \\
&C_{tl}(z)= -r_{16}e^{-r_{17}z}+r_{18}  \label{pf_9} \\
&R_s(z) =r_{19}e^{-r_{20}z}+r_{21} = x_4(z) \label{pf9}.
\end{align}
The parameters $r_1, \cdots,r_{21}$ used in the circuit elements in equation (\ref{pf6})-(\ref{pf9}) are constant real numbers.

\subsection{Universal Adaptive Stabilization}

The UAS based strategy has been employed in \cite{VCollapse} for fast error convergence. This motivated us to employ the UAS based adaptive estimation method for quick \cite{VCollapse} and yet accurate \cite{Daniyal,Access,usman2019} Li-ion battery parameters ($r_1,\cdots, r_{21}$) estimation. The implementation of a UAS based technique requires a switching function with high growth rate \cite{Ilchmann}. A Nussbaum function is a switching function, which is defined as a piecewise right continuous function $N(\cdot):[k^{'},\infty)\to\mathbb{R}$, $k_0>k^{'}$, that satisfies (\ref{pf25}) and (\ref{pf26}).
\begin{align}
\underset{k>k_0}{\mathrm{sup}}\,\,\frac{1}{k-k_0} \int \limits_{k_0}^k N(\tau)d\tau=+\infty, \label{pf25} \\
\underset{k>k_0}{\mathrm{inf}}\,\,\frac{1}{k-k_0} \int \limits_{k_0}^k N(\tau)d\tau=-\infty \label{pf26}.
\end{align}
Here, $k_o\in(k^{'},\infty)$. In this work a Nussbaum type switching function has been implemented using the Mittag-Leffler (ML) function, described by (\ref{ml}).  
\begin{align}
E_{\alpha}(\rho)=\sum_{k=0}^{\infty}\frac{\rho^k}{\Gamma(k\alpha+1)}, \label{ml}
\end{align}
%{[Search for different letters for z here, change modified observer to proposed observer] DONE!}
Here $\Gamma(\rho+1)=\rho\Gamma(\rho), \rho>0$ is the standard Gamma function. The Nussbaum switching function of ML type is employed in this work and in \cite{Daniyal,Access} for UAS based adaptation strategy. If $\alpha\in(2,3]$ and $\lambda>0$ then the ML function $E_{\alpha}(-\lambda t^{\alpha})$ is a Nussbaum function \cite{Nussbaum}. The MATLAB implementation of an ML type Nussbaum switching function can be found in \cite{ML}. In the section \ref{sec3}, a proposed UAS observer-based Li-ion battery model parameter estimator is described for accurate estimation of battery model parameters $r_1,\cdots, r_{21}$. 

\section{Proposed Adaptive Parameters Estimation Methodology of a Li-on Battery Model} \label{sec3}
This section first provides the formulation details and Algorithm to implement UAS based APE strategy. Whereas, the second section describes the operational flow of our proposed methodology.

\subsection{Proposed UAS based battery parameters estimation methodology}
A High gain adaptive estimator for a Li-ion battery model, based on (\ref{pf1})-(\ref{pf4}), is described by (\ref{pf11})-(\ref{pf14}). 
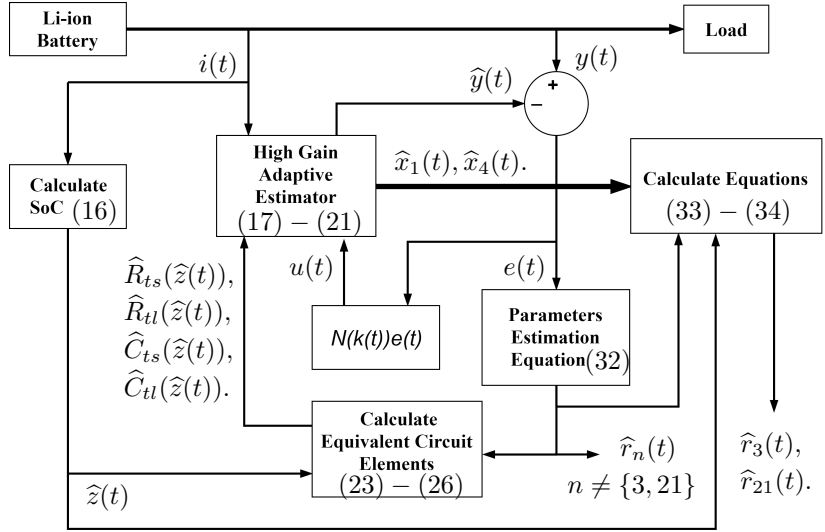
\begin{figure*}[t!]
	\centering
	\begin{overpic}[scale=0.1, unit=0.5mm]{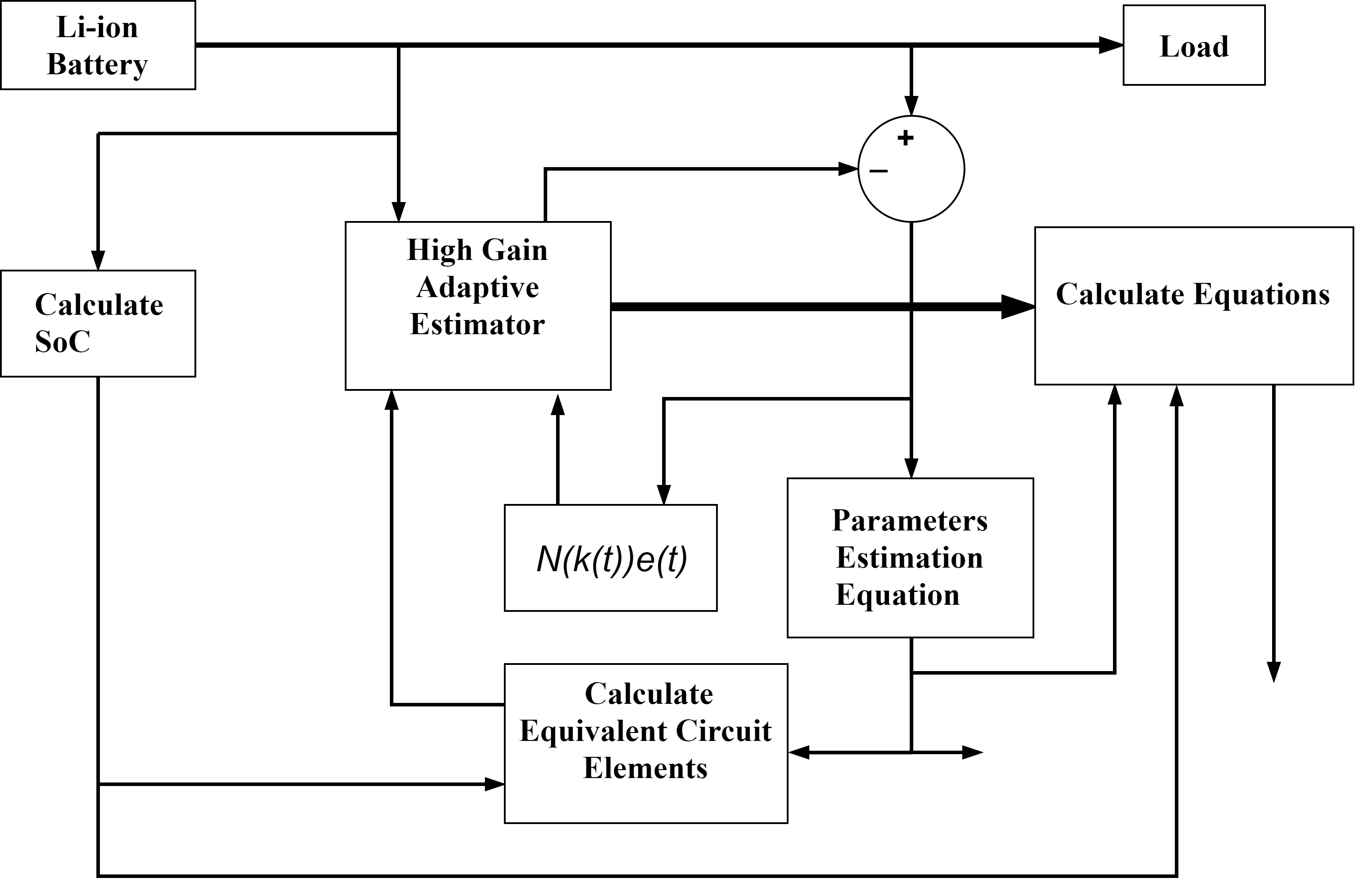}
		
		\put(14,85){ $(\ref{pf11})$ }
		
		\put(150.5,45){ $(\ref{adeq})$ }
		
		\put(58,82){ $(\ref{pf15})-(\ref{pf14})$ }
		
		\put(172,85){ $(\ref{ls1})-(\ref{ls2})$ }
		
		\put(85,12){ $(\ref{pf16})-(\ref{pf19})$ }
		
		%%%%%%%%%%%%%%%%%%%%%%%%%%%%%%%
		
		\put(48,124){ $i(t)$ }
		
		\put(72,70){ $u(t)$ }
		
		\put(120,120){ $\widehat{y}(t)$ }
		
		\put(148,126){ $y(t)$ }
		
		\put(18,9){ $\widehat{z}(t)$ }
		
		\put(28,68){ $\widehat{R}_{ts}(\widehat{z}(t)),$ }
		
		\put(28,58){ $\widehat{R}_{tl}(\widehat{z}(t)),$ }
		
		\put(28,48){ $\widehat{C}_{ts}(\widehat{z}(t)),$ }
		
		\put(28,38){ $\widehat{C}_{tl}(\widehat{z}(t)).$ }
		
		\put(129,70){ $e(t)$ }
		
		\put(160,22){ $\widehat{r}_n(t)$ }
		
		\put(146,12){ $n \neq \{3,21\}$ }
		
		\put(192,23){ $\widehat{r}_3(t),$ }
		
		\put(192,13){ $\widehat{r}_{21}(t).$ }
		
		\put(100,98){ $\widehat{x}_1(t), \widehat{x}_4(t).$ }

	\end{overpic}
	\caption{Flowchart of online UAS based adaptive parameters estimation of a Li-ion battery model.}
	\label{figBD}
\end{figure*}

\begin{align}
\dot{\widehat{z}}(t)& =-\frac{1}{C_c}i(t) , \,\,\,\, C_c=3600Cf_1 f_2 f_3 \label{pf11} \\
\dot{\widehat{x}}_1(t)& = \frac{\partial \widehat{x}_1(\widehat{z})}{\partial \widehat{z}(t)}\dot{\widehat{z}}(t) - u(t), \widehat{x}_1(t) \geq 0 \mbox{, therefore giving}\nonumber \\
\dot{\widehat{x}}_1(t)& = -\bigg(\widehat{r}_1\widehat{r}_2e^{-\widehat{r}_2\widehat{z}}+\widehat{r}_4-2\widehat{r}_5\widehat{z}+3\widehat{r}_6\widehat{z}^2\bigg)\frac{i(t)}{C_c} - u(t), \,\,\,\,  \label{pf15} \\
\dot{\widehat{x}}_2(t)&=-\frac{\widehat{x}_2(t)}{\widehat{R}_{ts}(\widehat{z})\widehat{C}_{ts}(\widehat{z})} +\frac{i(t)}{\widehat{C}_{ts}(z)}~
+u(t), \,\,\,\, \widehat{x}_2(t) \geq 0 \label{pf12}
\end{align}
\begin{align}
\dot{\widehat{x}}_3(t) &=-\frac{\widehat{x}_3(t)}{\widehat{R}_{tl}(\widehat{z})\widehat{C}_{tl}(\widehat{z})} +\frac{i(t)}{\widehat{C}_{tl}(z)}~
+u(t), \,\,\,\, \widehat{x}_3(t) \geq 0 \label{pf13} \\
\dot{\widehat{x}}_4(t)& = \frac{\partial \widehat{x}_4(\widehat{z}(t))}{\partial \widehat{z}(t)}\dot{\widehat{z}}(t) + u(t)\mbox{, therefore giving}\nonumber \\
\dot{\widehat{x}}_4(t) & = \bigg(\widehat{r}_{19}\widehat{r}_{20}e^{-\widehat{r}_{20}\widehat{z}} \bigg)\frac{i(t)}{C_c} + u(t), \,\,\,\, \widehat{x}_4(t) \geq 0 \label{pf20}\\
\widehat{y}(t)&=\widehat{x}_1(t)-\widehat{x}_2(t)-\widehat{x}_3(t)-i(t)\widehat{x}_4(t) \label{pf14}
\end{align}
Here $i(t)$ is the actual battery current and $\widehat{z}(t)$ is the estimated SOC, which is the same as $z(t)$ in (\ref{pf1}). The states $\widehat{x}_1$, $\widehat{x}_2$, $\widehat{x}_3$, and $\widehat{x}_4$ denote the estimates of open circuit voltage, voltage across $\widehat{R}_{ts}||\widehat{C}_{ts}$, $\widehat{R}_{tl}||\widehat{C}_{tl}$, and estimated series resistance respectively. For simplicity, the values of $f_1$, $f_2$, $f_3$ are taken as 1 in this work. The estimated voltage is represented by $\widehat{y}(t)$, whereas the estimated circuit elements are given by (\ref{pf161})-(\ref{pf191}).
\begin{align}
&\widehat{E}_o(\widehat{z}) = -\widehat{r}_1e^{-\widehat{r}_2\widehat{z}}+\widehat{r}_3+\widehat{r}_4\widehat{z}-\widehat{r}_5\widehat{z}^2+\widehat{r}_6\widehat{z}^3 = \widehat{x}_1(\widehat{z})  \label{pf161}\\
&\widehat{R}_{ts}(\widehat{z})=\widehat{r}_7e^{-\widehat{r}_8\widehat{z}}+\widehat{r}_9 \label{pf16}\\
&\widehat{R}_{tl}(\widehat{z})=\widehat{r}_{10}e^{-\widehat{r}_{11}\widehat{z}}+\widehat{r}_{12} \label{pf17} \\
&\widehat{C}_{ts}(\widehat{z})=-\widehat{r}_{13}e^{-\widehat{r}_{14}\widehat{z}}+\widehat{r}_{15} \label{pf18} \\
&\widehat{C}_{tl}(\widehat{z})=-\widehat{r}_{16}e^{-\widehat{r}_{17}\widehat{z}}+\widehat{r}_{18} \label{pf19} \\
&\widehat{R}_s(\widehat{z}) = \widehat{r}_{19}e^{-\widehat{r}_{20}\widehat{z}}+\widehat{r}_{21} = \widehat{x}_4(\widehat{z}). \label{pf191}
\end{align}
The control input $u(t)$ of UAS based-observer is designed by employing (\ref{pf21})-(\ref{pf24}).
\begin{align}
&e(t)=y(t)-\widehat{y}(t), \label{pf21} \\
&\dot{k}(t)=e^2(t), \,\,\,\, k(t_0)=k_0 \label{pf22} \\
&N(k(t))=E_{\alpha}(-\lambda k(t)^{\alpha}), \label{pf23} \\
&u(t)=-N(k(t))e(t). \label{pf24}
\end{align}
{In this work, the value of $\alpha=2.5$, and $\lambda=1$ are chosen by inspection.} The adaptive equation for battery parameters estimation from \cite{Daniyal,Access}, is given by (\ref{adeq}).
\begin{align}
&\dot{\widehat{r}}_n(t)=e^2(t)+ \lambda_{x_n}(r_{n_u}-\widehat{r}_n(t))+ \lambda_{y_n}(r_{n_l}-\widehat{r}_n(t)). \label{adeq}
\end{align}

\begin{algorithm*}[!t]
	\caption{Online UAS based algorithm for real-time adaptive parameters estimation of a Li-ion battery.}
	\label{algo2}
	\textbf{Requirements:} Data acquisition circuit to measure the terminal voltage $y(t)$ and current $i(t)$ of a Li-ion battery.  \\
	\textbf{Data:} 
	Initial values $\widehat{r}_n(0)$, upper bounds $r_{n_u}$, lower bounds $r_{n_l}$, confidence levels $\lambda_{x_n}$, and $\lambda_{y_n}$ for $n \in \{1,2,\cdots,21\} \backslash \{3,21\}$. Satisfying \textit{Lemma 4.1}. Initial states $\widehat{x}_1(0) = E_o(0)$, $\widehat{x}_2(0) = 0$, $\widehat{x}_3(0) = 0$, $\widehat{x}_4(0) = 0$, and $\widehat{y}(0) = y(0)$. A small positive tracking error bound $\epsilon$. Battery capacity value $C_c$(Ah).
	\newline
	\textbf{Output:} Estimated Li-ion battery model parameters $\widehat{r}_1(t)$, $\widehat{r}_2(t),\cdots,$ $\widehat{r}_{21}(t)$. \\
	\noindent\rule{17.5cm}{0.4pt}
	
	\begin{algorithmic}[1]
		
		\For{\textit{$t = t_0:t_{step}:t_{end}$}} 
		\State Read battery terminal voltage $y(t)$ and current $i(t)$.
		\State Update the error $e(t)$ using (\ref{pf21}).
		\State Estimate battery SoC value $\widehat{z}(t)$ using (\ref{pf11}).
		\State Evaluate (\ref{adeq}) for $\widehat{r}_n(t)$ estimation, where $n \in \{1,2,\cdots,21\} \backslash \{3,21\}$. 
		\State Calculate equivalent circuit element $\widehat{R}_{ts}(\widehat{z})$, $\widehat{R}_{tl}(\widehat{z})$, $\widehat{C}_{ts}(\widehat{z})$, $\widehat{C}_{tl}(\widehat{z})$ using (\ref{pf16})-(\ref{pf19}).
		\State Find $u(t)$ using (\ref{pf24}).
		\State Estimate the states $\widehat{x}_1(\widehat{t})$, $\widehat{x}_2(t)$, $\widehat{x}_3(t)$, $\widehat{x}_4(\widehat{t})$ using (\ref{pf15})-(\ref{pf20}).
		\State Estimate the terminal voltage $\widehat{y}(t)$ using (\ref{pf14}).
		\State Update the error $e(t)$ using (\ref{pf21}).
		\If{($|e(t)| < \epsilon$)}
		\If{$\Bigg[\widehat{r}_{14}(t) > -\dfrac{1}{\widehat{z}(t)} \ln{\bigg(\dfrac{\widehat{r}_{15}(t)}{\widehat{r}_{13}(t)}\bigg)}\Bigg]$ \textbf{and} $\Bigg[\widehat{r}_{17}(t) > -\dfrac{1}{\widehat{z}(t)} \ln{\bigg(\dfrac{\widehat{r}_{18}(t)}{\widehat{r}_{16}(t)}\bigg)}\Bigg]$}
		\State  Solve (\ref{ls1}) and (\ref{ls2}) to get $\widehat{r}_3(t)$ and $\widehat{r}_{21}(t)$.
		\State \textbf{Return} $\big[\widehat{r}_1(t), \widehat{r}_2(t),\cdots,\widehat{r}_{21}(t)\big]$.
		\Else
		\State \textbf{Continue} loop execution.
		\EndIf
		\Else
		\State \textbf{Continue} loop execution.
		\EndIf
		\EndFor
	\end{algorithmic}
\end{algorithm*}
The adaptive equation (\ref{adeq}) requires a steady-state upper bound $r_{n_u}$ and a lower bound $r_{n_l}$ for each estimated parameter $\widehat{r}_n(t)$, $n \in \{1,2,\cdots,21\} \backslash \{3,21\}$, and user's confidence levels, $\lambda_{x_n}$ and $\lambda_{y_n}$, on the upper and lower bounds respectively. It is shown in Lemma \ref{lemma4.3} that the positive real values of $r_{n_u}$, $r_{n_l}$, $\lambda_{x_n}$, and $\lambda_{y_n}$ leads to $\widehat{r}(t) > 0$, for $t>t_0$. The flowchart of proposed online APE method for Li-ion battery parameters estimation is shown in Figure \ref{figBD}. Note that the UAS based parameters estimation method, explained above, is capable of estimating the battery parameters $n \in \{1,2,\cdots,21\} \backslash \{3,21\}$. The estimates of $\widehat{r}_3$ and $\widehat{r}_{21}$ can be obtained, during or after the adaptation process, by applying the least squares estimation or curve fitting techniques on (\ref{pf161}) and (\ref{pf191}) respectively. However, this work uses a direct approach to estimate $\widehat{r}_3$ and $\widehat{r}_{21}$, during the adaptation process.
Our approach to estimate $\widehat{r}_3$ and $\widehat{r}_{21}$ is based on the results of \textit{Theorem 4.2}. In \textit{Theorem 4.2}, it is shown that
$\widehat{x}_1(t) \to x_1(t)$ and $\widehat{x}_4(t) \to x_4(t)$ as $t \to \infty$, and convergence of $\widehat{r}_n \to r_n$, where $n \in \{1,2,\cdots,21\} \backslash \{3,21\}$ as $t \to \infty$ respectively. Thus, $\widehat{x}_1(t) \to x_1(t)$ and $\widehat{x}_4(t) \to x_4(t)$ at $t \to \infty$ lets us write the equations (\ref{pf161}) and (\ref{pf191}) into (\ref{ls1}) and (\ref{ls2}) form to estimate $\widehat{r}_3$ and $\widehat{r}_{21}$ respectively.
\begin{align}
&\widehat{r}_3 = x_1(t) + \widehat{r}_1e^{-\widehat{r}_2\widehat{z}} -\widehat{r}_4\widehat{z} +\widehat{r}_5\widehat{z}^2 -\widehat{r}_6\widehat{z}^3, \label{ls1} \\
&\widehat{r}_{21} = x_4(t) -\widehat{r}_{19}e^{-\widehat{r}_{20}\widehat{z}}. \label{ls2}
\end{align}

The steps to implement UAS based adaptation methodology for battery model parameters estimation are described in algorithm 1. In the following subsection, the flowchart of algorithm 1 is presented and transcribed.

\subsection{Proposed algorithm for on-line Li-ion battery model parameters estimation}

This section provides the details of our proposed UAS based adaptation algorithm to estimate Li-ion battery model parameters. The flowchart of the algorithm 1 is shown in Figure \ref{figBD}. The UAS based adaptation process begins with the measurement of current and voltage of a Li-ion battery. A small positive current needs to be maintained during the adaptation, as per Theorem 4.2, for accurate results. The error between actual and estimated terminal voltages is used by UAS and the adaptive estimation equation in (\ref{adeq}) to identify $\widehat{r}_n(t)$, where $n \in \{1,2,\cdots,21\} \backslash \{3,21\}$. These estimated parameters are employed to calculate the equivalent circuit elements. Next, the equivalent circuit elements' estimates, together with the output of UAS and current are input to high gain adaptive estimator. The adaptation process ends with estimation of the states $\widehat{x}_1(\widehat{t})$, $\widehat{x}_2(t)$, $\widehat{x}_3(t)$, $\widehat{x}_4(\widehat{t})$, followed by terminal voltage estimation error update defined by \eqref{pf21}. When the error magnitude goes below the user's defined threshold during the adaptation, the estimated states approach to actual states of a Li-ion battery model, as per Theorem 4.2. Thereafter, the convergence of estimated states to their actual values allows us use equation (\ref{ls1}) and (\ref{ls2}) for identification of $\widehat{r}_3(t)$ and $\widehat{r}_{21}(t)$. In the following section, we provide mathematical justification of our proposed online UAS based adaptation strategy for a Li-ion battery model parameters estimation.

\section{Mathematical Justification} \label{sec4}
This section first proves the convergence of the terminal voltage estimation error $e(t)$ to zero. The proof of $e(t) \to 0$ as $t \to \infty$ provides the following results: $\widehat{x}_1(t) \to x_1(t)$, $\widehat{R}_{ts}(\widehat{z})\widehat{C}_{ts}(\widehat{z}) \to R_{ts}(z)C_{ts}(z)$, $\widehat{R}_{tl}(\widehat{z})\widehat{C}_{tl}(\widehat{z}) \to R_{tl}(z)C_{tl}(z)$, and $\widehat{x}_4(t) \to x_4(t)$ as $t \to \infty$. Further analysis of the  results above, leads to the conclusion that the proposed method can accurately estimate the Li-ion battery model parameters. Before proving the above results, some criteria for $\lambda_{x_n}, \lambda_{y_n}, r_{n_u},$ and $r_{n_l}$ selection needs to be established in Lemma \ref{lemma4.1}.

\begin{lem}\label{lemma4.1}
	Suppose $\lambda_{x_n}, \lambda_{y_n}, r_{n_u}$, and $r_{n_l}$ are the positive real numbers for $n=\{13,15,16,18\}$, and $\widehat{z}(t) \in (0,1]$, then the following conditions hold for all $t > t_0$.
	\begin{itemize}
		\item \textbf{If} $\widehat{r}_{13}(t_0) > \widehat{r}_{15}(t_0) > 0$, $\lambda_{x_{15}} + \lambda_{y_{15}} > \lambda_{x_{13}} + \lambda_{y_{13}}$, $\lambda_{x_{15}}\widehat{r}_{15_u} + \lambda_{y_{15}}\widehat{r}_{15_l} <  \lambda_{x_{13}}\widehat{r}_{13_u} + \lambda_{y_{13}}\widehat{r}_{13_l}$, and $\widehat{r}_{14}(t) > -\dfrac{1}{\widehat{z}(t)} \ln{\bigg(\dfrac{\widehat{r}_{15}(t)}{\widehat{r}_{13}(t)}\bigg)}$, \textbf{then} $\widehat{C}_{ts}({\widehat{z}(t)}) > 0$.
		
		\item \textbf{If} $\widehat{r}_{16}(t_0) > \widehat{r}_{18}(t_0) > 0$, $\lambda_{x_{18}} + \lambda_{y_{18}} > \lambda_{x_{16}} + \lambda_{y_{16}}$, $\lambda_{x_{18}}\widehat{r}_{18_u} + \lambda_{y_{18}}\widehat{r}_{18_l} <  \lambda_{x_{16}}\widehat{r}_{16_u} + \lambda_{y_{16}}\widehat{r}_{16_l}$, and $\widehat{r}_{17}(t) > -\dfrac{1}{\widehat{z}(t)} \ln{\bigg(\dfrac{\widehat{r}_{18}(t)}{\widehat{r}_{16}(t)}\bigg)}$ \textbf{then} $\widehat{C}_{tl}({\widehat{z}(t)}) > 0$.
	\end{itemize}
\end{lem}
The detailed proof of Lemma \ref{lemma4.1} is available in \cite{Daniyal}. The conditions established in Lemma \ref{lemma4.1} are utilized in the following theorem to prove the convergence of terminal voltage error $e(t)$ to zero, which leads to the convergence of estimated values of the circuit elements to actual ones. \qed
\begin{remark}\label{prevwork}
	\normalfont The novelty of the mathematical development presented in this work  in comparison to the earlier work \cite{Daniyal} is the following. In \cite{Daniyal}, the battery series resistance and battery open circuit voltage are not included as states in the observer. As a result, in \cite{Daniyal} parameters related to the battery series resistance and battery open circuit voltage cannot be estimated online but needs pre/post processing of data. In the current work, not only are the battery series resistance and battery open circuit voltage included as states in the proposed observer, but also the parameters $\hat{r}_{1},\cdots,\hat{r}_{6}$ of the battery series resistance in \eqref{pf16} and parameters $\hat{r}_{19},\cdots,\hat{r}_{21}$ of the open circuit voltage, in \eqref{pf191} are estimated online. This requires introducing additional states in the proposed observer formulation, and makes the mathematics in this work, much more involved compared to \cite{Daniyal}.
\end{remark}	

\begin{thm}\label{thm4.2}
	Let 	$A=\begin{bmatrix}
	1 & -1 & -1 & -1
	\end{bmatrix}$, $\mathbf{x}= \left(
	\begin{bmatrix}
	E_o(z(t)) & x_2(t) & x_3(t) & i(t)R_s(z(t))
	\end{bmatrix}^T - \right.$\\
	$\left. \begin{bmatrix}
	\widehat{x}_1(\widehat{z}(t)) & \widehat{x}_2(t) & \widehat{x}_3(t) & i(t)\widehat{x}_4(\widehat{z}(t))
	\end{bmatrix}^T \right)$. Suppose that conditions needed for Lemma \ref{lemma4.1} to hold are satisfied, and assuming there is no non-zero vector $\mathbf{x}$ in the nullspace of $A$. \textbf{If} the Li-ion battery discharge current $i(t)$ is a small positive value for $t>t_0$ \textbf{then} the following are obtained as $t \to \infty$
	\begin{itemize}
		\item $\widehat{x}_1(t) = x_1(t)$, 
		\item $\widehat{R}_{ts}(\widehat{z})\widehat{C}_{ts}(\widehat{z}) = R_{ts}(\widehat{z})C_{ts}(\widehat{z})$,
		\item $\widehat{R}_{tl}(\widehat{z})\widehat{C}_{tl}(\widehat{z}) = R_{tl}(\widehat{z})C_{tl}(\widehat{z})$,
		\item  $\widehat{x}_4(t) = x_4(t)$.
	\end{itemize} 
\end{thm}
\myproof
Suppose the assumptions mentioned above are satisfied.
Take the time derivative of (\ref{pf21}) to get
\begin{align}
\dot{e}(t) = \dot{y}(t)-\dot{\widehat{y}}(t), \label{pf29}
\end{align}
Addition and subtraction of $e(t)$ to R.H.S of (\ref{pf29}), and recognizing that $e(t) = y(t)-\widehat{y}(t)$ provides
\begin{align}
\dot{e}(t) = -e(t) +y(t) - \widehat{y}(t) + \dot{y}(t)-\dot{\widehat{y}}(t). \label{pf31}
\end{align}
Now, substitution of $-\widehat{y}(t)$ and $\dot{\widehat{y}}(t)$ from (\ref{pf14}) in (\ref{pf31}) provides
%\begin{equation}
\begin{align}
\dot{e}(t) =& -e(t) +y(t) + \dot{y}(t) - \widehat{x}_1(\widehat{z}(t))+\widehat{x}_2(t)+\widehat{x}_3(t)\nonumber\\
&+i(t)\widehat{x}_4(\widehat{z}(t))-\dot{\widehat{x}}_1(\widehat{z}(t))+\dot{\widehat{x}}_2(t)+\dot{\widehat{x}}_3(t)\nonumber\\
&+\frac{di(t)}{dt}\widehat{x}_4(\widehat{z}(t)) + i(t)\dot{\widehat{x}}_4(\widehat{z}(t))\label{pf32}
\end{align}
%\end{equation}
Using (\ref{pf12}) and (\ref{pf13}) in (\ref{pf32}) gives
%\begin{equation}
\begin{align}
\dot{e}(t) =& -e(t) +y(t) + \dot{y}(t) - \widehat{x}_1(\widehat{z}(t))+\widehat{x}_2(t)+\widehat{x}_3(t)\nonumber\\
&+ i(t)\widehat{x}_4(\widehat{z}(t))- \dot{\widehat{x}}_1(\widehat{z}(t))-\frac{\widehat{x}_2(t)}{\widehat{R}_{ts}(\widehat{z}(t))\widehat{C}_{ts}(\widehat{z}(t))}\nonumber\\
&-\frac{\widehat{x}_3(t)}{\widehat{R}_{tl}(\widehat{z}(t))\widehat{C}_{tl}(\widehat{z}(t))}   + \frac{i(t)}{\widehat{C}_{ts}(\widehat{z}(t))} + \frac{i(t)}{\widehat{C}_{tl}(\widehat{z}(t))} \nonumber\\
&+2u(t) +\frac{di(t)}{dt}\widehat{x}_4(\widehat{z}(t)) + i(t)\dot{\widehat{x}}_4(\widehat{z}(t)). \label{pf33}
\end{align}
%\end{equation}
Re-arrangement of  (\ref{pf33}) yields the following
\begin{small}
	%\begin{equation}
	\begin{align}
	\dot{e}(t) =& -e(t) +y(t) + \dot{y}(t)+\widehat{x}_2(t)\left(1-\frac{1}{\widehat{R}_{ts}(\widehat{z}(t))\widehat{C}_{ts}(\widehat{z}(t))}\right)\nonumber\\
	& +\widehat{x}_3(t)\bigg(1-\frac{1}{\widehat{R}_{tl}(\widehat{z}(t))\widehat{C}_{tl}(\widehat{z}(t))}\bigg)\nonumber \\& - \widehat{x}_1(\widehat{z}(t))+i(t)\widehat{x}_4(\widehat{z}(t))-\dot{\widehat{x}}_1(\widehat{z}(t))\nonumber\\
	&+i(t)\bigg(\frac{1}{\widehat{C}_{ts}(\widehat{z}(t))}+\frac{1}{\widehat{C}_{tl}(\widehat{z}(t))}\bigg) +\frac{di(t)}{dt}\widehat{x}_4(\widehat{z}(t))\nonumber \\
	&+ i(t)\dot{\widehat{x}}_4(\widehat{z}(t)) + 2u(t).\label{pf34}
	\end{align}
	%\end{equation}
\end{small}
Since by definition of (\ref{pf16}) and (\ref{pf17}), $\widehat{R}_{ts}(\widehat{z}(t)) > 0$, $\widehat{R}_{tl}(\widehat{z}(t)) > 0$ for all $t> t_0$. Also by Lemma \ref{lemma4.1}, we know that $\widehat{C}_{ts}(\widehat{z}(t)) > 0$ and $\widehat{C}_{tl}(\widehat{z}(t)) > 0$ for all $t> t_0$. Therefore, $\widehat{R}_{ts}(\widehat{z}(t))\widehat{C}_{ts}(\widehat{z}(t)) > 0$ and $\widehat{R}_{tl}(\widehat{z}(t))\widehat{C}_{tl}(\widehat{z}(t)) > 0$.
\begin{align}
\text{which implies} \,\,\, &1-\frac{1}{\widehat{R}_{ts}(\widehat{z}(t))\widehat{C}_{ts}(\widehat{z}(t))} < 1, \label{pf36}
\end{align}
From (\ref{pf12}), $\widehat{x}_2(t) \geq 0$, Thus,
\begin{align} &\widehat{x}_2(t)\bigg(1-\frac{1}{\widehat{R}_{ts}(\widehat{z}(t))\widehat{C}_{ts}(\widehat{z}(t))}\bigg) \leq \widehat{x}_2(t), \label{pf37}
\end{align}
Similarly using $\widehat{x}_3(t) \geq 0$ from (\ref{pf13}) provides
\begin{align}
\widehat{x}_3(t)\bigg(1-\frac{1}{\widehat{R}_{tl}(\widehat{z}(t))\widehat{C}_{tl}(\widehat{z}(t))}\bigg) \leq \widehat{x}_3(t),  \label{pf38}
\end{align}
From (\ref{pf37}) and (\ref{pf38}) we get
\begin{align}  \label{pf39}
\widehat{x}_2(t) +\widehat{x}_3(t) \geq &
{x}_2(t)\bigg(1-\frac{1}{\widehat{R}_{ts}(\widehat{z}(t))\widehat{C}_{ts}(\widehat{z}(t))}\bigg) \nonumber\\ &+\widehat{x}_3(t)\bigg(1-\frac{1}{\widehat{R}_{tl}(\widehat{z}(t))\widehat{C}_{tl}(\widehat{z}(t))}\bigg). 
\end{align}
Using (\ref{pf39}) in (\ref{pf34}) and re-arrangement of terms provides the following
%\begin{equation}
\begin{align}
\dot{e}(t) \leq& -e(t) +y(t) + \dot{y}(t) - \widehat{x}_1(\widehat{z}(t))+\widehat{x}_2(t)+\widehat{x}_3(t)\nonumber\\
&+i(t)\widehat{x}_4(\widehat{z}(t))- \dot{\widehat{x}}_1(\widehat{z}(t))+\frac{di(t)}{dt}\widehat{x}_4(\widehat{z}(t))\nonumber\\
&+ i(t)\dot{\widehat{x}}_4(\widehat{z}(t))+i(t)\bigg(\frac{1}{\widehat{C}_{ts}(\widehat{z}(t))}+\frac{1}{\widehat{C}_{tl}(\widehat{z}(t))}\bigg)\nonumber\\
&+ 2u(t). \label{pf40}
\end{align}
%\end{equation}
Simplification of (\ref{pf40}) using (\ref{pf14}) and (\ref{pf21}) gives
\begin{align}
\dot{e}(t) \leq& -y(t) + \widehat{y}(t) +y(t) + \dot{y}(t) -\widehat{y}(t)-\dot{\widehat{x}}_1(\widehat{z}(t))\nonumber\\ &+\frac{di(t)}{dt}\widehat{x}_4(\widehat{z}(t)) + i(t)\dot{\widehat{x}}_4(\widehat{z}(t))\nonumber\\
&+i(t)\bigg(\frac{1}{\widehat{C}_{ts}(\widehat{z}(t))}+\frac{1}{\widehat{C}_{tl}(\widehat{z}(t))}\bigg) +2u(t), \mbox{ i.e.}\label{pf41}
\end{align}
\begin{align}
\dot{e}(t) \leq&~ \dot{y}(t)-\dot{\widehat{x}}_1(\widehat{z}(t))+\frac{di(t)}{dt}\widehat{x}_4(\widehat{z}(t)) + i(t)\dot{\widehat{x}}_4(\widehat{z}(t))\nonumber \\
&+i(t)\bigg(\frac{1}{\widehat{C}_{ts}(\widehat{z}(t))}+\frac{1}{\widehat{C}_{tl}(\widehat{z}(t))}\bigg) +2u(t). \label{pf42}
\end{align}
Using (\ref{pf15}) and (\ref{pf20}) in (\ref{pf42}) gives

\begin{align}
\dot{e}(t) \leq &~ \dot{y}(t) - \frac{\partial \widehat{x}_1(\widehat{z}(t))}{\partial \widehat{z}(t)}\dot{\widehat{z}}(t) +\frac{di(t)}{dt}\widehat{x}_4(\widehat{z}(t))\nonumber\\
&+ i(t)\frac{\partial \widehat{x}_4(\widehat{z}(t))}{\partial \widehat{z}(t)}\dot{\widehat{z}}(t) + i(t)u(t)\nonumber\\ &+i(t)\bigg(\frac{1}{\widehat{C}_{ts}(\widehat{z}(t))}+\frac{1}{\widehat{C}_{tl}(\widehat{z}(t))}\bigg) + 3u(t).\label{pf_42}
\end{align}
Following this, the proof of error $e(t)$ convergence to zero is derived from the equation (\ref{pf_42}). We will consider two cases of error, i.e. $e(t)$ can be either positive or negative, and each case produces a particular form. This particular form in both cases lets us show that $e(t) \to 0$ as $t \to \infty$. Prior to considering the case of positive or negative error, some inequalities are required to be established. \\
Consider the following inequality related to $e(t)$ and the first term of R.H.S of (\ref{pf_42}),
%Expansion of (\ref{pf45})-(\ref{pf488}) gives
\begin{flalign}
\bigg(e(t)-\dot{y}(t)\bigg)^2 &\geq 0,   \nonumber \\
\frac{1}{2}e^2(t) + \frac{1}{2}\dot{y}^2(t) &\geq e(t)\dot{y}(t).  \label{pf49}
\end{flalign}
The inequality related to $e(t)$ and the second term of R.H.S of (\ref{pf_42}) is as follows, 
\begin{small}
	\begin{flalign}
	\bigg(e(t)+\frac{\partial \widehat{x}_1(\widehat{z}(t))}{\partial \widehat{z}(t)}\dot{\widehat{z}}(t)\bigg)^2 &\geq 0,  \nonumber \\
	\frac{1}{2}e^2(t) + \frac{1}{2}\bigg(\frac{\partial \widehat{x}_1(\widehat{z}(t))}{\partial \widehat{z}(t)}\dot{\widehat{z}}(t)\bigg)^2 &\geq -e(t)\frac{\partial \widehat{x}_1(\widehat{z}(t))}{\partial \widehat{z}(t)}\dot{\widehat{z}}(t).  \label{pf50}
	\end{flalign}
\end{small}
The inequality related to $e(t)$ and the third term of R.H.S of (\ref{pf_42}) is given as, 
\begin{flalign}
\bigg(e(t)-\frac{di(t)}{dt}\widehat{x}_4(\widehat{z}(t))\bigg)^2 &\geq 0,  \nonumber \\
\frac{1}{2}e^2(t) + \frac{1}{2}\bigg(\frac{di(t)}{dt}\bigg)^2\widehat{x}^2_4(\widehat{z}(t)) &\geq e(t)\frac{di(t)}{dt}\widehat{x}_4(\widehat{z}(t)).  \label{pf51}
\end{flalign} 
The inequality related to $e(t)$ and the fourth term of R.H.S of (\ref{pf_42}) is as follows, 
\begin{small}
	\begin{align} 
	\bigg(e(t)-i(t)\frac{\partial \widehat{x}_4(\widehat{z}(t))}{\partial \widehat{z}(t)}\dot{\widehat{z}}(t)\bigg)^2 &\geq 0,   \nonumber \\
	\frac{1}{2}e^2(t) + \frac{1}{2}i^2(t)\bigg(\frac{\partial \widehat{x}_4(\widehat{z}(t))}{\partial \widehat{z}(t)}\dot{\widehat{z}}(t)\bigg)^2 &\geq e(t)i(t)\frac{\partial \widehat{x}_4(\widehat{z}(t))}{\partial \widehat{z}(t)}\dot{\widehat{z}}(t).  \label{pf52}
	\end{align}
\end{small} %\vspace{-1em}
The inequality related to $e(t)$ and the sixth term of R.H.S of (\ref{pf_42}) is given below, 
\begin{small}
	\begin{flalign} 
	\bigg(e(t)-i(t)\bigg(\frac{1}{\widehat{C}_{ts}(\widehat{z}(t))}+\frac{1}{\widehat{C}_{tl}(\widehat{z}(t))}\bigg)\bigg)^2 &\geq 0,   \nonumber \\
	\frac{1}{2}e^2(t) + \frac{1}{2}i^2(t)\bigg(\frac{1}{\widehat{C}_{ts}(\widehat{z}(t))}+\frac{1}{\widehat{C}_{tl}(\widehat{z}(t))}\bigg)^2 &\geq e(t)i(t)\times\nonumber\\ \bigg(\frac{1}{\widehat{C}_{ts}(\widehat{z}(t))}&+\frac{1}{\widehat{C}_{tl}(\widehat{z}(t))}\bigg). \label{pf522}
	\end{flalign}
\end{small}
From (\ref{pf49}), (\ref{pf50}), (\ref{pf51}), (\ref{pf52}), and (\ref{pf522}), we get (\ref{pfadd})
\begin{small}
	\begin{align}
	\frac{5}{2}e^2(t) + \frac{1}{2}\dot{y}^2(t) + \frac{1}{2}\bigg(\frac{\partial \widehat{x}_1(\widehat{z}(t))}{\partial \widehat{z}(t)}\dot{\widehat{z}}(t)\bigg)^2 + \frac{1}{2}\bigg(\frac{di(t)}{dt}\bigg)^2\widehat{x}^2_4(\widehat{z}(t))\nonumber\\ +\frac{1}{2}i^2(t)\bigg(\frac{\partial \widehat{x}_4(\widehat{z}(t))}{\partial \widehat{z}(t)}\dot{\widehat{z}}(t)\bigg)^2 +  \frac{1}{2}i^2(t)\bigg(\frac{1}{\widehat{C}_{ts}(\widehat{z}(t))}+\frac{1}{\widehat{C}_{tl}(\widehat{z}(t))}\bigg)^2\nonumber \\   \geq  \left(e(t)\dot{y}(t) -e(t)\frac{\partial \widehat{x}_1(\widehat{z}(t))}{\partial \widehat{z}(t)}\dot{\widehat{z}}(t)
	+ e(t)\frac{di(t)}{dt}\widehat{x}_4(\widehat{z}(t))\right.\nonumber \\\left. + e(t)i(t)\frac{\partial \widehat{x}_4(\widehat{z}(t))}{\partial \widehat{z}(t)}\dot{\widehat{z}}(t)
	+ e(t)i(t)\bigg(\frac{1}{\widehat{C}_{ts}(\widehat{z}(t))}+\frac{1}{\widehat{C}_{tl}(\widehat{z}(t))}\bigg)\right).  \label{pfadd} 
	\end{align}
\end{small}
Similarly, consider the following inequalities related to $e(t)$ and the first term of R.H.S of (\ref{pf_42}),
%Expansion of (\ref{pf45n})-(\ref{pf488n}) gives
\begin{flalign}
-\bigg(e(t)+\dot{y}(t)\bigg)^2 &\leq 0,   \nonumber \\
-\frac{1}{2}e^2(t) - \frac{1}{2}\dot{y}^2(t) &\leq e(t)\dot{y}(t). \label{pf45nn}
\end{flalign}
The inequality related to $e(t)$ and the second term of R.H.S of (\ref{pf_42}) is as follows,
\begin{small}
	\begin{flalign}
	-\bigg(e(t)-\frac{\partial \widehat{x}_1(\widehat{z}(t))}{\partial \widehat{z}(t)}\dot{\widehat{z}}(t)\bigg)^2 &\leq 0,   \nonumber \\
	-\frac{1}{2}e^2(t) - \frac{1}{2}\bigg(\frac{\partial \widehat{x}_1(\widehat{z}(t))}{\partial \widehat{z}(t)}\dot{\widehat{z}}(t)\bigg)^2 &\leq -e(t)\frac{\partial \widehat{x}_1(\widehat{z}(t))}{\partial \widehat{z}(t)}\dot{\widehat{z}}(t). \label{pf46nn}
	\end{flalign}
\end{small}
The inequality related to $e(t)$ and the third term of R.H.S of (\ref{pf_42}) is given as,
\begin{flalign}
-\bigg(e(t)+\frac{di(t)}{dt}\widehat{x}_4(\widehat{z}(t))\bigg)^2 &\leq 0,  \nonumber \\
-\frac{1}{2}e^2(t)  -\frac{1}{2}\bigg(\frac{di(t)}{dt}\bigg)^2\widehat{x}^2_4(\widehat{z}(t)) &\leq e(t)\frac{di(t)}{dt}\widehat{x}_4(\widehat{z}(t)). \label{pf47nn}
\end{flalign}
The inequality related to $e(t)$ and the fourth term of R.H.S of (\ref{pf_42}) is as follows,
\begin{small}
	\begin{flalign}
	-\bigg(e(t)+i(t)\frac{\partial \widehat{x}_4(\widehat{z}(t))}{\partial \widehat{z}(t)}\dot{\widehat{z}}(t)\bigg)^2 &\leq 0,  \nonumber \\
	-\frac{1}{2}e^2(t) - \frac{1}{2}i^2(t)\bigg(\frac{\partial \widehat{x}_4(\widehat{z}(t))}{\partial \widehat{z}(t)}\dot{\widehat{z}}(t)\bigg)^2 &\leq e(t)i(t)\frac{\partial \widehat{x}_4(\widehat{z}(t))}{\partial \widehat{z}(t)}\dot{\widehat{z}}(t).  \label{pf48nn}
	\end{flalign}
\end{small}
The inequality related to $e(t)$ and the sixth term of R.H.S of (\ref{pf_42}) is given below,
\begin{small}
	\begin{flalign}
	-\bigg(e(t)+i(t)\bigg(\frac{1}{\widehat{C}_{ts}(\widehat{z}(t))}+\frac{1}{\widehat{C}_{tl}(\widehat{z}(t))}\bigg)\bigg)^2 &\leq 0,  \nonumber\\
	-\frac{1}{2}e^2(t) - \frac{1}{2}i^2(t)\bigg(\frac{1}{\widehat{C}_{ts}(\widehat{z}(t))}+\frac{1}{\widehat{C}_{tl}(\widehat{z}(t))}\bigg)^2 &\leq e(t)i(t)\times \nonumber \\ \bigg(\frac{1}{\widehat{C}_{ts}(\widehat{z}(t))}+\frac{1}{\widehat{C}_{tl}(\widehat{z}(t))}\bigg).\label{pf488nn}
	\end{flalign}
\end{small}
From (\ref{pf45nn}), (\ref{pf46nn}), (\ref{pf47nn}), (\ref{pf48nn}), and (\ref{pf488nn}), we get (\ref{pfadd2}).

\begin{small}
	\begin{align}
	-\frac{5}{2}e^2(t) - \frac{1}{2}\dot{y}^2(t) - \frac{1}{2}\bigg(\frac{\partial \widehat{x}_1(\widehat{z}(t))}{\partial \widehat{z}(t)}\dot{\widehat{z}}(t)\bigg)^2 - \frac{1}{2}\bigg(\frac{di(t)}{dt}\bigg)^2\widehat{x}^2_4(\widehat{z}(t))  \nonumber \\ -\frac{1}{2}i^2(t)\bigg(\frac{\partial \widehat{x}_4(\widehat{z}(t))}{\partial \widehat{z}(t)}\dot{\widehat{z}}(t)\bigg)^2 - \frac{1}{2}i^2(t)\bigg(\frac{1}{\widehat{C}_{ts}(\widehat{z}(t))}+\frac{1}{\widehat{C}_{tl}(\widehat{z}(t))}\bigg)^2 \nonumber \\  \leq  \left(e(t)\dot{y}(t) -e(t)\frac{\partial \widehat{x}_1(\widehat{z}(t))}{\partial \widehat{z}(t)}\dot{\widehat{z}}(t)  + e(t)\frac{di(t)}{dt}\widehat{x}_4(\widehat{z}(t))\right. \nonumber \\ \left. + e(t)i(t)\frac{\partial \widehat{x}_4(\widehat{z}(t))}{\partial \widehat{z}(t)}\dot{\widehat{z}}(t)  + e(t)i(t)\bigg(\frac{1}{\widehat{C}_{ts}(\widehat{z}(t))}+\frac{1}{\widehat{C}_{tl}(\widehat{z}(t))}\bigg)\right). \label{pfadd2} 
	\end{align} 
\end{small}
In the following part, we will consider (\ref{pf_42}) with two cases of error, i.e. error being positive and negative, and utilize (\ref{pfadd}) and (\ref{pfadd2}) for the positive and negative error cases respectively to show the convergence of error $e(t)$.

\noindent \textbf{Case 1.} Consider $e(t) > 0$, at some instant $t > t_0$. Multiplying (\ref{pf_42}) by $e(t)$ and using (\ref{pf24}) gives

\begin{small}
	\begin{flalign}
	e(t)\dot{e}(t) \leq &e(t)\dot{y}(t) - e(t)\frac{\partial \widehat{x}_1(\widehat{z}(t))}{\partial \widehat{z}(t)}\dot{\widehat{z}}(t) +e(t)\frac{di(t)}{dt}\widehat{x}_4(\widehat{z}(t)) \nonumber \\ &+ e(t)i(t)\frac{\partial \widehat{x}_4(\widehat{z}(t))}{\partial \widehat{z}(t)}\dot{\widehat{z}}(t)  + e(t)i(t)\times \nonumber \\ &\bigg(\frac{1}{\widehat{C}_{ts}(\widehat{z}(t))} +\frac{1}{\widehat{C}_{tl}(\widehat{z}(t))}\bigg) -(3+i(t))N(k(t))e^2(t), \label{pf43new}
	\end{flalign}
\end{small}	
Now use (\ref{pfadd}) in (\ref{pf43new}) to get the following
\begin{small}
	\begin{flalign}
	e(t)\dot{e}(t) \leq  &\frac{5}{2}e^2(t) + \frac{1}{2}\dot{y}^2(t) + \frac{1}{2}\bigg(\frac{\partial \widehat{x}_1(\widehat{z}(t))}{\partial \widehat{z}(t)}\dot{\widehat{z}}(t)\bigg)^2 \nonumber \\&+ \frac{1}{2}\bigg(\frac{di(t)}{dt}\bigg)^2\widehat{x}^2_4(\widehat{z}(t))+ \frac{1}{2}i^2(t)\bigg(\frac{\partial \widehat{x}_4(\widehat{z}(t))}{\partial \widehat{z}(t)}\dot{\widehat{z}}(t)\bigg)^2  \nonumber \\ &+ \frac{1}{2}i^2(t)\bigg(\frac{1}{\widehat{C}_{ts}(\widehat{z}(t))}+\frac{1}{\widehat{C}_{tl}(\widehat{z}(t))}\bigg)^2 \nonumber \\ &- (3+i(t))N(k(t))e^2(t). \label{pf53new}
	\end{flalign}
\end{small}
Since $\frac{d}{dt}\big(\frac{1}{2}e^2(t)\big) = e(t)\dot{e}(t)$, thus integrating (\ref{pf53new}) from $t_0$ to $t$, and using (\ref{pf22}) provides

\begin{small}
	\begin{flalign} 
	\frac{1}{2}e^2(t) \leq& \frac{5}{2}(k(t)-k(t_0)) + \frac{1}{2}\int_{t_0}^{t}\dot{y}^2(\tau) d\tau \nonumber \\ &+ \frac{1}{2}\int_{t_0}^{t}\bigg(\frac{\partial \widehat{x}_1(\widehat{z}(\tau))}{\partial \widehat{z}(\tau)}\dot{\widehat{z}}(\tau)\bigg)^2 d\tau \nonumber \\& + \frac{1}{2}\int_{t_0}^{t}\bigg(\frac{di(\tau)}{d\tau}\bigg)^2\widehat{x}^2_4(\widehat{z}(\tau)) d\tau \nonumber \\&+ \frac{1}{2}\int_{t_0}^{t}i^2(\tau)\bigg(\frac{\partial \widehat{x}_4(\widehat{z}(\tau))}{\partial \widehat{z}(\tau)}\dot{\widehat{z}}(\tau)\bigg)^2 d\tau \nonumber \\&+ \frac{1}{2}\int_{t_0}^{t}i^2(\tau)\bigg(\frac{1}{\widehat{C}_{ts}(\widehat{z}(\tau))}+\frac{1}{\widehat{C}_{tl}(\widehat{z}(\tau))}\bigg)^2 d\tau \nonumber \\&   - 3\int_{t_0}^{t}N(k(\tau))\dot{k}(\tau) d\tau 
	-\int_{t_0}^{t}i(\tau)N(k(\tau))\dot{k}(\tau) d\tau,  \label{pf54new}
	\end{flalign} 
\end{small}
Let $\widetilde{k}(t) = k(t) - k(t_0)$. Dividing (\ref{pf54new}) by $\widetilde{k}(t)$ and recognizing that $\dot{\widehat{z}}(t) = -\dfrac{i(t)}{C_c}$, $\int_{t_0}^{t}N(k(\tau))\dot{k}(\tau) d\tau = \int_{k(t_0)}^{k(t)}N(k) dk$ and $\int_{t_0}^{t}i(\tau)N(k(\tau))\dot{k}(\tau) d\tau = i(t)\int_{k(t_0)}^{k(t)}N(k) dk$ gives

\begin{small}
	\begin{flalign} 
	\frac{e^2(t)}{2\widetilde{k}(t)} \leq& \frac{5}{2} + \frac{1}{2\widetilde{k}(t)}\int_{t_0}^{t}\dot{y}^2(\tau) d\tau \nonumber \\ &+ \frac{1}{2\widetilde{k}(t)}\int_{t_0}^{t}\bigg(\frac{i(\tau)}{C_c}\frac{\partial \widehat{x}_1(\widehat{z}(\tau))}{\partial \widehat{z}(\tau)}\bigg)^2 d\tau \nonumber \\ &+ \frac{1}{2\widetilde{k}(t)}\int_{t_0}^{t}\bigg(\frac{di(\tau)}{d\tau}\bigg)^2\widehat{x}^2_4(\widehat{z}(\tau)) d\tau \nonumber \\& + \frac{1}{2\widetilde{k}(t)}\int_{t_0}^{t}\bigg(\frac{i^2(\tau)}{C_c}\frac{\partial \widehat{x}_4(\widehat{z}(\tau))}{\partial \widehat{z}(\tau)}\bigg)^2 d\tau  \nonumber \\ &+ \frac{1}{2\widetilde{k}(t)}\int_{t_0}^{t}i^2(\tau)\bigg(\frac{1}{\widehat{C}_{ts}(\widehat{z}(\tau))}+\frac{1}{\widehat{C}_{tl}(\widehat{z}(\tau))}\bigg)^2 d\tau \nonumber \\& - \frac{3}{\widetilde{k}(t)}\int_{k(t_0)}^{k(t)}N(k) dk -\frac{i(t)}{\widetilde{k}(t)}\int_{k(t_0)}^{k(t)}N(k) dk.  \label{pf_54new}
	\end{flalign} 
\end{small}
The equation (\ref{pf_54new}) is the result established for $e(t) >0$ case. Now, the Case 2, i.e. for $e(t) <0$, is considered and an inequality having a form similar to (\ref{pf_54new}) will be derived. The results of both Case 1 and Case 2 will be discussed together after establishing the required equation for Case 2.

\noindent \textbf{Case 2.} Consider $e(t) < 0$, at some instant $t > t_0$. Multiplying (\ref{pf_42}) by $e(t)$ and using (\ref{pf24}) gives

\begin{small}
	\begin{flalign}
	e(t)\dot{e}(t) \geq& e(t)\dot{y}(t) - e(t)\frac{\partial \widehat{x}_1(\widehat{z}(t))}{\partial \widehat{z}(t)}\dot{\widehat{z}}(t)+e(t)\frac{di(t)}{dt}\widehat{x}_4(\widehat{z}(t)) \nonumber \\ & +e(t)i(t)\frac{\partial \widehat{x}_4(\widehat{z}(t))}{\partial \widehat{z}(t)}\dot{\widehat{z}}(t)  + e(t)i(t)\times \nonumber \\
	&\bigg(\frac{1}{\widehat{C}_{ts}(\widehat{z}(t))}+\frac{1}{\widehat{C}_{tl}(\widehat{z}(t))}\bigg) -(3+i(t))N(k(t))e^2(t), \label{pf43}
	\end{flalign} 
\end{small}
Now use (\ref{pfadd2}) in (\ref{pf43}) to get the following

\begin{small}
	\begin{flalign} 
	e(t)\dot{e}(t) \geq& -\frac{5}{2}e^2(t) - \frac{1}{2}\dot{y}^2(t) - \frac{1}{2}\bigg(\frac{\partial \widehat{x}_1(\widehat{z}(t))}{\partial \widehat{z}(t)}\dot{\widehat{z}}(t)\bigg)^2 \nonumber \\ &- \frac{1}{2}\bigg(\frac{di(t)}{dt}\bigg)^2\widehat{x}^2_4(\widehat{z}(t))- \frac{1}{2}i^2(t)\bigg(\frac{\partial \widehat{x}_4(\widehat{z}(t))}{\partial \widehat{z}(t)}\dot{\widehat{z}}(t)\bigg)^2 \nonumber \\& - \frac{1}{2}i^2(t)\bigg(\frac{1}{\widehat{C}_{ts}(\widehat{z}(t))}-\frac{1}{\widehat{C}_{tl}(\widehat{z}(t))}\bigg)^2 \nonumber \\ &- (3+i(t))N(k(t))e^2(t). \label{pf53}
	\end{flalign}
\end{small}
Since $\frac{d}{dt}\big(\frac{1}{2}e^2(t)\big) = e(t)\dot{e}(t)$, thus integrating (\ref{pf53}) from $t_0$ to $t$, and using (\ref{pf22}) provides
\begin{small}
	\begin{flalign}
	\frac{1}{2}e^2(t) \geq& -\frac{5}{2}(k(t)-k(t_0)) - \frac{1}{2}\int_{t_0}^{t}\dot{y}^2(\tau) d\tau \nonumber \\&- \frac{1}{2}\int_{t_0}^{t}\bigg(\frac{\partial \widehat{x}_1(\widehat{z}(\tau))}{\partial \widehat{z}(\tau)}\dot{\widehat{z}}(\tau)\bigg)^2 d\tau \nonumber \\&- \frac{1}{2}\int_{t_0}^{t}\bigg(\frac{di(\tau)}{d\tau}\bigg)^2\widehat{x}^2_4(\widehat{z}(\tau)) d\tau \nonumber \\& - \frac{1}{2}\int_{t_0}^{t}i^2(\tau)\bigg(\frac{\partial \widehat{x}_4(\widehat{z}(\tau))}{\partial \widehat{z}(\tau)}\dot{\widehat{z}}(\tau)\bigg)^2 d\tau \nonumber \\&- \frac{1}{2}\int_{t_0}^{t}i^2(\tau)\bigg(\frac{1}{\widehat{C}_{ts}(\widehat{z}(\tau))}-\frac{1}{\widehat{C}_{tl}(\widehat{z}(\tau))}\bigg)^2 d\tau  \nonumber \\& - 3\int_{t_0}^{t}N(k(\tau))\dot{k}(\tau) d\tau 
	-\int_{t_0}^{t}i(\tau)N(k(\tau))\dot{k}(\tau) d\tau,  \label{pf54}
	\end{flalign}
\end{small}
Let $\widetilde{k}(t) = k(t) - k(t_0)$. Dividing (\ref{pf54}) by $\widetilde{k}(t)$ and recognizing that $\dot{\widehat{z}}(t) = -\dfrac{i(t)}{C_c}$, $\int_{t_0}^{t}N(k(\tau))\dot{k}(\tau) d\tau = \int_{k(t_0)}^{k(t)}N(k) dk$ and $\int_{t_0}^{t}i(\tau)N(k(\tau))\dot{k}(\tau) d\tau = i(t)\int_{k(t_0)}^{k(t)}N(k) dk$ gives us

\begin{small}
	\begin{flalign}
	\frac{e^2(t)}{2\widetilde{k}(t)} \geq& -\frac{5}{2} - \frac{1}{2\widetilde{k}(t)}\int_{t_0}^{t}\dot{y}^2(\tau) d\tau \nonumber \\&- \frac{1}{2\widetilde{k}(t)}\int_{t_0}^{t}\bigg(\frac{i(\tau)}{C_c}\frac{\partial \widehat{x}_1(\widehat{z}(\tau))}{\partial \widehat{z}(\tau)}\bigg)^2 d\tau \nonumber \\&- \frac{1}{2\widetilde{k}(t)}\int_{t_0}^{t}\bigg(\frac{di(\tau)}{d\tau}\bigg)^2\widehat{x}^2_4(\widehat{z}(\tau)) d\tau \nonumber \\& - \frac{1}{2\widetilde{k}(t)}\int_{t_0}^{t}\bigg(\frac{i^2(\tau)}{C_c}\frac{\partial \widehat{x}_4(\widehat{z}(\tau))}{\partial \widehat{z}(\tau)}\bigg)^2 d\tau \nonumber \\&- \frac{1}{2\widetilde{k}(t)}\int_{t_0}^{t}i^2(\tau)\bigg(\frac{1}{\widehat{C}_{ts}(\widehat{z}(\tau))}-\frac{1}{\widehat{C}_{tl}(\widehat{z}(\tau))}\bigg)^2 d\tau \nonumber \\& - \frac{3}{\widetilde{k}(t)}\int_{k(t_0)}^{k(t)}N(k) dk 
	-\frac{i(t)}{\widetilde{k}(t)}\int_{k(t_0)}^{k(t)}N(k) dk,  \label{pf_54}
	\end{flalign}
\end{small}
Notice that the (\ref{pf_54}) and (\ref{pf_54new}) have similar form. The differences between them are the sign of inequalities and the terms on R.H.S of (\ref{pf_54}) are negative. The reciprocal of (\ref{pf_54}) provides the following

\begin{small}
	\begin{flalign}
	\frac{2\widetilde{k}(t)}{e^2(t)} \leq& \bigg[-\frac{5}{2} - \frac{1}{2\widetilde{k}(t)}\int_{t_0}^{t}\dot{y}^2(\tau) d\tau \nonumber \\&- \frac{1}{2\widetilde{k}(t)}\int_{t_0}^{t}\bigg(\frac{i(\tau)}{C_c}\frac{\partial \widehat{x}_1(\widehat{z}(\tau))}{\partial \widehat{z}(\tau)}\bigg)^2 d\tau \nonumber \\&- \frac{1}{2\widetilde{k}(t)}\int_{t_0}^{t}\bigg(\frac{di(\tau)}{d\tau}\bigg)^2\widehat{x}^2_4(\widehat{z}(\tau)) d\tau \nonumber \\& - \frac{1}{2\widetilde{k}(t)}\int_{t_0}^{t}\bigg(\frac{i^2(\tau)}{C_c}\frac{\partial \widehat{x}_4(\widehat{z}(\tau))}{\partial \widehat{z}(\tau)}\bigg)^2 d\tau \nonumber \\&- \frac{1}{2\widetilde{k}(t)}\int_{t_0}^{t}i^2(\tau)\bigg(\frac{1}{\widehat{C}_{ts}(\widehat{z}(\tau))}-\frac{1}{\widehat{C}_{tl}(\widehat{z}(\tau))}\bigg)^2 d\tau \nonumber \\& - \frac{3}{\widetilde{k}(t)}\int_{k(t_0)}^{k(t)}N(k) dk   
	-\frac{i(t)}{\widetilde{k}(t)}\int_{k(t_0)}^{k(t)}N(k) dk\bigg]^{-1}.  \label{pf_544}
	\end{flalign}
\end{small}
Any battery can only be discharged for a certain interval of time, say $T > t_0$. After time $t > T$, the following occurs:
$i(t)=0$, $y(t) = 0$, $z(t) = 0$, because all the charge in the battery is exhausted. Therefore, as $t \to \infty$, $\dot{y}(t) = 0$, and $\frac{d(i)}{dt} = 0$. Thus, from these facts, we can conclude that the terms $\int_{t_0}^{t}\dot{y}^2(\tau) d\tau$, $\int_{t_0}^{t}\bigg(\frac{i(\tau)}{C_c}\frac{\partial \widehat{x}_1(\widehat{z}(\tau))}{\partial \widehat{z}(\tau)}\bigg)^2 d\tau$, $\int_{t_0}^{t}\bigg(\frac{di(\tau)}{d\tau}\bigg)^2\widehat{x}^2_4(\widehat{z}(\tau)) d\tau$, $\int_{t_0}^{t}\bigg(\frac{i^2(\tau)}{C_c}\frac{\partial \widehat{x}_4(\widehat{z}(\tau))}{\partial \widehat{z}(\tau)}\bigg)^2 d\tau$, and $\frac{1}{2}\int_{t_0}^{t}i^2(\tau)\bigg(\frac{1}{\widehat{C}_{ts}(\widehat{z}(\tau))}-\frac{1}{\widehat{C}_{tl}(\widehat{z}(\tau))}\bigg)^2 d\tau$ are bounded in (\ref{pf_54new}) and (\ref{pf_544}) as $t \to \infty$. Now suppose that $k(t) \to \infty$ as $t \to \infty$, then the above discussion lets us write as $t \to \infty$ for (\ref{pf_54new}),
\begin{small}
	\begin{align}
	\lim\limits_{t\to \infty} \frac{e^2(t)}{2\widetilde{k}(t)} \leq \frac{5}{2} - \frac{3}{\widetilde{k}(t)}\int_{k(t_0)}^{k(t)}N(k) dk 
	-\frac{i(t)}{\widetilde{k}(t)}\int_{k(t_0)}^{k(t)}N(k) dk, \label{pf57}
	\end{align} 
\end{small}
And from (\ref{pf_544}), we can write the following

\begin{small}
	\begin{align}
	\lim\limits_{t\to \infty} \frac{2\widetilde{k}(t)}{e^2(t)} \leq \dfrac{1}{-\frac{5}{2} - \frac{3}{\widetilde{k}(t)}\int_{k(t_0)}^{k(t)}N(k) dk  
		-\frac{i(t)}{\widetilde{k}(t)}\int_{k(t_0)}^{k(t)}N(k) dk}.  \label{pf_57}
	\end{align}
\end{small}
Now if $k(t) \to \infty$ as $t \to \infty$ then by the definition of a Nussbaum function in (\ref{pf25}), the term $+\frac{1}{k(t)-k(t_0)}\int_{k(t_0)}^{k(t)}N(k) dk$, in (\ref{pf57}) and (\ref{pf_57}), can take values approaching $+\infty$, and therefore this will violate the positiveness of the LHS of  (\ref{pf57}) and (\ref{pf_57}).
By this contradiction, the assumption that $k(t) \to \infty$ is false and therefore $k(t)$ is bounded. However $\dot{k}(t)$ is an increasing function by definition and $k(t)$ is bounded, this implies that $k(t) \to k_\infty$ as $t \to \infty$ which further implies that $\dot{k}(t) \to 0$ as $t \to \infty$, i.e. $e^2(t) \to 0$ as $t \to \infty$ or $e(t) \to 0$ as $t \to \infty$, i.e. $y(t) \to \widehat{y}(t)$ as $t \to \infty$.
Consider now that  $y(t) \to \widehat{y}(t)$, which implies that

\begin{align}
E_o(z(t))-x_2(t)-x_3(t)-i(t)R_s(z(t)) =\widehat{x}_1(\widehat{z}(t))-\widehat{x}_2(t) \nonumber \\ -\widehat{x}_3(t)-i(t)\widehat{x}_4(\widehat{z}(t)), \label{pf58}
\end{align}
Re-arrangement of (\ref{pf58}) yields (\ref{pf59})
\begin{small}
	\begin{align}
	\begin{bmatrix}
	1 & -1 & -1 & -1
	\end{bmatrix} \left(
	\begin{bmatrix}
	E_o(z(t)) \\ x_2(t) \\ x_3(t) \\ i(t)R_s(z(t))
	\end{bmatrix} -
	\begin{bmatrix}
	\widehat{x}_1(\widehat{z}(t)) \\ \widehat{x}_2(t) \\ \widehat{x}_3(t) \\ i(t)\widehat{x}_4(\widehat{z}(t))
	\end{bmatrix} \right)
	= 0. \label{pf59}
	\end{align}
\end{small}
Considering the assumptions of this theorem, \eqref{pf59} is of the form $A\mathbf{x}=0$, and there is no non-zero vector $\mathbf{x}$ in the nullspace of $A$. This implies that $\widehat{x}_1(\widehat{z}(t)) = E_o(z(t))$, and $\widehat{x}_4(\widehat{z}(t)) = R_s(z(t))$.
Equation (\ref{pf59}) also implies $\widehat{x}_2(t) = x_2(t)$, and $\widehat{x}_3(t) = x_3(t)$, which means that $\dot{\widehat{x}}_2(t) = \dot{x}_2(t)$, and $\dot{\widehat{x}}_3(t) = \dot{x}_3(t)$. Let us consider $\dot{\widehat{x}}_2(t) = \dot{x}_2(t)$, the following can be written using (\ref{pf2}) and (\ref{pf12})

\begin{align}
-\frac{\widehat{x}_2(t)}{\widehat{R}_{ts}(\widehat{z}(t))\widehat{C}_{ts}(\widehat{z}(t))} +\frac{i(t)}{\widehat{C}_{ts}(z)}+~
u(t) = \nonumber \\ -\frac{x_2(t)}{R_{ts}(z(t))C_{ts}(z(t))}+\frac{i(t)}{C_{ts}(z(t))}. \label{pf60}
\end{align}
Since it is proved above that $e(t) \to 0$ as $t \to \infty$, $u(t)=-N(k(t))e(t)$, $i(t)$ are infinitesimally small, and $\widehat{x}_2(t) = x_2(t)$, therefore (\ref{pf60}) provides 
\vspace{-0.5em}
\begin{align}
\widehat{R}_{ts}(\widehat{z}(t)) \widehat{C}_{ts}(\widehat{z}(t)) = R_{ts}(z(t))C_{ts}(z(t)) \label{pf61}
\end{align}
Considering $\dot{\widehat{x}}_3(t) = \dot{x}_3(t)$ and following the exact same arguments as above, it is similarly possible to conclude that $\widehat{R}_{tl}(\widehat{z}(t)) \widehat{C}_{tl}(\widehat{z}(t)) = R_{tl}(z(t))C_{tl}(z(t))$.
This completes the proof.	
\qed
\begin{remark}\label{r2} \normalfont The results proved in Theorem \ref{thm4.2} hold valid provided that the battery discharging current remains small, i.e. $i(t) \to 0$ as $t \to \infty$ and the conditions in Lemma \ref{lemma4.1} are satisfied. Also note that the assumptions related to the nullspace are not necessarily restrictive. This is because, the nullspace of $A=\begin{bmatrix}
	1 & -1 & -1 & -1
	\end{bmatrix}$ is easy to calculate, which allows immediate verification if the vector $\left. \begin{bmatrix}
	\widehat{x}_1(\widehat{z}(t)) & \widehat{x}_2(t) & \widehat{x}_3(t) & i(t)\widehat{x}_4(\widehat{z}(t))
	\end{bmatrix}^T \right)\pm \Delta$, $(\Delta \in\mathbb{R}_{4 \times 1})$ %\mathbf{x}= \left(
	%\begin{bmatrix}
	%E_o(z(t)) & x_2(t) & x_3(t) & i(t)R_s(z(t))
	%\end{bmatrix}^T - \right.$\\
	%$\left. \begin{bmatrix}
	%\widehat{x}_1(\widehat{z}(t)) & \widehat{x}_2(t) & \widehat{x}_3(t) & i(t)\widehat{x}_4(\widehat{z}(t))
	%\end{bmatrix}^T \right)$ 
	is infact in the the nullspace of $A$, if yes, Algorithm \ref{algo2} can simply be run and data corresponding to a different instant of time, that satisfies line 11 in Algorithm \ref{algo2} can be used for parameters estimation. Also, the quantities $E_o(z(t)), x_2(t), x_3(t), i(t)R_s(z(t))$ all have distinct convergence times to their respective equilibrium/zero, and $E_o$ is non-zero for a healthy battery, and $i$ can be made zero as desired. This further implies that the equality in \eqref{pf59} can simply be considered term by term if one waits until all terms have zero-ed out and first achieves convergence of $E_o(z(t))$ to $\widehat{z}(t))$. This can be then further used to cancel these terms out in a next round of observation, and acquire convergence of other items in \eqref{pf59} by following this procedure in a loop. Note that this doesn't need any more experimental data, and doesn't necessarily need re-running the estimator, but simply needs one to wait for the appropriate moment to observe convergence. Which; as shown in the appendix occurs very rapidly.
	
	Next, we will show the convergence of some Li-ion battery model parameters $\widehat{r}_n(t)$ as $t \to \infty$, where $n \in \{1,2,\cdots,21\} \backslash \{3,21\}$.
\end{remark}

\begin{lem} \label{lemma4.3}
	Suppose $\lambda_{xn}, \lambda_{yn}, r_{nu}$ and $r_{nl}$ are the positive real numbers for $n \in \{1,2,\cdots,21\} \backslash \{3,21\}$. \textbf{If} the conditions required for Theorem \ref{thm4.2} are satisfied, \textbf{then} $\widehat{r}_n(t)$ converges to some constant $r_{\infty}$ as $t \to \infty$.
\end{lem}
\myproof
The solution of (\ref{adeq}) with $e^2(t) + \lambda_{xn}r_{nu}+\lambda_{yn}r_{nl}$ as an input is as follows
\begin{align}
\widehat{r}_n(t) = &\,\, \widehat{r}_n(t_0)e^{-(\lambda_{xn}+\lambda_{yn})t}\nonumber \\
&+ \bigg((\lambda_{xn}r_{nu}+\lambda_{yn}r_{nl})\times \int_{t_0}^{t}e^{-(\lambda_{xn}+\lambda_{yn})\tau}d\tau\bigg)\nonumber \\
& + \int_{t_0}^{t}e^2(t-\tau)e^{-(\lambda_{xn}+\lambda_{yn})\tau}d\tau \label{pf62}
\end{align}
Because $e^{-(\lambda_{xn}+\lambda_{yn})t} \to 0$ as $t \to \infty$, and from Theorem \ref{thm4.2}, $e(t) \to 0$ as $t \to \infty$. So $e^{-(\lambda_{xn}+\lambda_{yn})t}$ and $e^2(t)$ remain positive and approach to zero as $t \to \infty$. Thus, on the R.H.S of (\ref{pf62}), the first term will go to zero, the second and third terms will be bounded and approach to a constant term as $t \to \infty$. Hence, $\widehat{r}_n(t)$ converges as $t \to \infty$ for $n \in \{1,2,\cdots,21\} \backslash \{3,21\}$.\qed

\begin{figure*}[b]
	\rule{18cm}{0.4pt}
	\begin{align}
	\begin{bmatrix}
	(-\widehat{r}_1e^{-\widehat{r}_2{z}} + r_1e^{-r_2z}) & (\widehat{r}_3 - r_3) & (\widehat{r}_4 - r_4) & (-\widehat{r}_5 + r_5) & (\widehat{r}_6 - r_6) 
	\end{bmatrix}
	\begin{bmatrix}
	1 \\ 1 \\ z(t) \\ z^2(t) \\ z^3(t)
	\end{bmatrix}
	= 0, \label{pf65}
	\end{align}
\end{figure*}

\subsection{Accuracy analysis of estimated Li-ion battery model parameters}\label{saa}
\quad In this section, we will first demonstrate that the parameters of $\widehat{x}_1(\widehat{z}(t))$, and $\widehat{x}_4(\widehat{z}(t))$ converges to their actual values based on the results derived in Theorem \ref{thm4.2}. Afterward, the accuracy analysis of $R_{ts}(\widehat{z}(t))$, and $R_{tl}(\widehat{z}(t))$ will lead us to show the convergence of these circuit elements parameters to their actual values.
As per the results derived in Theorem \ref{thm4.2}, $\widehat{x}_1(\widehat{z}(t)) = x_1(z(t))$, and $\widehat{x}_4(\widehat{z}(t)) = x_4(z(t))$ as $t \to \infty$. Using (\ref{pf6}), (\ref{pf9}), (\ref{pf161}), and (\ref{pf191}), the above two results can be written as follows
\begin{align}
-\widehat{r}_1e^{-\widehat{r}_2\widehat{z}}+\widehat{r}_3+\widehat{r}_4\widehat{z}-\widehat{r}_5\widehat{z}^2+\widehat{r}_6\widehat{z}^3 = \nonumber \\  -r_1e^{-r_2z}+r_3+r_4z-r_5z^2+r_6z^3, \label{pf63}
\end{align}
and
\begin{align}
\widehat{r}_{19}e^{-\widehat{r}_{20}\widehat{z}}+\widehat{r}_{21} = r_{19}e^{-r_{20}z}+r_{21}. \label{pf64}
\end{align}
Since $ \widehat{z}(t) = z(t) $, thus the equation (\ref{pf63}) can be rewritten as (\ref{pf65}). Similarly, the equation (\ref{pf64}) can be represented by (\ref{pf66}),

\begin{align}
\begin{bmatrix}
(\widehat{r}_{19}e^{-\widehat{r}_{20}{z}} - r_{19}e^{-r_{20}z}) & (\widehat{r}_{21} - r_{21}) 
\end{bmatrix}
\begin{bmatrix}
1 \\ 1
\end{bmatrix}
= 0. \label{pf66}
\end{align} 

At $ \widehat{z}(t) \to 0 $ as $ t \to \infty $, and $ \widehat{z}(t) \neq 0 $ as $ t \to \infty $, assuming that no non-zero vector $\begin{small}\begin{bmatrix}
(r_1-\widehat{r}_1) & (\widehat{r}_3-r_3) & (\widehat{r}_4-r_4) & (-\widehat{r}_5+r_5) & (\widehat{r}_6-r_6) 
\end{bmatrix}\end{small}$ is in the left-nullspace of $\begin{bmatrix}
1 & 1 & z(t) & z^2(t) & z^3(t)
\end{bmatrix}^T$, the equation (\ref{pf65}) implies that $\widehat{r}_1 \to r_1$, $\widehat{r}_3 \to r_3$, $\widehat{r}_4 \to r_4$, $\widehat{r}_5 \to r_5$, and $\widehat{r}_6 \to r_6$. Using $\widehat{r}_1 \to r_1$ in $\widehat{r}_1e^{-\widehat{r}_2\widehat{z}} = r_1e^{-r_2z}$ provides $\widehat{r}_2 \to r_2$. Similarly, using the same arguments as above but assuming that no non-zero vector $\begin{small}\begin{bmatrix}
(\widehat{r}_{19}-r_{19}) & (\widehat{r}_{21}-r_{21}) 
\end{bmatrix}\end{small}$ is in the left-nullspace of $\begin{bmatrix}
1 & 1
\end{bmatrix}^T$, we can infer from (\ref{pf66}) that $\widehat{r}_{19} \to r_{19}$, $\widehat{r}_{20} \to r_{20}$ and $\widehat{r}_{21} \to r_{21}$. Please note that the imposition of such assumptions is not necessarily restrictive. As discussed in Remark \ref{r2}, the nullspaces may be computed and thus the bounds on the estimated values ($\hat{r}$) can be found such that they do not cause the estimates to violate the condition related to nullpaces. Now consider $\widehat{C}_{ts}(\widehat{z}(t))=C_{ts}(z(t)) + \Delta$, where $\Delta$ is the estimation error due to inappropriate selection of parameters such as $\lambda_{xn}, \lambda_{yn}, r_{nu}$ and $r_{nl}$ for $n \in \{1,2,\cdots,21\} \backslash \{3,21\}$, and violation of condition $i(t) \to \infty$. Since $ \widehat{z}(t) = z(t) $, and $\widehat{R}_{ts}(z(t)) \widehat{C}_{ts}(z(t)) = R_{ts}(z(t))C_{ts}(z(t))$, from Theorem \ref{thm4.2}, leads to the following
\begin{align}
\widehat{R}_{ts}(z(t)) = \frac{R_{ts}(z(t))C_{ts}(z(t))}{C_{ts}(z(t))+ \Delta} = \frac{R_{ts}(z(t))}{1 + \dfrac{\Delta}{C_{ts}(z(t))}}. \label{pf68}
\end{align}
Because the value of $C_{ts}(z(t))$ ranges in the order of hundred or thousand Farads, the magnitude of $\Delta$ is expected to be much smaller than the magnitude of $C_{ts}(z(t))$. The above assumption results in $ \widehat{R}_{ts}(z(t)) \to R_{ts}(z(t)) $ from (\ref{pf68}). Now using (\ref{pf_6}) and (\ref{pf16}), we can write the following
\begin{align}
\begin{bmatrix}
(\widehat{r}_{7}e^{-\widehat{r}_{8}\widehat{z}} - r_{7}e^{-r_{7}z}) & (\widehat{r}_{9} - r_{9}) 
\end{bmatrix}
\begin{bmatrix}
1 \\ 1
\end{bmatrix}
= 0. \label{pf67}
\end{align}
Recalling the similar arguments and assumptions described earlier for the convergence of estimated parameters as described in the second paragraph of Section \ref{saa}, it is possible to present that $\widehat{r}_7 \to r_7$, $\widehat{r}_8 \to r_8$, and $\widehat{r}_9 \to r_9$. 
Similarly, by considering $\widehat{R}_{tl}(z(t)) \widehat{C}_{tl}(z(t)) = R_{tl}(z(t))\times$ $C_{tl}(z(t))$ from Theorem \ref{thm4.2}, we can conclude that $ \widehat{R}_{tl}(z(t)) \to R_{tl}(z(t)) $ and $\widehat{r}_{10} \to r_{10}$, $\widehat{r}_{11} \to r_{11}$, and $\widehat{r}_{12} \to r_{12}$. The accuracy analysis shows the convergence of \textit{fifteen} parameters to their actual values except the parameters of $C_{ts}$ and $C_{tl}$, which are due to the aforementioned  reasons. The results derived in this section will be discussed and validated through simulation in the next section.

\section{Simulation Results at the cell level} \label{sec5}

\begin{table*}[!t]
	\centering 
	\vspace{1em}
	\caption{Simulation results of a 4.1 V, 270 mAh Li-ion battery model parameters.} 
	\begin{adjustbox}{width=0.8\linewidth} 
		\begin{tabular}{c c c c c c c c c} 
			\hline\hline 
			Parameter & \thead{Upper bound \\$ (r_{nu}) $} & \thead{Lower bound \\$ (r_{nl}) $} & $ \lambda_{x_n} $ & $ \lambda_{y_n} $ & \thead{Initial\\ value} & \thead{Estimated\\ value} & \thead{Desired\\ value} & \thead{Estimation\\ error (\%)}  \\ [1ex] 
			\hline
			
			$\widehat{r}_{1}$ & 4 & 0.1 & 20 & 65 & 100 & 1.0176 & 1.031 & 1.3 \\ 
			$\widehat{r}_{2}$ & 50 & 25 & 50 & 70 & 2000 & 35.4167 & 35 & 1.2 \\ 
			$\widehat{r}_{3}$ & -- & -- & -- & -- & -- & 3.6855 & 3.685 & 0.014 \\ 
			$\widehat{r}_{4}$ & 0.5 & 0.1 & 30 & 70 & 50 & 0.22 & 0.2156 & 2.04 \\ 
			$\widehat{r}_{5}$ & 0.5 & 0.01 & 20 & 70 & 30 & 0.1189 & 0.1178 & 0.934 \\ 
			$\widehat{r}_{6}$ & 0.5 & 0.1 & 60 & 50 & 200 & 0.3182 & 0.3201 & 0.594 \\
			$\widehat{r}_{7}$ & 1 & 0.1 & 50 & 50 & 180 & 0.3002 & 0.3208 & 6.42 \\ 
			$\widehat{r}_{8}$ & 50 & 10 & 50 & 50 & 1700 & 30 & 29.14 & 2.95 \\ 
			$\widehat{r}_{9}$ & 0.1 & 0.01 & 50 & 50 & 240 & 0.055 & 0.04669 & 17.79 \\ 
			$\widehat{r}_{10}$ & 10 & 1 & 70 & 50 & 3600 & 6.2533 & 6.603 & 5.3 \\ 
			$\widehat{r}_{11}$ & 200 & 100 & 50 & 50 & 9300 & 149.9 & 155.2 & 3.41 \\ 
			$\widehat{r}_{12}$ & 0.1 & 0.01 & 50 & 50 & 264 & 0.0553 & 0.04984 & 10.95 \\ 
			$\widehat{r}_{13}$ & 1000 & 500 & 60 & 55 & 50000 & 760.869 & 752.9 & 1.06 \\ 
			$\widehat{r}_{14}$ & 30 & 1 & 5 & 10 & 1000 & 10.6672 & 13.51 & 21.04 \\
			$\widehat{r}_{15}$ & 800 & 500 & 80 & 50 & 50000 & 684.62 & 703.6 & 2.69 \\ 
			$\widehat{r}_{16}$ & 7000 & 5000 & 10 & 10 & 50000 & 6000 & 6056 & 0.92 \\
			$\widehat{r}_{17}$ & 50 & 5 & 50 & 50 & 1000 & 27.5 & 27.12 & 1.40 \\
			$\widehat{r}_{18}$ & 5000 & 3000 & 50 & 50 & 50000 & 4500 & 4475 & 0.558 \\ 
			$\widehat{r}_{19}$ & 0.5 & 0.01 & 20 & 50 & 60 & 0.15 & 0.1562 & 3.97 \\
			$\widehat{r}_{20}$ & 50 & 15 & 30 & 80 & 1200 & 24.5455 & 24.37 & 0.72 \\ 
			$\widehat{r}_{21}$ & -- & -- & -- & -- & -- & 0.0826 & 0.07446 & 10.93 \\
			
			\hline 
		\end{tabular}
		\label{tabpf1}
	\end{adjustbox} 
\end{table*}

\begin{figure*}[!t]
	\centering
	\begin{subfigure}[t]{0.5\textwidth}
		\centering
		%\captionsetup{justification=centering,margin=2cm}
		\includegraphics[scale=0.4]{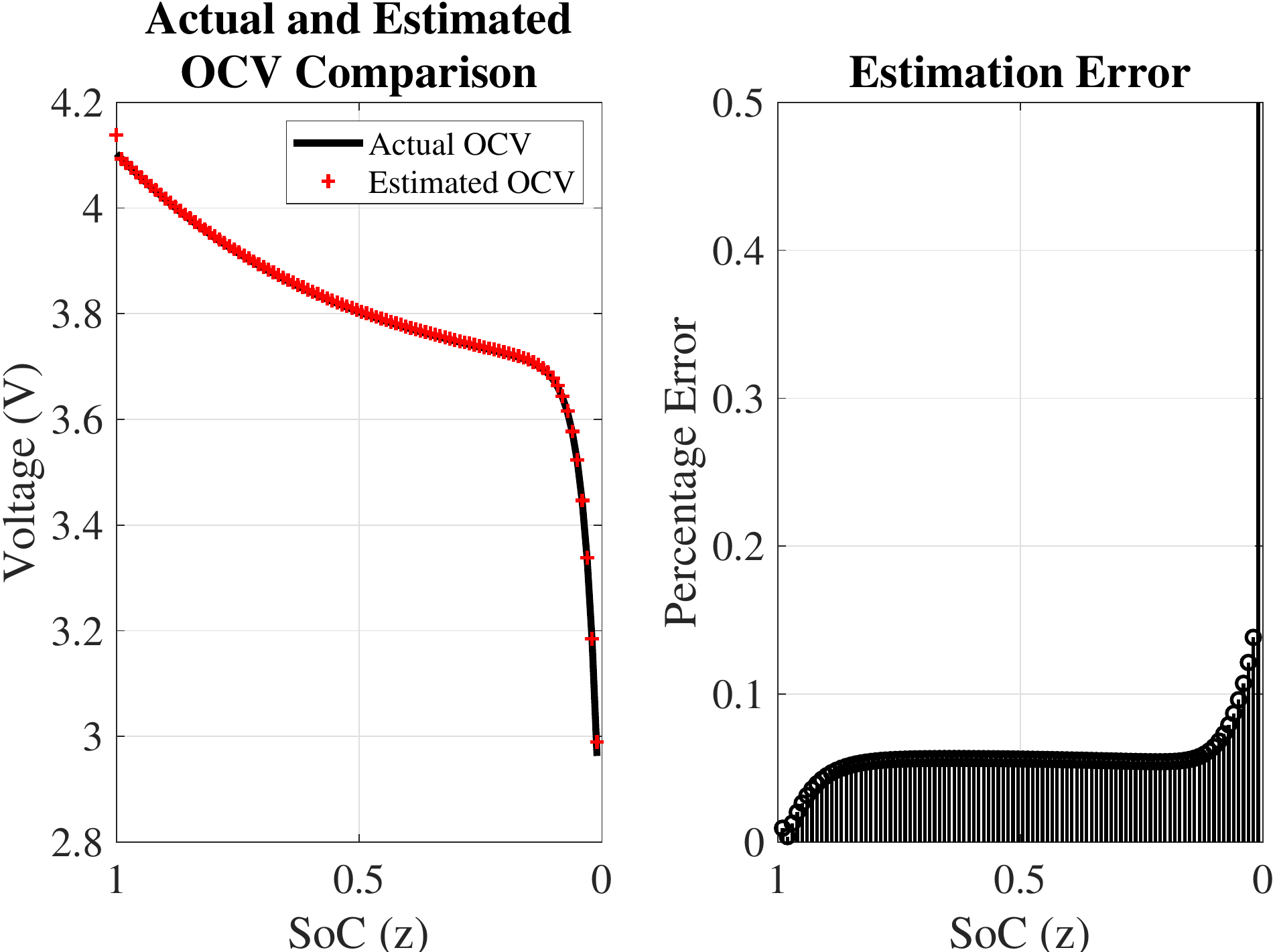}
		\caption{Open circuit voltage $E_o$ VS SoC}
		\label{fig:p1}
	\end{subfigure}~
	\begin{subfigure}[t]{0.5\textwidth}
		\centering
		%\captionsetup{justification=centering,margin=2cm}
		\includegraphics[scale=0.4]{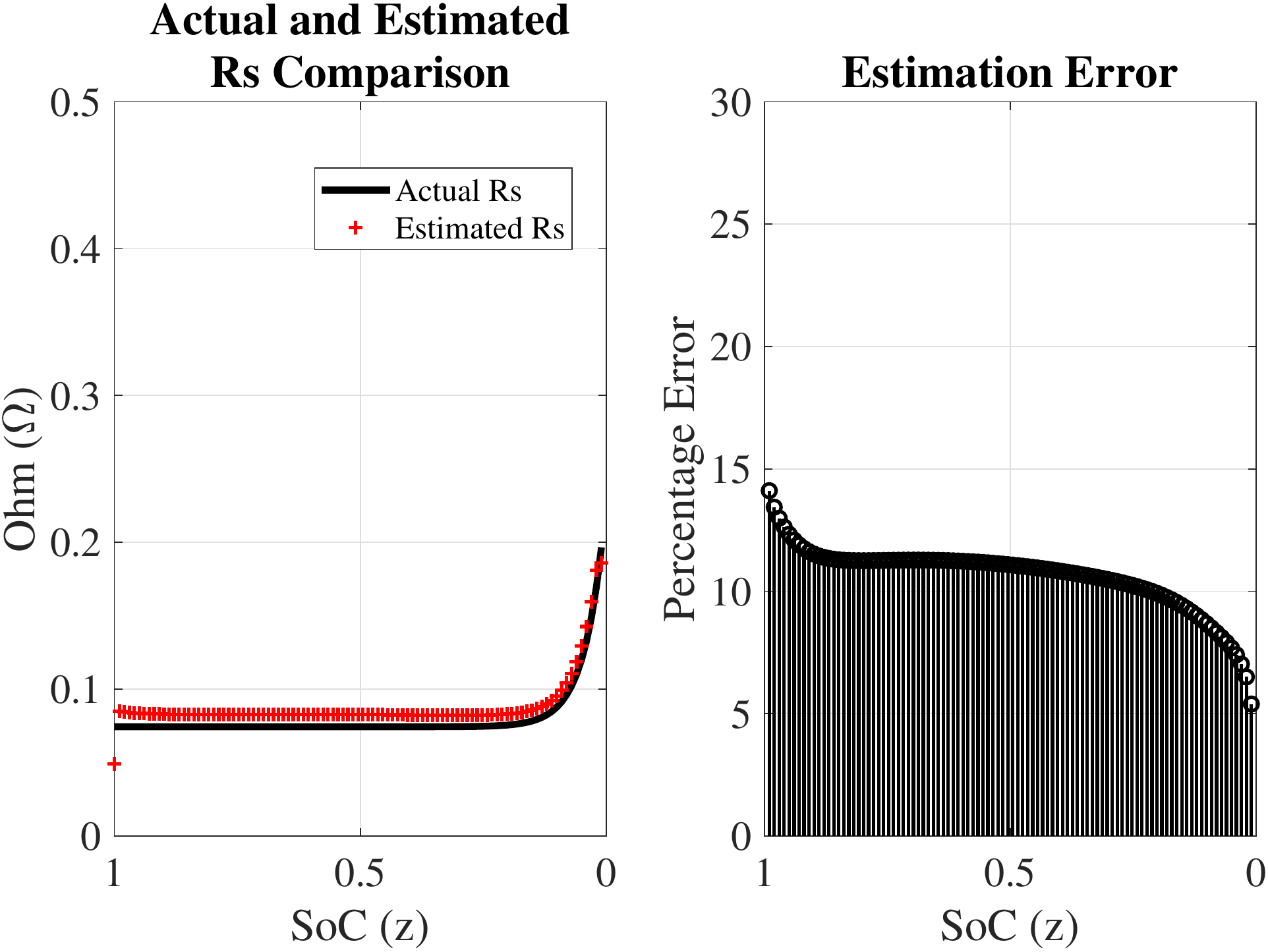}
		\caption{Series resistance $R_s$ VS SoC}
		\label{fig:p2}
	\end{subfigure}\\~\\
	\begin{subfigure}[t]{0.5\textwidth}
		\centering
		%\captionsetup{justification=centering,margin=2cm}
		\includegraphics[scale=0.395]{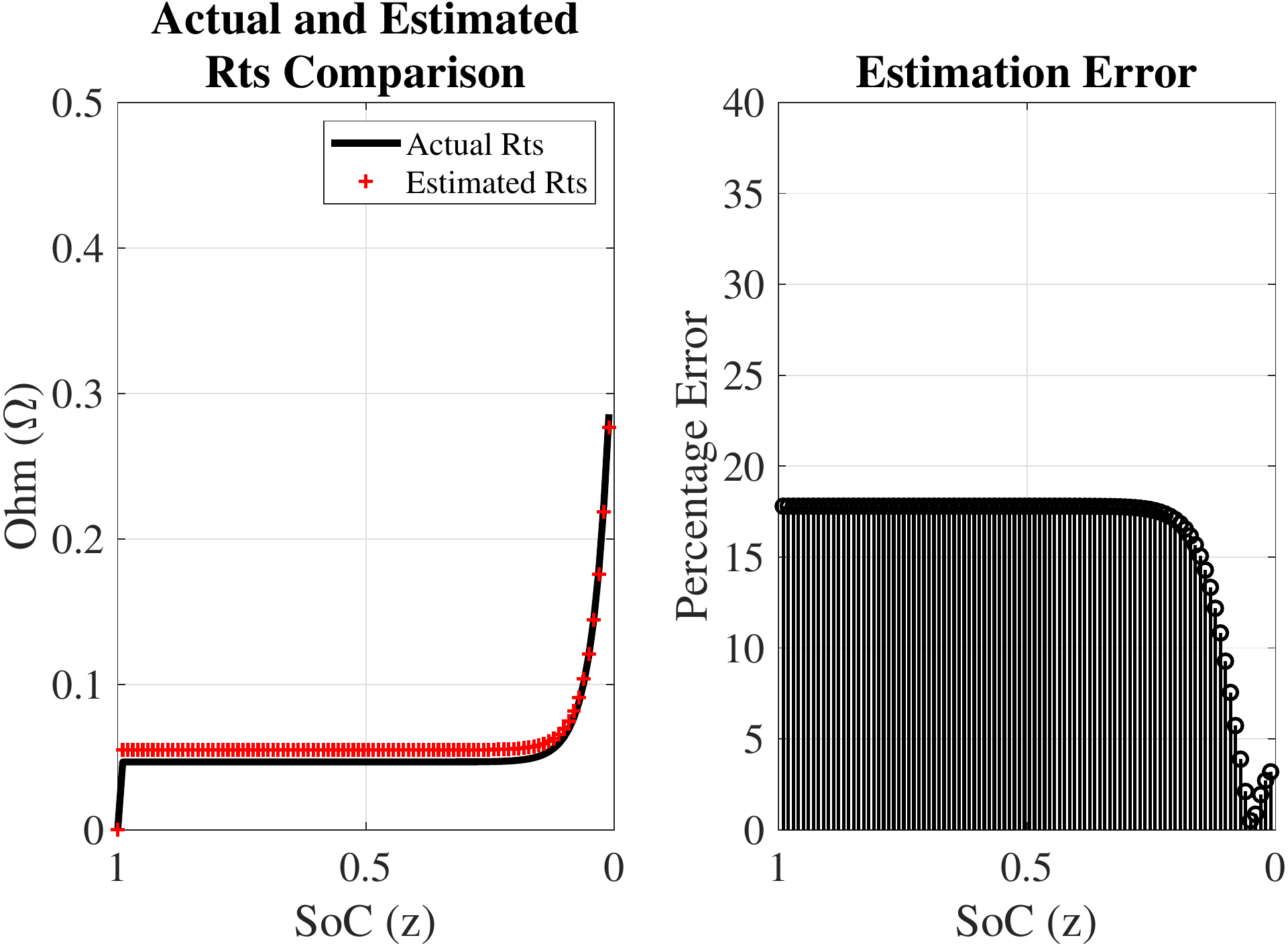}
		\caption{Short term resistance $R_{ts}$ VS SoC}
		\label{fig:p3}
	\end{subfigure}~
	\begin{subfigure}[t]{0.5\textwidth}
		\centering
		%\captionsetup{justification=centering,margin=2cm}
		\includegraphics[scale=0.4]{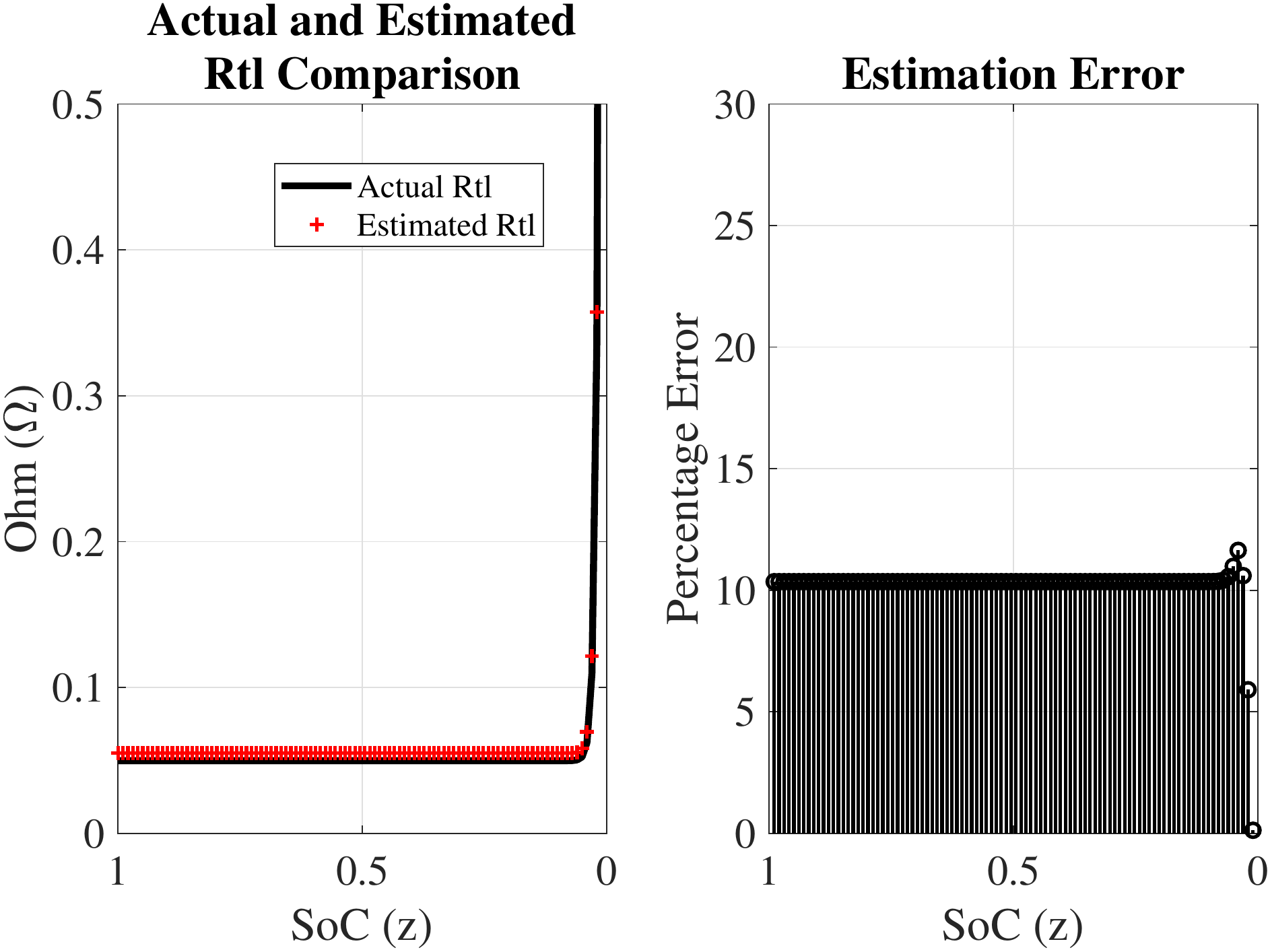}
		\caption{Long term resistance $R_{tl}$ VS SoC}
		\label{fig:p4}
	\end{subfigure}\\~\\
	\begin{subfigure}[t]{0.5\textwidth}
		\centering
		%\captionsetup{justification=centering,margin=2cm}
		\includegraphics[scale=0.4]{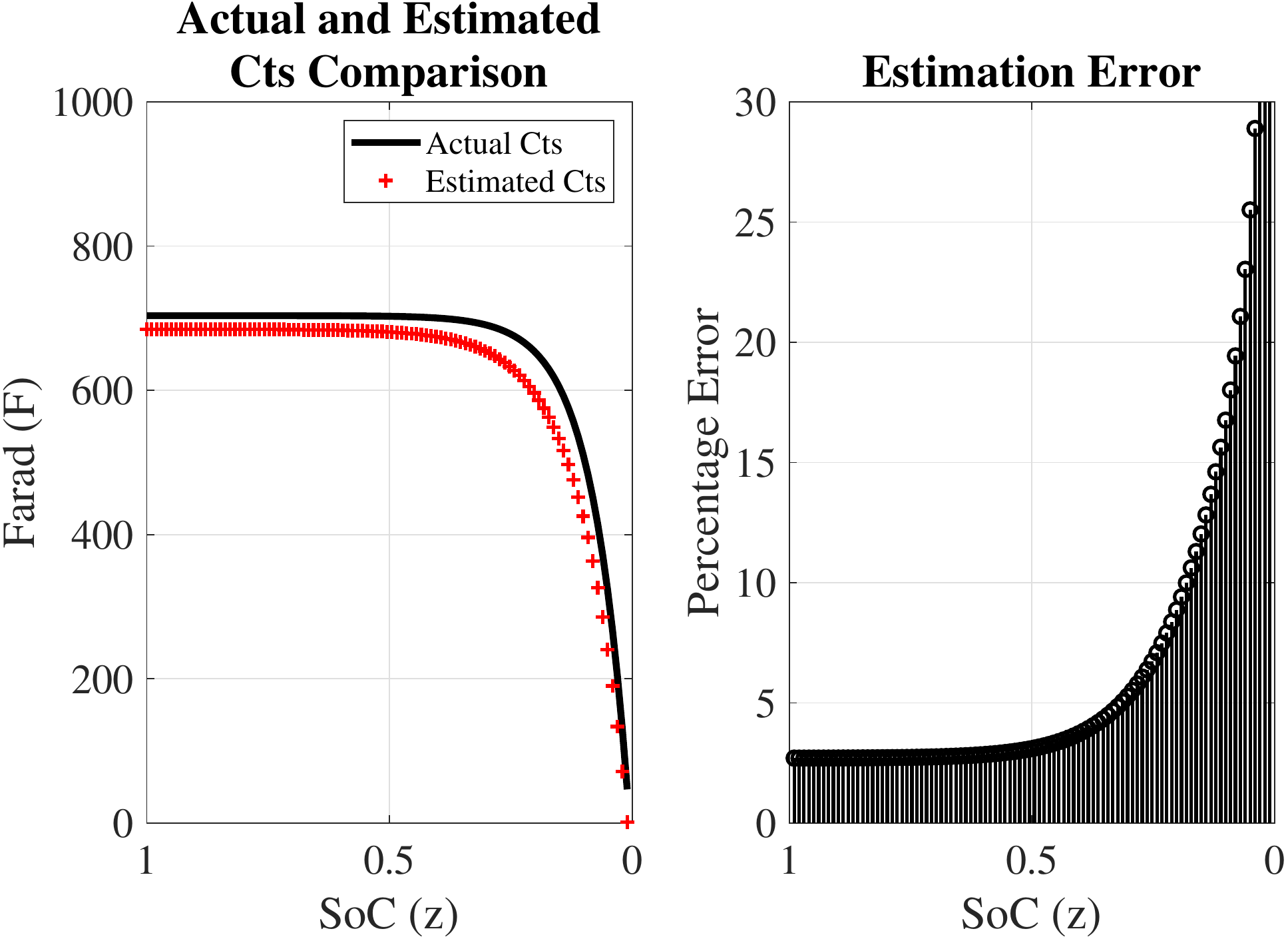}
		\caption{Short term capacitance $C_{ts}$ VS SoC}
		\label{fig:p5}
	\end{subfigure}~
	\begin{subfigure}[t]{0.5\textwidth}
		\centering
		%\captionsetup{justification=centering,margin=2cm}
		\includegraphics[scale=0.4]{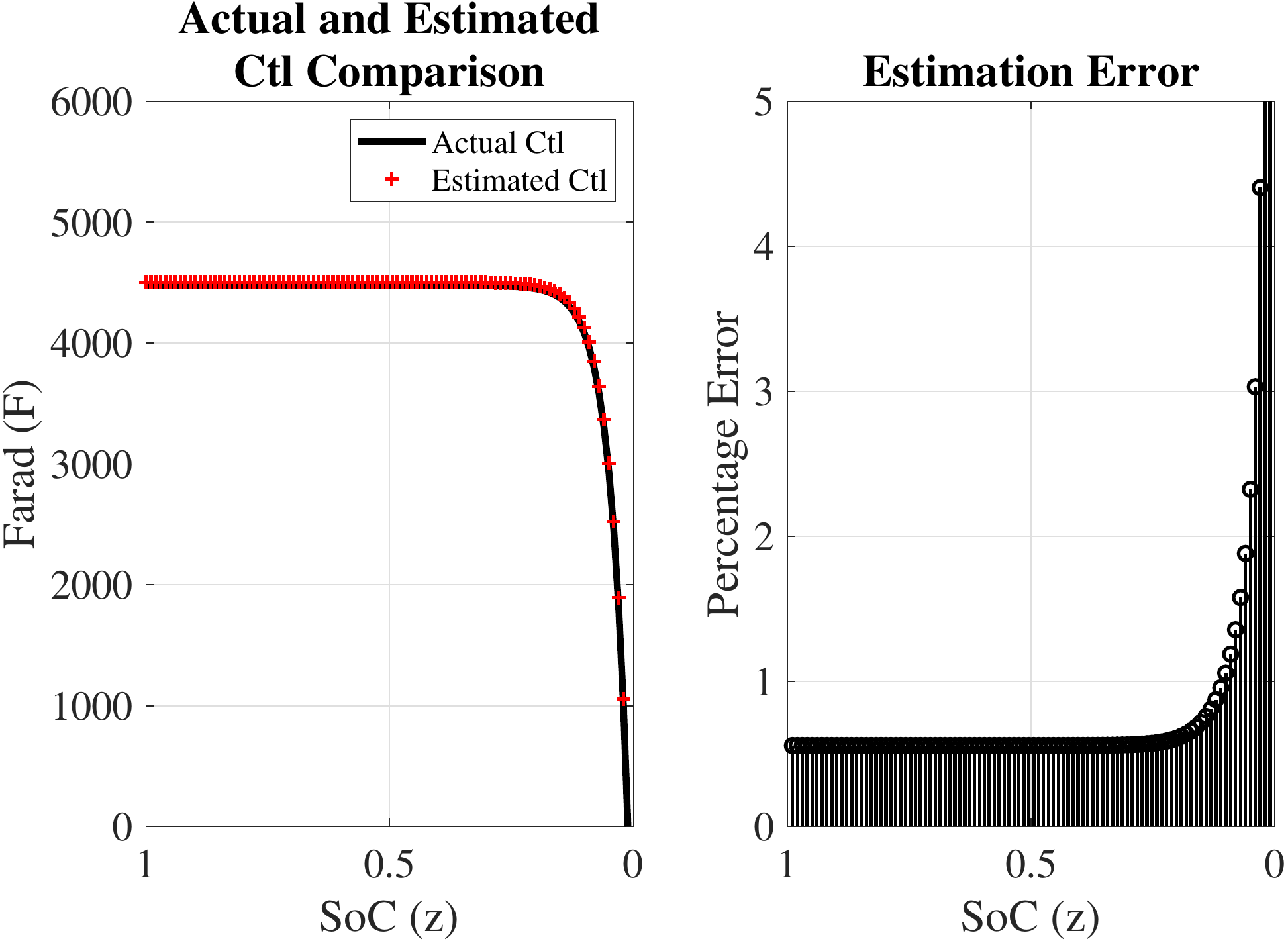}
		\caption{Long term capacitance $C_{tl}$ VS SoC}
		\label{fig:p6}
	\end{subfigure}%
	
	\caption{Comparison of actual and estimated circuit elements of Li-ion battery model during adaptive estimation process.}
	\label{fig3}
\end{figure*}

\begin{figure}[!t]
	\centering
	\includegraphics[scale=0.34]{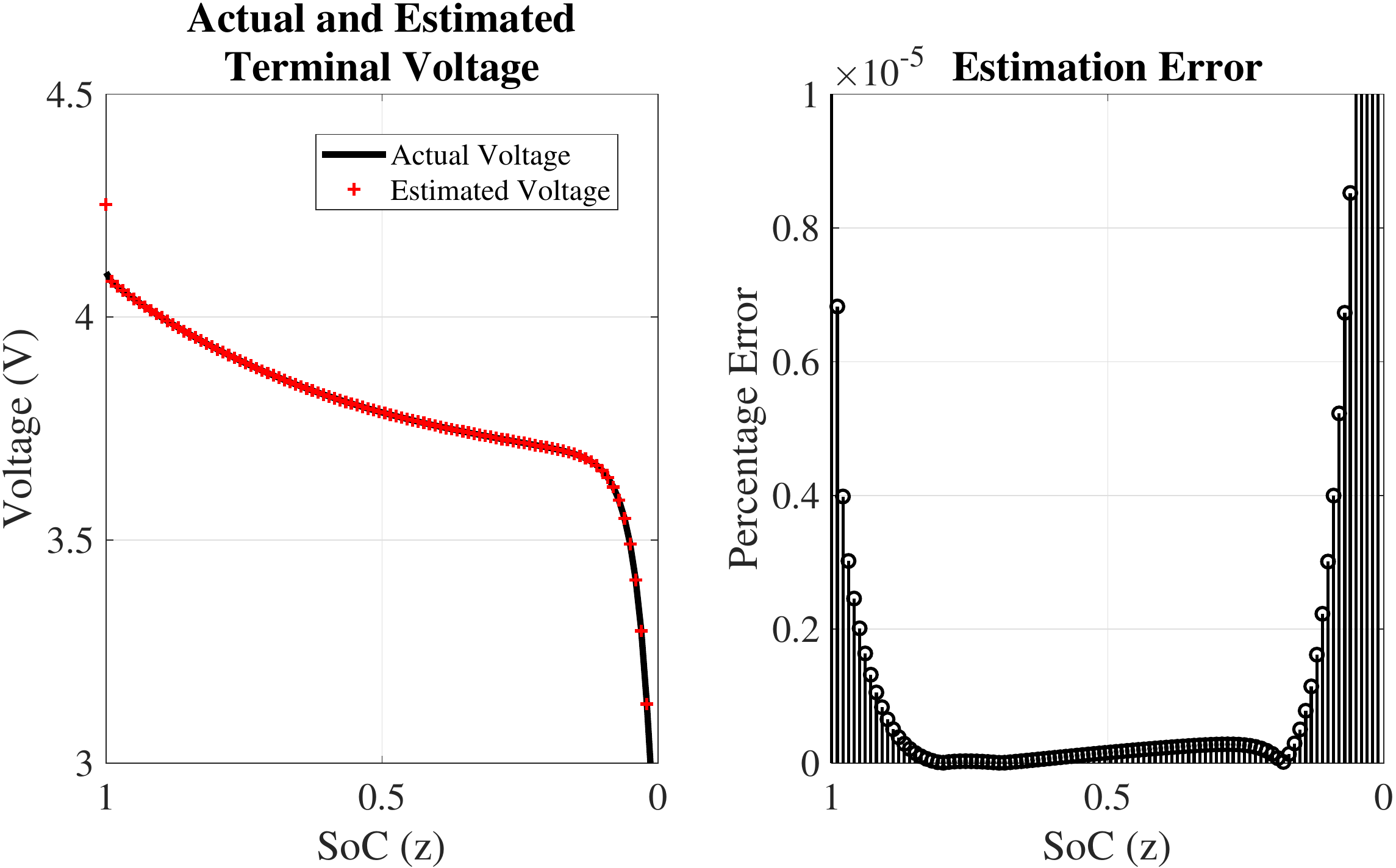}
	\caption{Li-ion battery output voltage VS SoC during adaptation process.}
	\label{fig2}
\end{figure}

The proposed methodology for the convergence of estimated values to their actual values, is verified by MATLAB simulation results. The accuracy of estimated circuit elements and their parameters is validated in simulation, by comparing the estimated values with the ones provided by Chen and Mora \cite{CM} for a 4.1 V, 850 mAh Li-ion battery. However in this work, the battery capacity is reduced to 270 mAh which decreases the simulation time by almost $66 \%$ as compared to the case of using the rated 850 mAh capacity. The reduction of battery capacity does not change the battery dynamics, because these parameters are independent of the battery capacity \cite{CM}, and influence the shape of voltage vs. time profile only. %This is because the battery parameters obtained in \cite{CM} are constant and  SoC, battery capacity, charging and discharging current, and battery temperature. 

{The results form Chen and Mora's (CM) work} \cite{CM} {are considered as actual values for the case of 4.1 V cell, in Table} \ref{tabpf1} {and Figure} \ref{fig3}. {The authors in} \cite{CM} {performed 40 experiments, ten discharging curves at 80, 160, 320, and 640 mA, to extract equivalent circuit elements of a Li-ion battery. These parameters are able to predict Li-ion battery voltage at any load profile within 0.4\% run-time error and 30 mV maximum voltage error} \cite{CM}. {Therefore, owing to high accuracy of CM work and its extensive utilization in many of the state-of-the-art research studies, we refer to equivalent circuit parameters from CM work as actual values, and use these parameter values as actual values for comparison purposes in this section.}
The parameters adaptation process begins with the appropriate choice of some constraints. These constraints include the selection of steady-state upper and lower bounds and their respective confidence levels for each parameter, described in Table \ref{tabpf1}, and initial values of state variables, provided in algorithm \ref{algo2}. Note that the selection of upper and lower bounds and their respective confidence levels for each parameter does not require a strenuous effort from a user with some knowledge and experience of Li-ion batteries. The selection rules for initial values of state variables have already been provided in algorithm \ref{algo2}. The battery discharge current needs to be kept very small during the adaptation process, as per Theorem \ref{thm4.2}, for the convergence of estimated battery parameters and state variables to the actual values. Algorithm \ref{algo2} is run in MATLAB for real-time parameters estimation of a Li-ion battery, and the results are provided in Table \ref{tabpf1}. Note that each estimated parameter is recorded in a separate array during the adaptation process, i.e. the estimated parameters results are recorded in \textit{twenty-one} arrays. The average value of each array (after convergence of the estimation algorithm) is considered as the estimated value of a corresponding parameter, and is shown in Table \ref{tabpf1}. The results in Table \ref{tabpf1} show that the estimation error is less than 5\% for most of the estimated parameters. This level of accuracy is achieved despite the selection of initial values of parameter estimates being far off from their actual values. However, an appropriate selection of upper and lower bounds can further reduce the estimation error of all parameters. The estimated parameters are then employed to analyze the variation of circuit elements values with SoC. The variation of the estimated and actual circuit elements values $E_0,R_s,R_{ts},R_{tl},C_{ts}$ and $C_{tl}$ versus SOC are shown in the left subplots of Figure \ref{fig3}(a)-\ref{fig3}(f) respectively. The right subplots of Figure \ref{fig3}(a)-\ref{fig3}(f) show the respective estimation errors. The Chen and Mora's results are used as \textit{actual values} in these simulation results.

All the circuit elements converged within a 10\% error bound, except $R_{ts}$ which can be further improved by fixing the upper and lower bounds appropriately. It can be noticed that estimation error of circuit elements is higher when SoC approaches zero.  A Li-ion battery becomes unstable when the SOC value becomes lower than a certain threshold \cite{VCollapse}, which causes the estimated parameters to diverge from their actual values. Therefore, in this work the battery model parameters are estimated until the SoC is reduced to 7\%, though the results in Figure \ref{fig3} are displayed until the SoC reaches 1\%. Furthermore, the comparison of actual and estimated terminal voltages during the online adaptation process is shown in Figure \ref{fig2}. In Figure \ref{fig2}, the estimated terminal voltage converges to the actual voltage with very low estimation error of about 10e-4 V. %[We are here 07/07/2020.]

\begin{figure}[!t]
	\centering
	\includegraphics[scale=0.4]{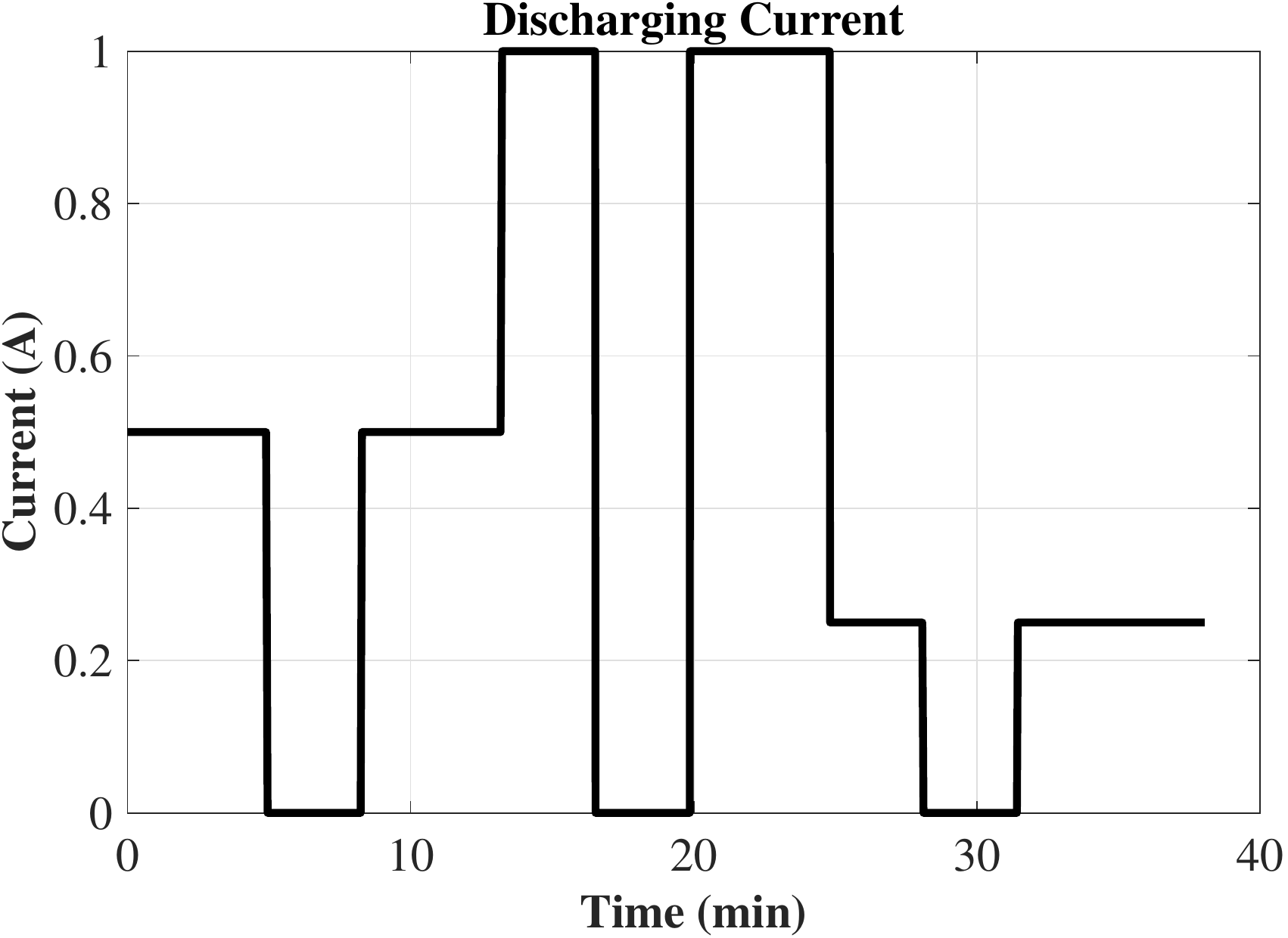}
	\caption{Variable current drawn from Li-ion battery.}
	\label{fig5_Current}
\end{figure}

\begin{figure*}[!t]
	\centering
	\begin{subfigure}[t]{0.5\textwidth}
		\centering
		%\captionsetup{justification=centering,margin=2cm}
		\includegraphics[scale=0.4]{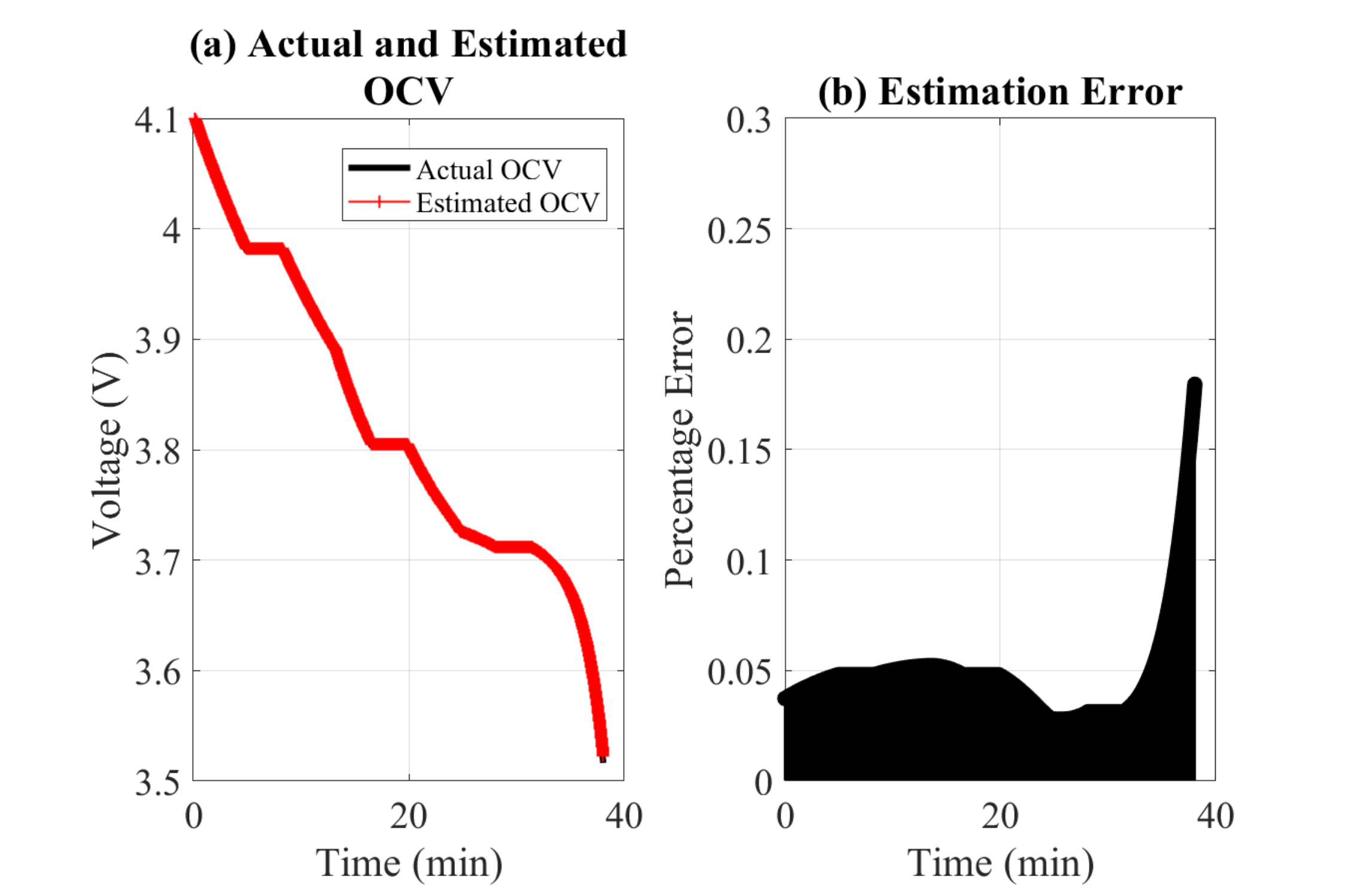}
		\label{fig:ocv}
	\end{subfigure}~
	\begin{subfigure}[t]{0.5\textwidth}
		\centering
		%\captionsetup{justification=centering,margin=2cm}
		\includegraphics[scale=0.41]{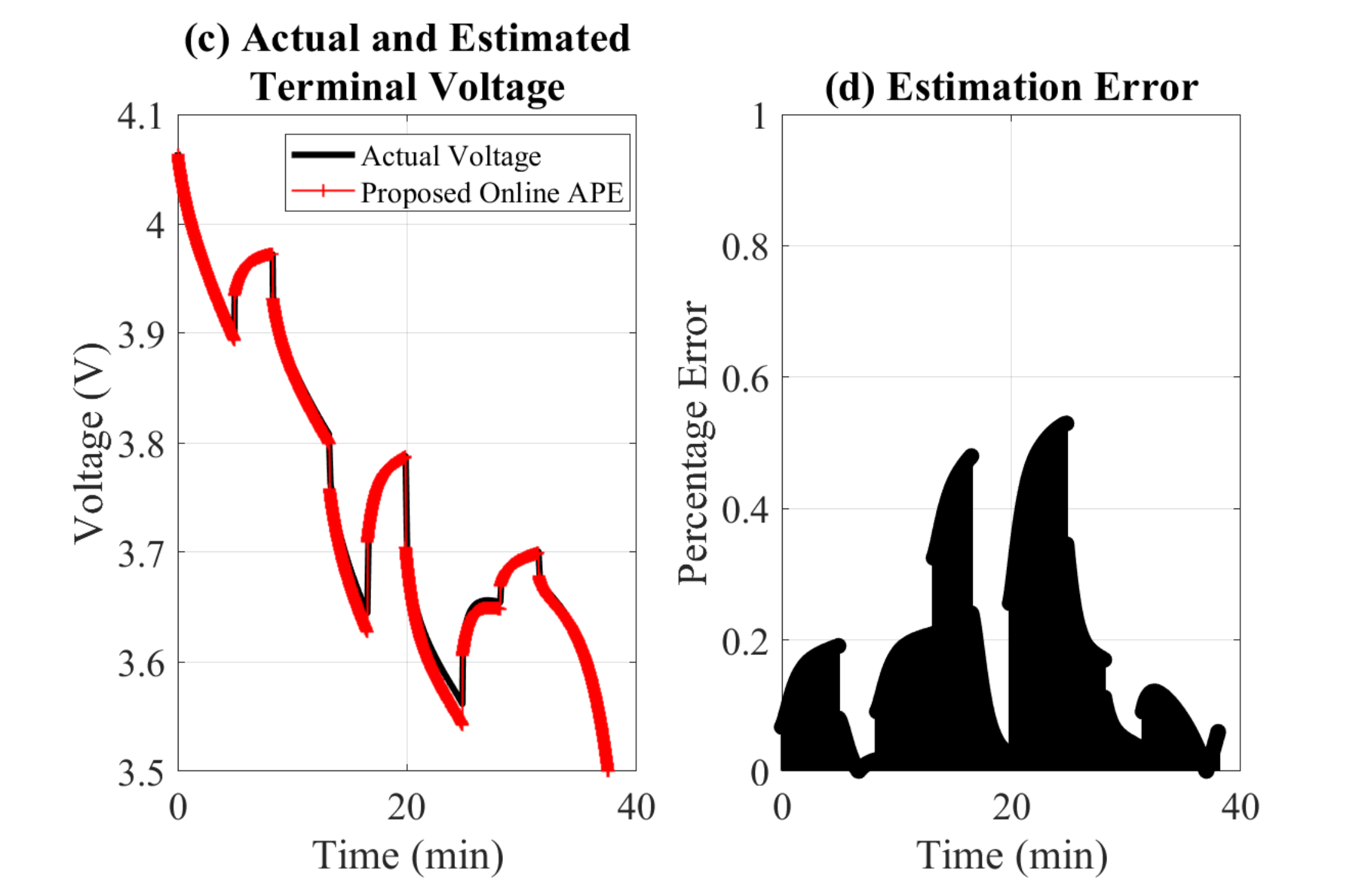}
		\label{fig:tv}
	\end{subfigure}\\
	\begin{subfigure}[t]{0.5\textwidth}
		\centering
		%\captionsetup{justification=centering,margin=2cm}
		\includegraphics[height=0.7\linewidth,width=1.02\linewidth]{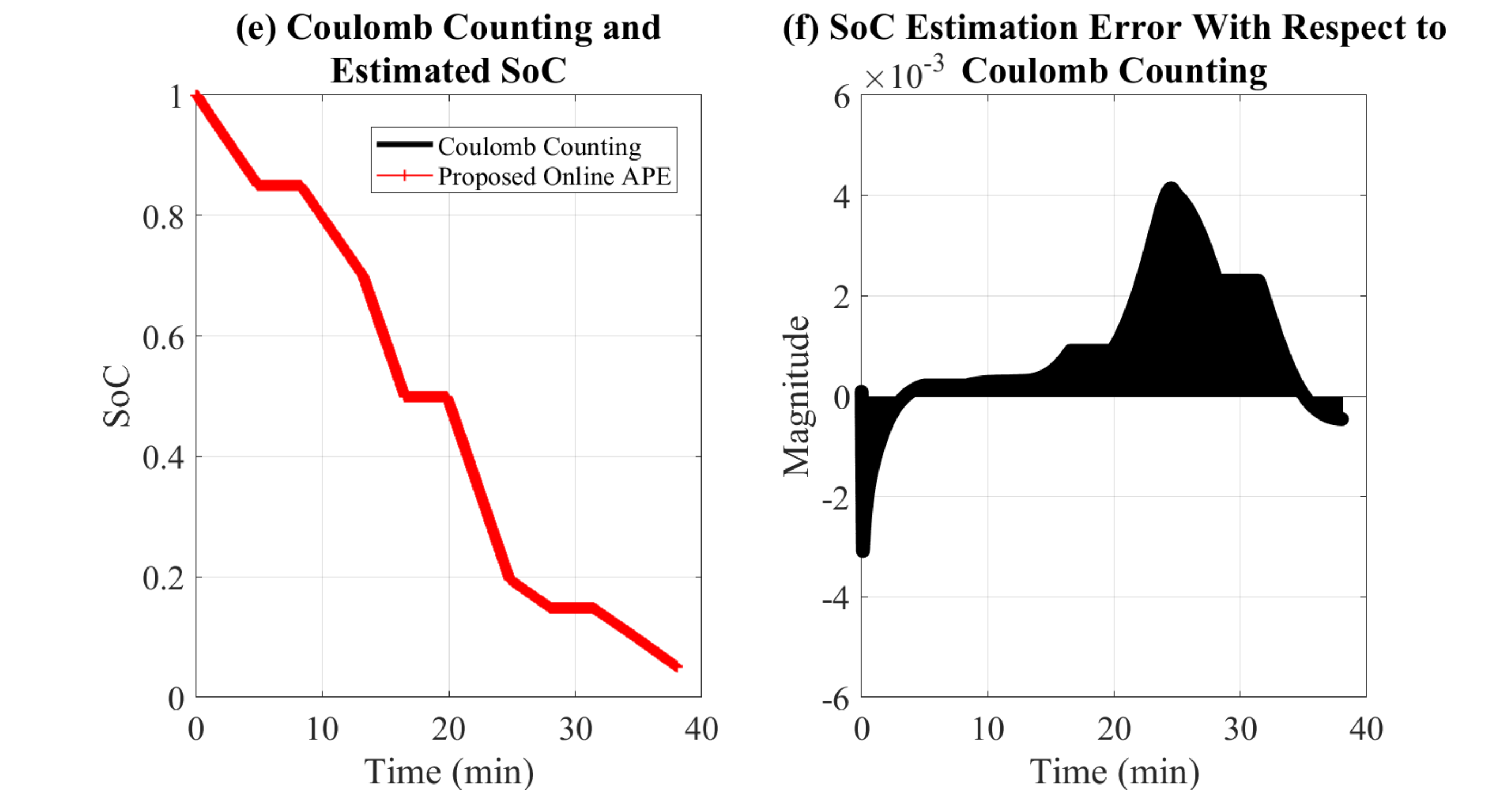}
		\label{fig:soc}
	\end{subfigure}~
	\begin{subfigure}[t]{0.5\textwidth}
		\centering
		%\captionsetup{justification=centering,margin=2cm}
		\includegraphics[scale=0.4]{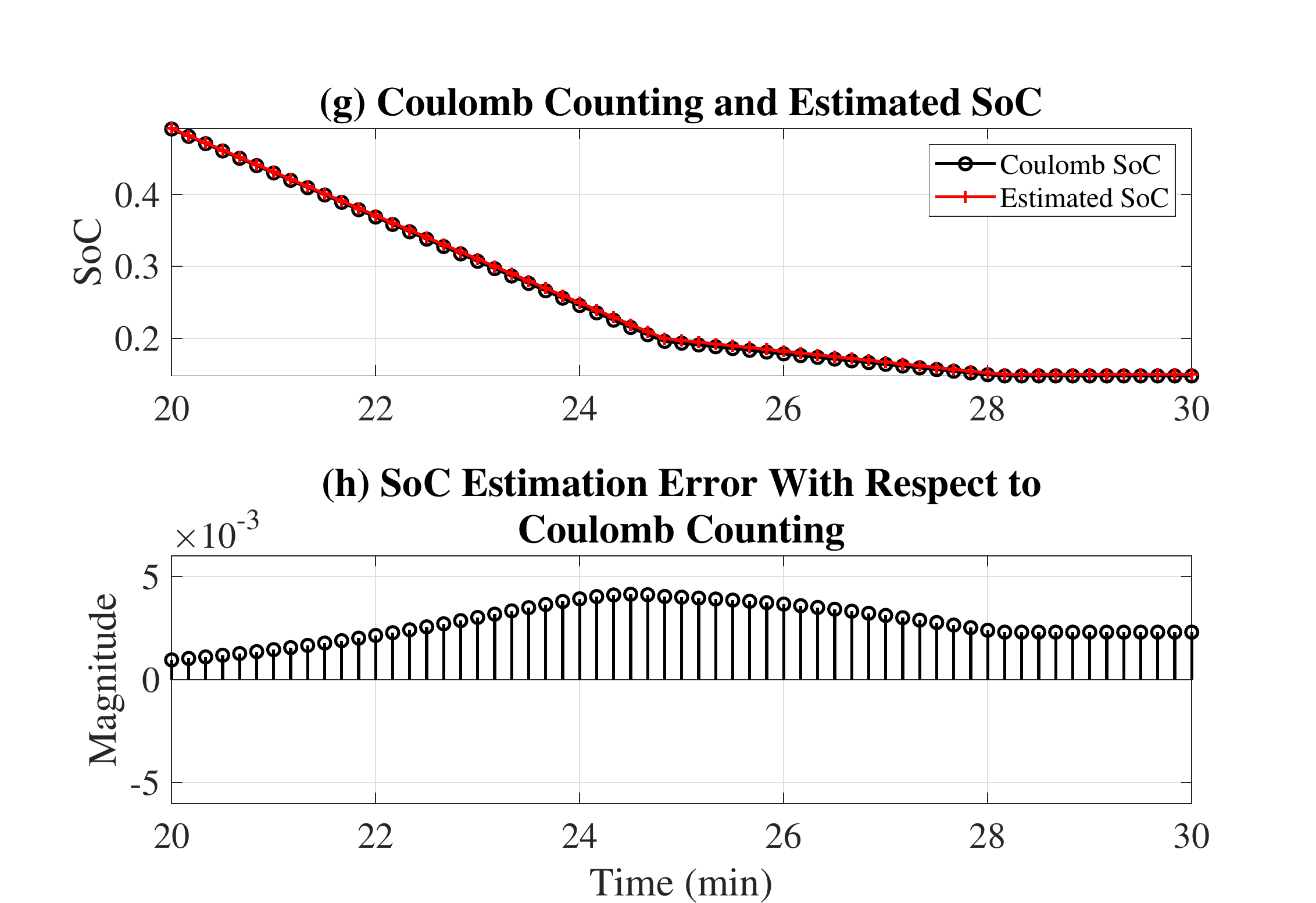}
		\label{fig:zoom}
	\end{subfigure}
	\caption{{Validation of estimated OCV and terminal voltage of Li-ion battery, and comparison of estimated SoC with Coulomb counting SoC when the battery is subjected to variable load. Sub-figures (g) and (h) show zoomed in view of portions of sub-figures (e) and (f) respectively.}}
	\label{fig4}
\end{figure*}

We construct and test two 4.1 V, 275 mAh Li-ion battery models in simulation for validating the estimated parameters results against those obtained by Chen and Mora \cite{CM}. The first model contains the parameters estimated by the proposed method, while the second one, set as a reference model, uses Chen and Mora's \cite{CM} parameters. Each battery model is subjected to a random discharge current as shown in Figure \ref{fig5_Current}, and their open circuit  and terminal voltages are compared in Figure \ref{fig4}(a)  and Figure \ref{fig4}(c), respectively. While their respective estimation errors are plotted in Figure \ref{fig4}(b) and \ref{fig4}(d), respectively. The low estimation error in both the open circuit and terminal voltage profiles in Figure \ref{fig4}(b) and Figure \ref{fig4}(d) show the accuracy of the proposed strategy. 

Finally, the estimated parameters are used to determine the SoC using the open circuit voltage via interpolation \cite{Daniyal}, with the discharge current shown in Figure \ref{fig5_Current}. The estimated SoC and the one obtained by conventional Coulomb counting method are plotted in Figure \ref{fig4}(e), while their difference is presented in figure \ref{fig4}(f). This difference becomes larger when a small current is drawn from the battery after 25 minutes. {Figures} \ref{fig4}(g) and \ref{fig4}(h) {show zoomed views of Figure} \ref{fig4}(e) and \ref{fig4}(f) {for the SoC estimation. Since the error in Figure} \ref{fig4}(e) and \ref{fig4}(f) {is relatively high in the 20 to 30 minutes interval, this range is selected zooming in the Figure} \ref{fig4}. In the Coulomb counting equation (\ref{pf1}), the small discharge current of the battery is divided by a comparatively much larger battery Ah capacity. Therefore, the Coulomb counting method does not capture small details of SoC when a low current is drawn from a battery. Thus the proposed methodology can improve the accuracy of SOC estimation.
{We would like to mention a few notable works that employ UAS-based strategies for robust control applications. For instance, the authors in} \cite{A1} {perform a series of rigorous tracking experiments using UAS for robust control applications. Theoretical justifications of these experiments are shared in} \cite{A2}. {Moreover, in} \cite{A3}, {a UAS-based strategy is used for robot motion control, and is tested by injecting noise. Recently, in} \cite{A4}, {a UAS-based strategy showed DC motor parameters estimation with good accuracy, in the presence of multiple sensor noises, i.e., current sensor and tachometer for current and speed measurements, respectively. The results in the above works that use UAS-based adaptation strategies for parameters estimation, not only ensure convergence but also guarantee accurate parameters estimation in the presence of sensor noise or external disturbances. Likewise, the proposed work also demonstrates the robustness of the UAS-based strategy by estimating the SoC over variable discharge current in Figure} \ref{fig4}(e)-\ref{fig4}(f) {in the manuscript - where the pattern of variable discharge current is shown in Figure} \ref{fig5_Current} {in the manuscript.}

\section{Experimental validation on a 22.2 V, 6.6 Ah Lithium-Polymer battery pack} \label{sec6}

\begin{figure}[!t]
	\centering
	\includegraphics[scale=0.40]{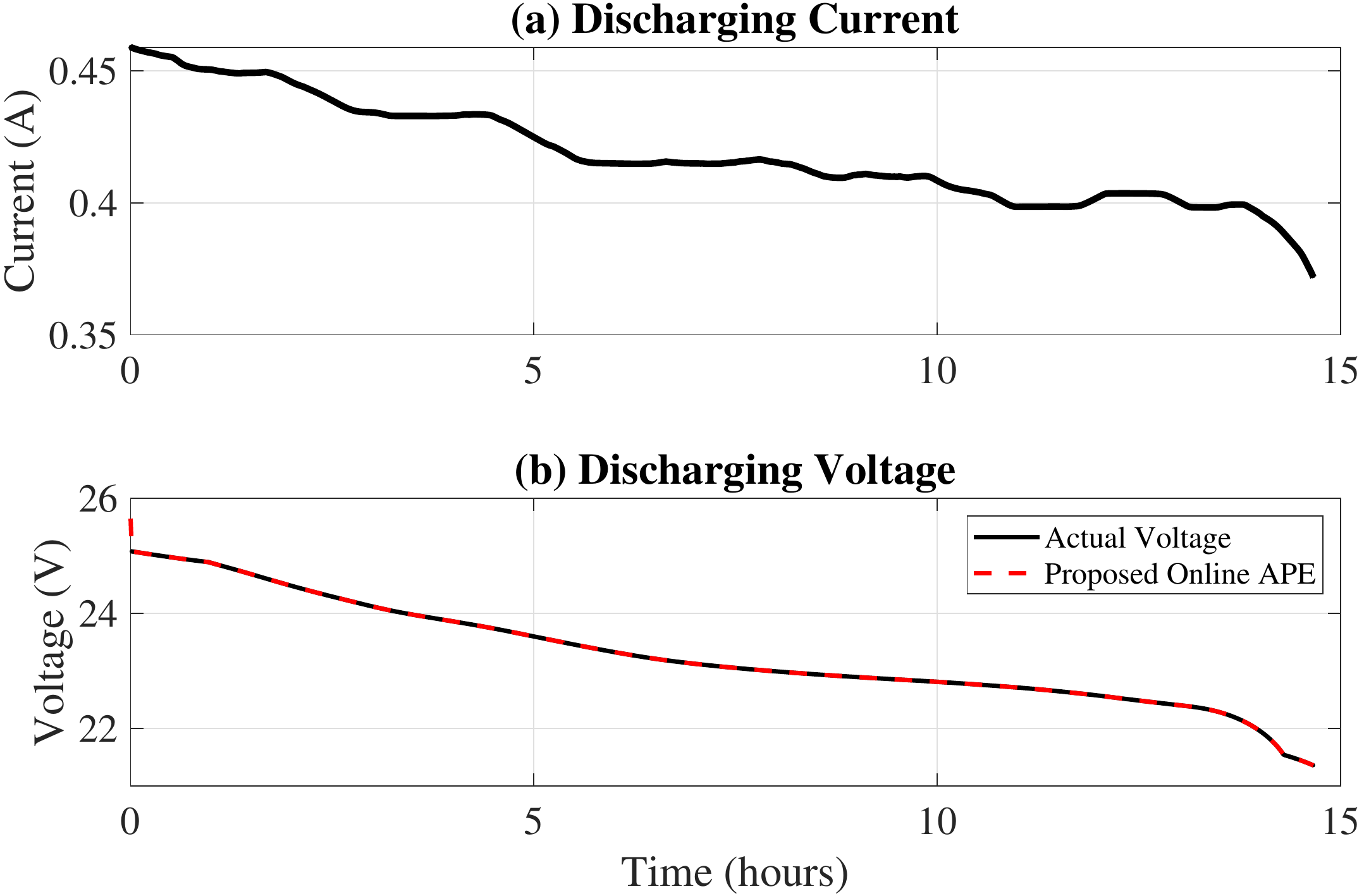}
	\caption{Lithium-Polymer battery discharging current and voltage profiles during adaptation process.}
	\label{figpf5}
\end{figure}

This section demonstrates rigorous experimental verification of the proposed online APE strategy on a 22.2 V, 6.6 Ah Li-ion battery, for sixteen different discharging and sixteen constant current charging profiles. The results of the proposed online APE strategy are comprehensively compared with the existing APE results \cite{Daniyal}, which is termed as `reference offline APE' in this work. The accuracies of the estimated terminal voltage obtained by the proposed online APE strategy and the reference offline APE strategy, are evaluated by comparison against the measured battery terminal voltage. Also, a comparative analysis is performed between the proposed online APE strategy and the reference offline APE strategy by considering the histogram and cumulative distribution of terminal voltage estimation errors. The results of the proposed online APE strategy are compared with the reference offline APE strategy. A fully charged 22.2 V, 6.6Ah TP6600 6S, 25C Lipo battery is connected with a resistive load of 50 ohms, which allows a small discharging current of about 0.4 amps. Note that a small discharging current ensures the convergence of estimated equivalent circuit elements to their actual values, as per the mathematical proof provided in section \ref{sec4}.
The battery is discharged  up to  7\% of its rated capacity in about 15 hours. The measured battery discharging current, measured and estimated terminal voltage using the proposed online APE strategy, are plotted in Figure \ref{figpf5} subplots (a) and (b) respectively.

\begin{table*}[!t]
	\centering
	\vspace{1em} 
	\caption{Experimental results of a 22 V, 6.6 Ah Li-ion battery model parameters.} 
	\begin{adjustbox}{width=0.8\linewidth} 
		\begin{tabular}{c c c c c c c c c} 
			\hline\hline 
			Parameter & \thead{Upper bound \\$ (r_{nu}) $} & \thead{Lower bound \\$ (r_{nl}) $} & $ \lambda_{x_n} $ & $ \lambda_{y_n} $ & \thead{Initial\\ value} & \thead{Estimated\\ value} & \thead{Estimated value\\ provided in \cite{Daniyal}} & \thead{Estimation\\ error (\%)} \\ [1ex] 
			\hline
			
			$\widehat{r}_{1}$ & 6 & 4 & 50 & 50 & 100 & 5 & 5.112 & 2.19 \\ 
			$\widehat{r}_{2}$ & 50 & 30 & 50 & 50 & 2000 & 40 & 40.955 & 2.33  \\ 
			$\widehat{r}_{3}$ & -- & -- & -- & -- & -- & 22.1782 & 22.195 & 0.07 \\ 
			$\widehat{r}_{4}$ & 3 & 1 & 50 & 50 & 50 & 2 & 1.9215 & 4.08  \\ 
			$\widehat{r}_{5}$ & 2.5 & 1 & 50 & 50 & 30 & 1.75 & 1.759 & 0.51  \\ 
			$\widehat{r}_{6}$ & 4 &2 & 50 & 50 & 200 & 3 & 3.0435 & 1.43 \\ 
			$\widehat{r}_{7}$ & 1 & 0.1 & 50 & 50 & 180 & 0.5505 & 0.5505 & 0  \\
			$\widehat{r}_{8}$ & 50 & 10 & 50 & 50 & 1700 & 30 & 30.0475 & 0.16 \\ 
			$\widehat{r}_{9}$ & 0.1 & 0.01 & 50 & 50 & 240 & 0.055 & 0.0551 & 0.18 \\ 
			$\widehat{r}_{10}$ & 10 & 1 & 70 & 50 & 3600 & 6.2557 & 6.2585 & 0.04 \\ 
			$\widehat{r}_{11}$ & 50 & 10 & 50 & 50 & 9300 & 30 & 30 & 0 \\ 
			$\widehat{r}_{12}$ & 0.1 & 0.01 & 50 & 50 & 264 & 0.0555 & 0.0551 & 0.73 \\ 
			$\widehat{r}_{13}$ & 1000 & 500 & 60 & 55 & 50000 & 760.8691 & 760.2266 & 0.08 \\ 
			$\widehat{r}_{14}$ & 15 & 5 & 70 & 50 & 1000 & 10.8334 & 10.7686 & 0.60 \\ 
			$\widehat{r}_{15}$ & 800 & 500 & 80 & 50 & 50000 & 684.615 & 685.7457 & 0.16 \\ 
			$\widehat{r}_{16}$ & 7000 & 5000 & 10 & 10 & 50000 & 6000 & 6036.4 & 0.60  \\ 
			$\widehat{r}_{17}$ & 50 & 5 & 50 & 50 & 1000 & 27.5 & 27.5422 & 0.15 \\ 
			$\widehat{r}_{18}$ & 5000 & 3000 & 10 & 20 & 50000 & 3667 & 3696 & 0.78  \\ 
			$\widehat{r}_{19}$ & 0.1 & 0.01 & 50 & 50 & 60 & 0.0551 & 0.0439 & 25.5  \\ 
			$\widehat{r}_{20}$ & 70 & 50 & 50 & 50 & 1200 & 60 & 59.07 & 1.57 \\ 
			$\widehat{r}_{21}$ & -- & -- & -- & -- & -- & 0.2408 & 0.2246 & 7.21 \\ 
			
			\hline 
		\end{tabular}
		\label{tab7}
	\end{adjustbox} 
\end{table*}

\begin{figure}[!t]
	\centering
	\includegraphics[scale=0.45]{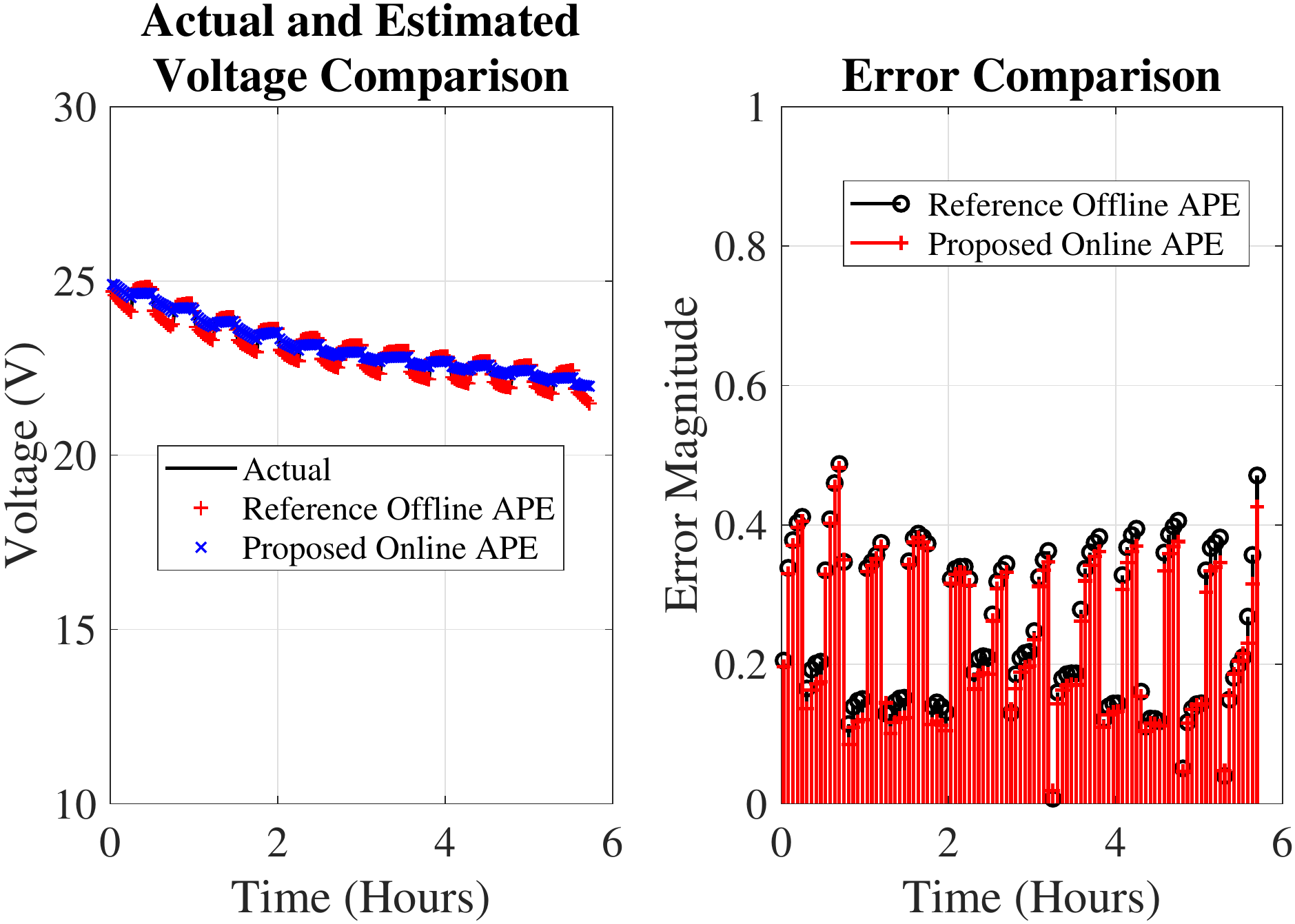}
	\caption{Terminal voltage estimation and absolute error $|e(t)|$ comparison for resistive load of 11.11 $\Omega$ with 15 minutes ON and 15 minutes OFF times.}
	\label{figpf6}
\end{figure}

\begin{figure}[!t]
	\centering
	\includegraphics[scale=0.45]{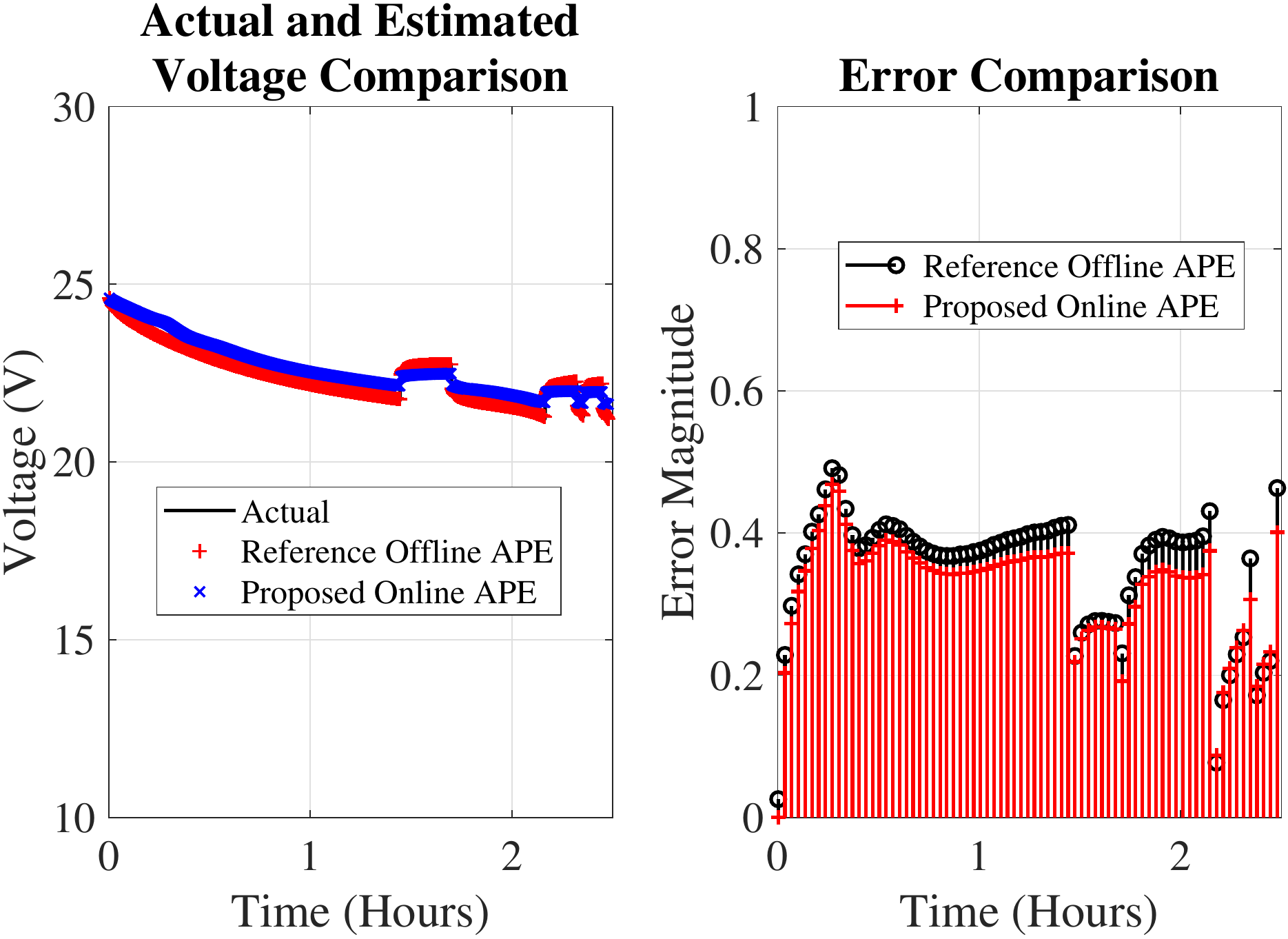}
	\caption{Terminal voltage estimation and absolute error $|e(t)|$ comparison for resistive load of 7.5 $\Omega$ with random time period.}
	\label{figpf7}
\end{figure}

\begin{figure}[!b]
	\centering
	\includegraphics[scale=0.4]{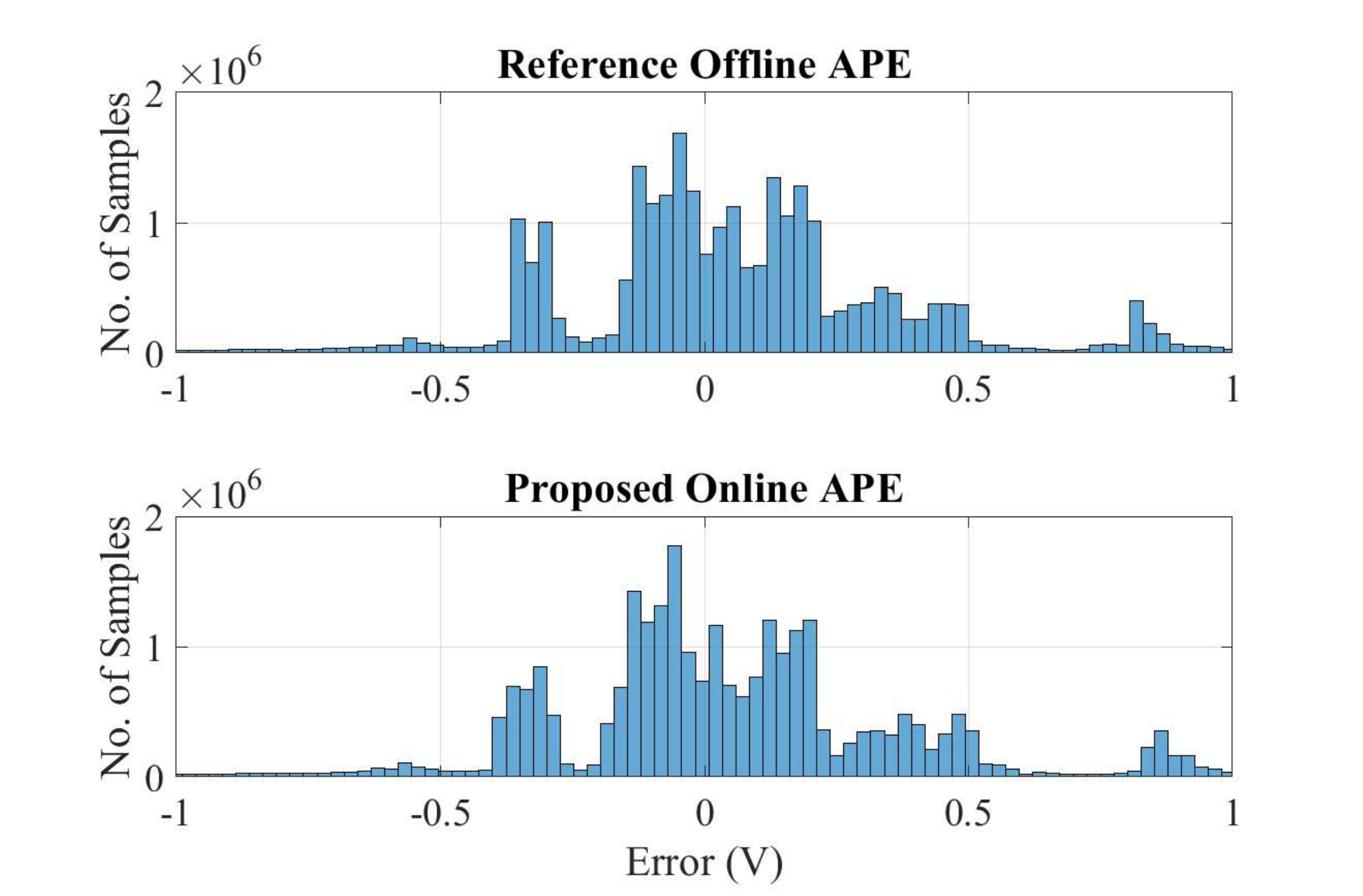}
	\caption{Histogram of terminal voltage estimation error for reference offline APE and proposed online APE under sixteen different discharging profiles.}
	\label{figpf8}
\end{figure}

\begin{figure}[!t]
	\centering
	\includegraphics[scale=0.45]{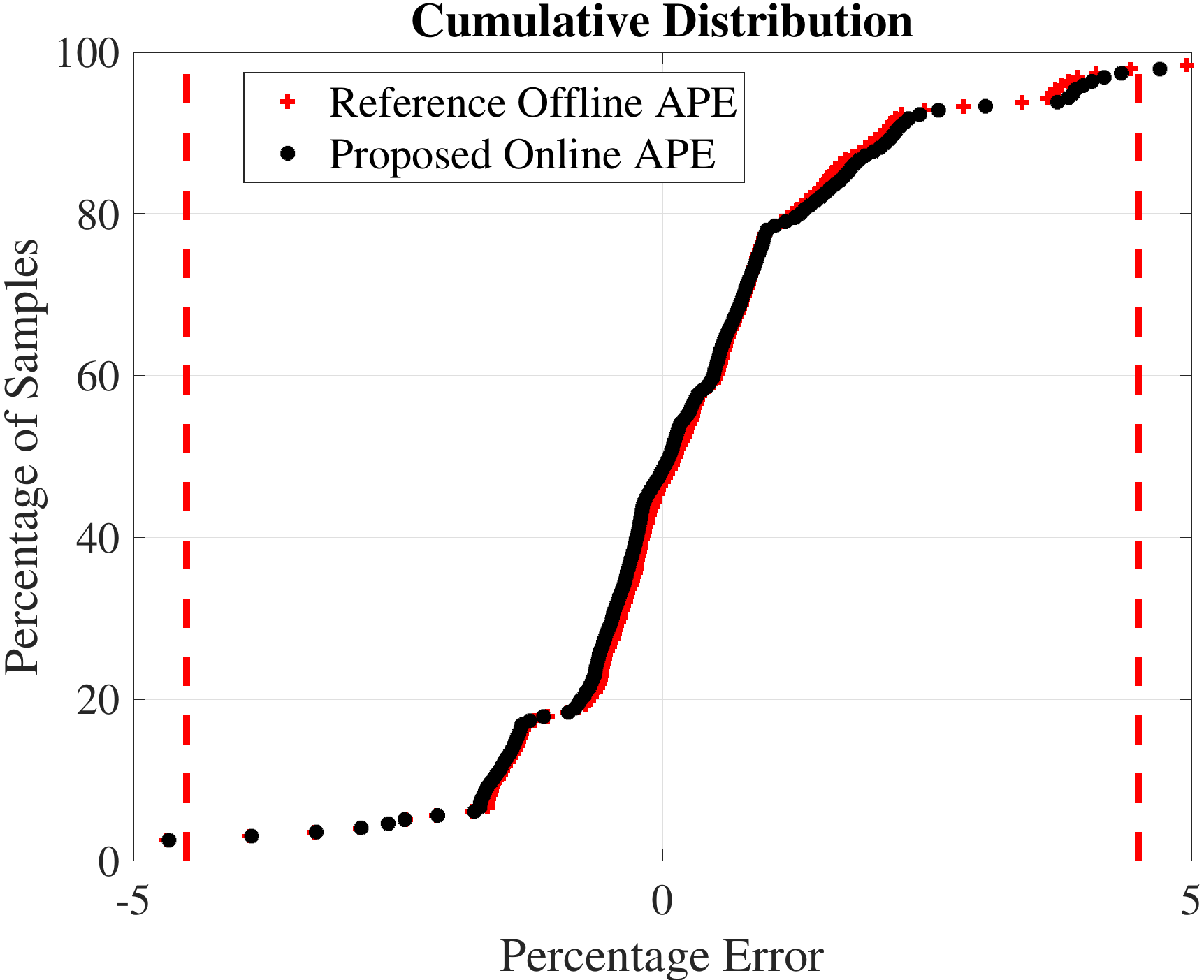}
	\caption{Cumulative distribution of terminal voltage estimation error for reference offline APE and proposed online APE under sixteen different discharging profiles.}
	\label{figpf9}
\end{figure}

The battery parameters, $\widehat{r}_{1},\cdots, \widehat{r}_{21}$ are estimated online using Algorithm \ref{algo2} and recorded in 21 separate arrays. The initial values, steady-state upper and lower bounds chosen for the estimated parameters satisfy the conditions mentioned in Lemmas \ref{lemma4.1}, \ref{lemma4.3}. The battery terminal voltage is estimated and compared with the actual voltage. The value of the estimated parameters is picked and considered a good estimate, when the terminal voltage estimation error gets below the user defined error bound. The average value of such good estimates of a particular parameter is finally considered as the estimated parameter's value, and the estimation process is ended. 

The sampling time of the discharging voltage and current is set to 0.01 seconds, and the final values of the estimated parameters are obtained in about two seconds when the battery terminal voltage estimation error decreased to $0.1\times10^{-5}$ as shown in Figure \ref{fig2}. The results of estimated parameters obtained from the proposed online APE strategy are provided in Table \ref{tab7}, and are compared with the results of the reference offline APE technique shared in \cite{Daniyal}.
{In Table} \ref{tab7}, {we use the values from} \cite{Daniyal} {as benchmark for 22.2 V, 6.6 Ah Lithium-ion battery. The authors in} \cite{Daniyal} {perform 32 experiments, sixteen different discharging and sixteen constant charging profiles. The average error for a set of sixteen different discharging profiles is 0.1\%, and 1.7\% for sixteen constant charging profiles. Therefore, owing to high accuracy of parameters provided in} \cite{Daniyal}, {we use them as reference in Table} \ref{tab7}.
Note that in Table \ref{tab7}, values related to parameters $\widehat{r}_3$ and $\widehat{r}_{21}$ are shown by dashes. This is because $\widehat{r}_3$ and $\widehat{r}_{21}$ disappear from the observer equations used in the proposed online APE strategy. So, parameters $\widehat{r}_3$ and $\widehat{r}_{21}$ are not estimated adaptively, but are calculated using equations (\ref{ls1})-(\ref{ls2}).  Also, the aim of the proposed online APE strategy is to reduce the experimental effort required compared to the reference offline APE strategy \cite{Daniyal} which performs offline estimation of open circuit voltage and series resistance.

{Note that the battery parameters and SoC level cannot be measured, but rather estimated using a battery model. The only possible way to quantify the accuracy of estimated battery parameters, based on physically measurable ground truth, is that these parameters should be able to predict the actual battery terminal voltage at any load, and this could be compared with the measured ground truth terminal voltage. Moreover, terminal voltage comparison is widely used in the literature to quantify the accuracy of a battery model parameters estimation} \cite{CM,B2,Access,Daniyal}.

Therefore, in the next section, the results of the proposed online APE strategy are comprehensively compared with the reference offline APE technique for sixteen different discharging load protocols and sixteen constant current charging protocols. The detailed description of the discharging load protocols is given in our previous work \cite{Access}. %{[We are here 16 July 2020]}.

\begin{table}[!t]
	\centering
	\caption{Terminal voltage estimation error statistics while discharging the battery with sixteen different load profiles for reference offline APE and proposed online APE.}
	\begin{adjustbox}{width=1\linewidth}
		\begin{tabular}{c c c c c} 
			\hline\hline
			\thead{Parameters estimation \\ methods}  & \thead{Mean of\\ error (V)} & \thead{Median of\\error (V)} & \thead{Mode of\\error (V)} & \thead{Standard deviation\\ of error (V)}  \\
			\hline \\       
			Reference Offline APE & 0.0211 & 0.027 & -0.4038 & 0.5026  \\    
			%\hline
			Proposed Online APE & 0.0218 & 0.0143 & -0.347 & 0.5139  \\   
			\hline
		\end{tabular}
		\label{tab8}
	\end{adjustbox}
\end{table}

\subsection{Parameters estimation accuracy assessment via battery discharging tests}
\quad Sixteen different discharging load profiles are successively applied to 22.2 V, 6.6 AH Li-ion battery and the battery terminal voltage is estimated online using the acquired battery model.
As a sample, the estimated and measured terminal voltages along with the absolute voltage estimation error for two of the sixteen discharging load profiles are shown in Figure \ref{figpf6} and Figure \ref{figpf7}. The voltage estimation error in Figure \ref{figpf6} and Figure \ref{figpf7} shows that the proposed online APE strategy produces similar results as compared to the reference offline APE technique. The terminal voltage estimation error data, for all sixteen discharging profiles, is stacked together to form a single large `error array' of $2.75\mathrm{e}{7}$ samples. The statistical analysis of terminal voltage estimation error array is performed to quantify the accuracy of the proposed online APE strategy against the reference offline APE technique. 
The mean, median, mode, and standard deviation analysis of the error array for proposed online APE and reference offline APE strategies are provided in Table \ref{tab8}. The mean value of the error array for proposed and reference APE methods are 0.0211 V and 0.0218 V, respectively. Whereas, the median value of the error array for proposed and reference APE methods are 0.027 V and 0.0143, respectively. Similarly, the mode value for proposed and reference APE methods are -0.4038 and -0.347, respectively. Likewise, the standard deviation value for proposed and reference APE methods are found to be 0.5026 and 0.5139, respectively.
The mean and standard deviation values for both techniques are very similar, while the median and mode values of proposed APE strategy slightly deviate from that of referenced APE technique \cite{Daniyal}.

\begin{figure}[!t]
	\centering
	\includegraphics[scale=0.45]{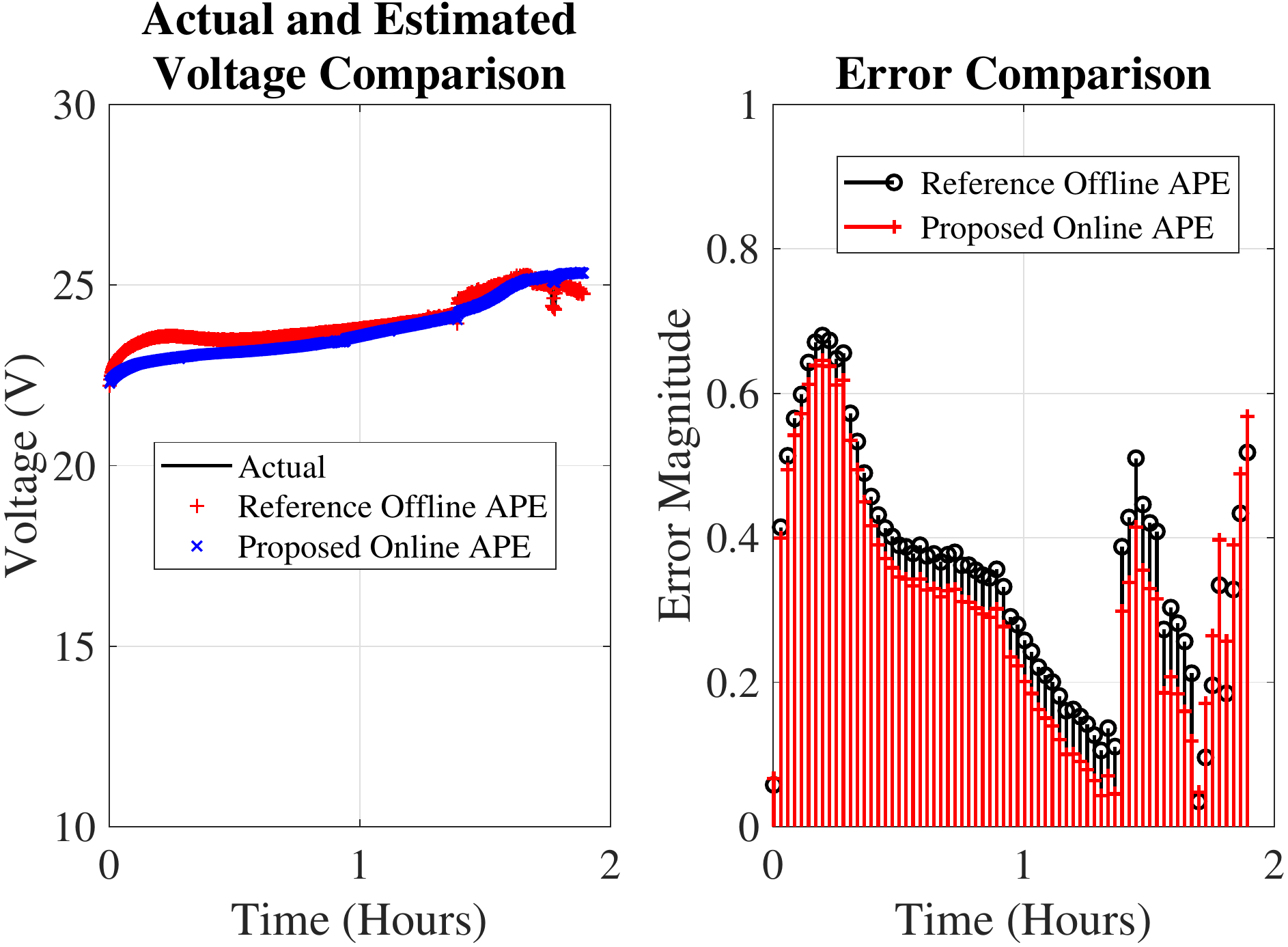}
	\caption{Terminal voltage estimation and absolute error $|e(t)|$ comparison while charging the 22.2 V, 6.6 Ah Li-Polymer battery.}
	\label{figpf10}
\end{figure}

An extensive investigation of the overall terminal voltage estimation error array is carried out by showing its histogram and cumulative distribution graphs in Figure \ref{figpf8} and Figure \ref{figpf9} respectively. Where, the red vertical lines in Figure \ref{figpf9} indicate the $\pm$ 4.5\% terminal voltage estimation error limits, i.e. $\pm$ 1 V. Figure \ref{figpf8} and Figure \ref{figpf9} show no significant difference between the proposed online APE results compared to the reference offline APE technique. %As mentioned in \cite{Access}, the accuracy of the proposed online APE and reference offline APE strategies can be further enhanced by subsequent incorporation of optimization strategies.

\subsection{Parameters estimation accuracy assessment via battery charging tests}

\begin{figure}[!t]
	\centering
	\includegraphics[scale=0.4]{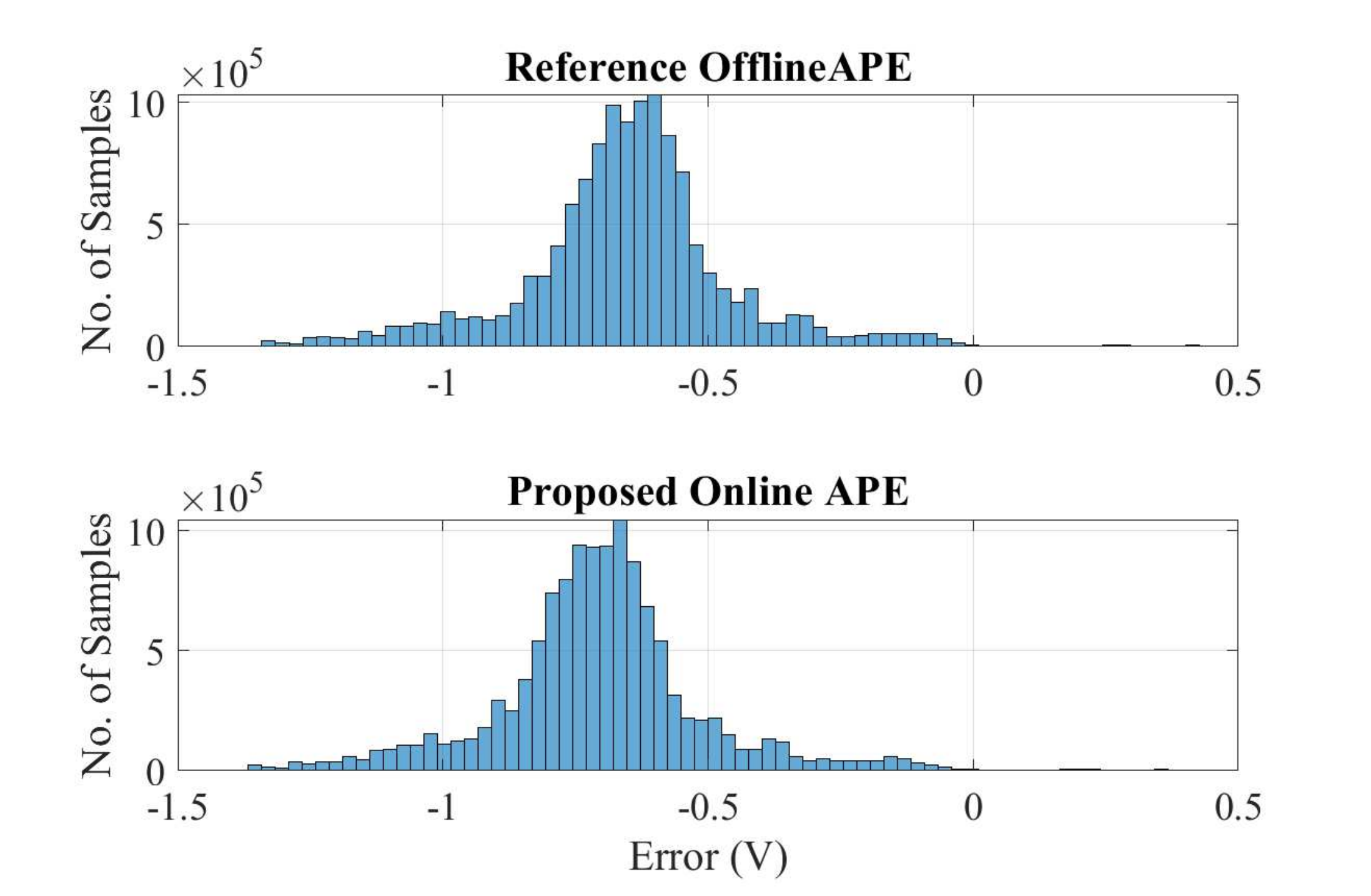}
	\caption{Histogram of terminal voltage estimation error for reference offline APE and proposed online APE techniques while charging sixteen individual batteries with a constant 2.5 A current.}
	\label{figpf11}
\end{figure}

\begin{figure}[!t]
	\centering
	\includegraphics[scale=0.45]{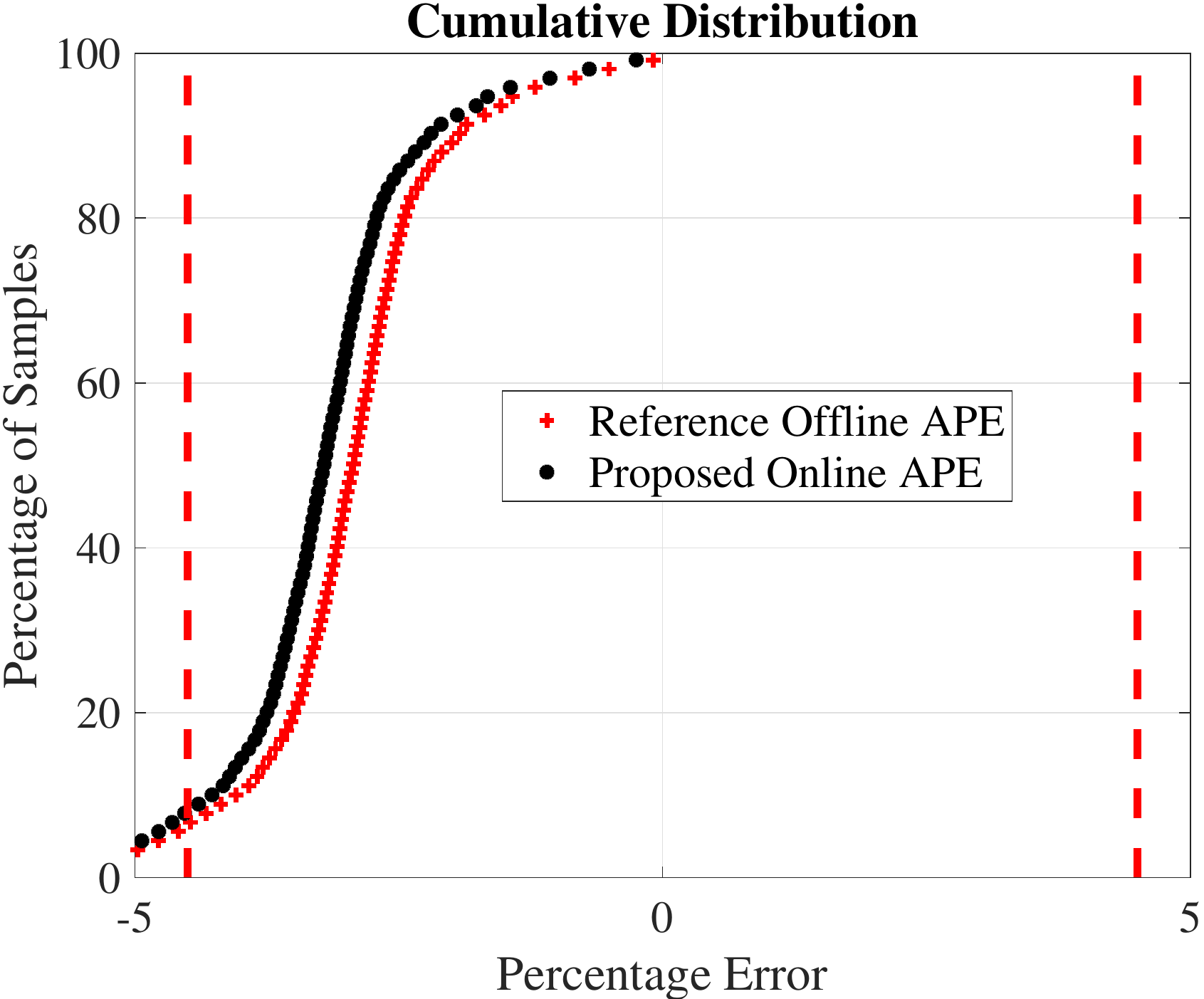}
	\caption{Cumulative distribution of terminal voltage estimation error for reference offline APE and proposed online APE techniques while charging sixteen individual batteries with a constant 2.5 A current.}
	\label{figpf12}
\end{figure}

\quad The estimated parameters obtained from the proposed online APE strategy are further assessed against the results obtained using the reference offline APE technique for sixteen constant current charging protocols. The actual Lipo battery is charged with a constant current of 2.5 amperes using the Thunder-Power charger (TP820CD). As a sample, the estimated and measured terminal voltages along with the absolute voltage estimation error for a single test are shown in Figure \ref{figpf10}. The statistical analysis, similar to discharging load protocols, is performed for comparing the terminal voltage estimation errors of both the proposed and reference APE strategies. The total number of samples collected in the terminal voltage estimation `error array' while charging the batteries are $1.258\mathrm{e}{7}$, for both the proposed and reference APE methods. The histogram and cumulative distribution graphs of the error array are shown Figure \ref{figpf11} and Figure \ref{figpf12}, respectively, for both the proposed and reference APE strategies. Moreover, the statistical analysis of terminal voltage estimation error is provided in Table \ref{tab9}. The mean value of the error array for proposed and reference APE methods are -0.6518 V and -0.7080 V, respectively. Whereas, the median value of the error array for proposed and reference APE methods are -0.6451 V and -0.7059, respectively. Similarly, the mode value for proposed and reference APE methods are -2.1223 and -2.1470, respectively. Likewise, the standard deviation value for proposed and reference APE methods are found to be 0.2271 and 0.2231, respectively.
The statistical analysis along with histogram and cumulative distribution graphs show that the proposed online APE strategy produces results that are comparable to the reference offline APE technique while charging a Lipo battery.

\begin{figure*}[!t]
	\centering
	\includegraphics[scale=0.55]{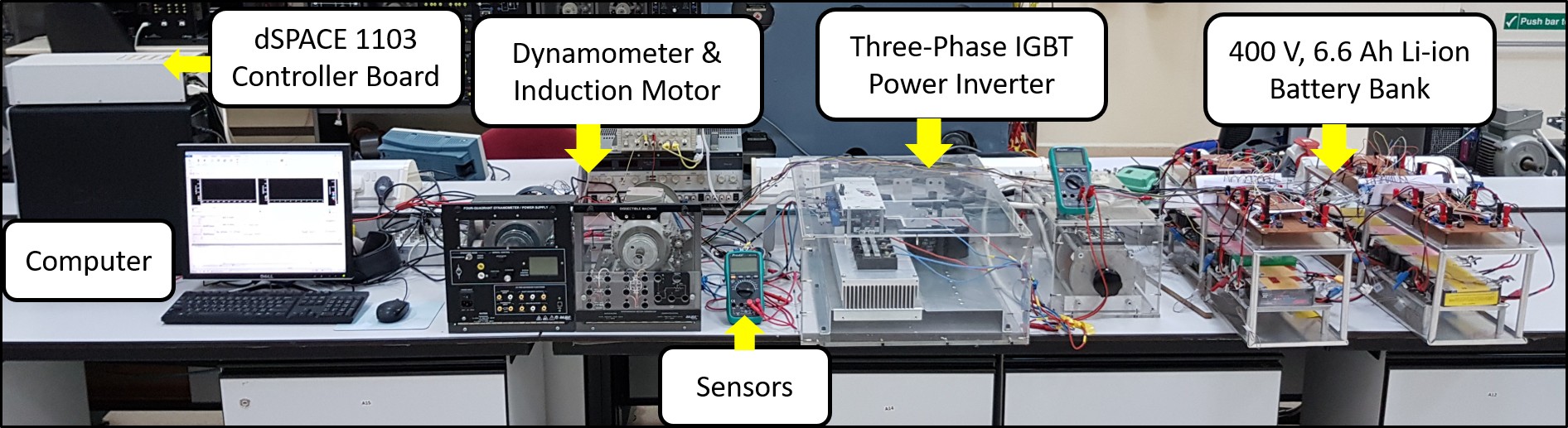}
	\caption{Li-ion battery bank powered EV traction system \cite{usman2019}.}
	\label{figsetup}
\end{figure*}

\begin{table}[!t]
	\centering
	\caption{Terminal voltage estimation error statistics while charging sixteen different batteries with a constant 2.5 Amperes for reference offline APE and proposed online APE.}
	\begin{adjustbox}{width=1\linewidth}
		\begin{tabular}{c c c c c} 
			\hline\hline
			\thead{Parameters estimation \\ Methods}  & \thead{Mean of\\error (V)} & \thead{Median of\\error (V)} & \thead{Mode of\\error (V)} & \thead{Standard deviation\\ of error (V)}  \\
			\hline \\       
			Reference Offline APE & -0.6518 & -0.6451 & -2.1223 & 0.2271  \\    
			%\hline
			Proposed Online APE & -0.7080 & -0.7059 & -2.1470 & 0.2231  \\   
			\hline
		\end{tabular}
		\label{tab9}
	\end{adjustbox}
\end{table}

\begin{table*}[!t]
	\centering 
	\caption{Experimental results of a 400 V, 6.6 Ah Li-ion battery bank model parameters.} 
	\begin{adjustbox}{width=0.8\linewidth} 
		\begin{tabular}{c c c c c c c c c} 
			\hline\hline 
			Parameter & \thead{Upper bound \\$ (r_{nu}) $} & \thead{Lower bound \\$ (r_{nl}) $} & $ \lambda_{x_n} $ & $ \lambda_{y_n} $ & \thead{Initial\\ value} & \thead{Estimated value\\ (Real-time)} & \thead{Estimated value\\ (Offline)} & \thead{Estimation\\ error (\%)} \\ [1ex] 
			\hline
			
			$\widehat{r}_{1}$ & 150 & 45 & 50 & 50 & 100 & 97.5 & 97.51 & 0.01 \\ 
			$\widehat{r}_{2}$ & 50 & 20 & 50 & 50 & 2000 & 35 & 35.01 & 0.03  \\ 
			$\widehat{r}_{3}$ & -- & -- & -- & -- & -- & 356.865 & 357.236 & 0.1 \\ 
			$\widehat{r}_{4}$ & 7.5 & 1.5 & 50 & 50 & 100 & 4.5 & 5.2 & 13.4  \\ 
			$\widehat{r}_{5}$ & 20 & 2 & 50 & 50 & 230 & 11 & 11.01 & 0.1  \\ 
			$\widehat{r}_{6}$ & 50 & 25 & 50 & 50 & 400 & 37.5 & 37.55 & 0.13 \\ 
			$\widehat{r}_{7}$ & 1 & 0.1 & 50 & 50 & 180 & 0.6125 & 0.5643 & 8.54  \\
			$\widehat{r}_{8}$ & 50 & 10 & 50 & 50 & 1700 & 30 & 30.01 & 0.03 \\ 
			$\widehat{r}_{9}$ & 0.1 & 0.01 & 50 & 50 & 240 & 0.0568 & 0.069 & 17.68 \\ 
			$\widehat{r}_{10}$ & 10 & 1 & 70 & 50 & 3600 & 6.4074 & 6.262 & 2.32 \\ 
			$\widehat{r}_{11}$ & 200 & 100 & 50 & 50 & 9300 & 150 & 150 & 0 \\ 
			$\widehat{r}_{12}$ & 0.1 & 0.01 & 50 & 50 & 264 & 0.0694 & 0.0693 & 0.14 \\ 
			$\widehat{r}_{13}$ & 1000 & 500 & 60 & 55 & 50000 & 760.8586 & 760.882 & 0.003 \\ 
			$\widehat{r}_{14}$ & 15 & 5 & 70 & 50 & 1000 & 10.8367 & 10.845 & 0.07 \\ 
			$\widehat{r}_{15}$ & 800 & 500 & 80 & 50 & 50000 & 684.6064 & 684.626 & 0.16 \\ 
			$\widehat{r}_{16}$ & 7000 & 5000 & 10 & 10 & 50000 & 5998.5 & 6000 & 0.025  \\ 
			$\widehat{r}_{17}$ & 50 & 5 & 50 & 50 & 1000 & 27.507 & 27.514 & 0.025 \\ 
			$\widehat{r}_{18}$ & 5000 & 3000 & 10 & 20 & 50000 & 3666 & 3666.71 & 0.02  \\ 
			$\widehat{r}_{19}$ & 25 & 5 & 50 & 50 & 100 & 15 & 15.014 & 0.1  \\ 
			$\widehat{r}_{20}$ & 40 & 15 & 50 & 50 & 1200 & 27.505 & 27.514 & 0.033 \\ 
			$\widehat{r}_{21}$ & -- & -- & -- & -- & -- & 5.01 & 5.428 & 7.7 \\ 
			
			\hline 
		\end{tabular}
		\label{tab400}
	\end{adjustbox} 
\end{table*}

In the next section, the proposed online APE strategy is employed for real-time parameters estimation of a 400 V, 6.6 Ah, Li-ion battery bank. The Li-ion battery bank is utilized to power an indirect field-oriented control based electric vehicle (EV) traction system. The real-time estimated parameters are also validated against the offline results on a 400 V, 6.6 Ah Li-ion battery bank. 

\section{Online Parameters Estimation of a 400 V, 6.6 Ah Lithium-Polymer Battery Bank : Powering a Prototype Electric Vehicle Traction System} \label{sec7}

\begin{figure}[!t]
	\centering
	\includegraphics[scale=0.42]{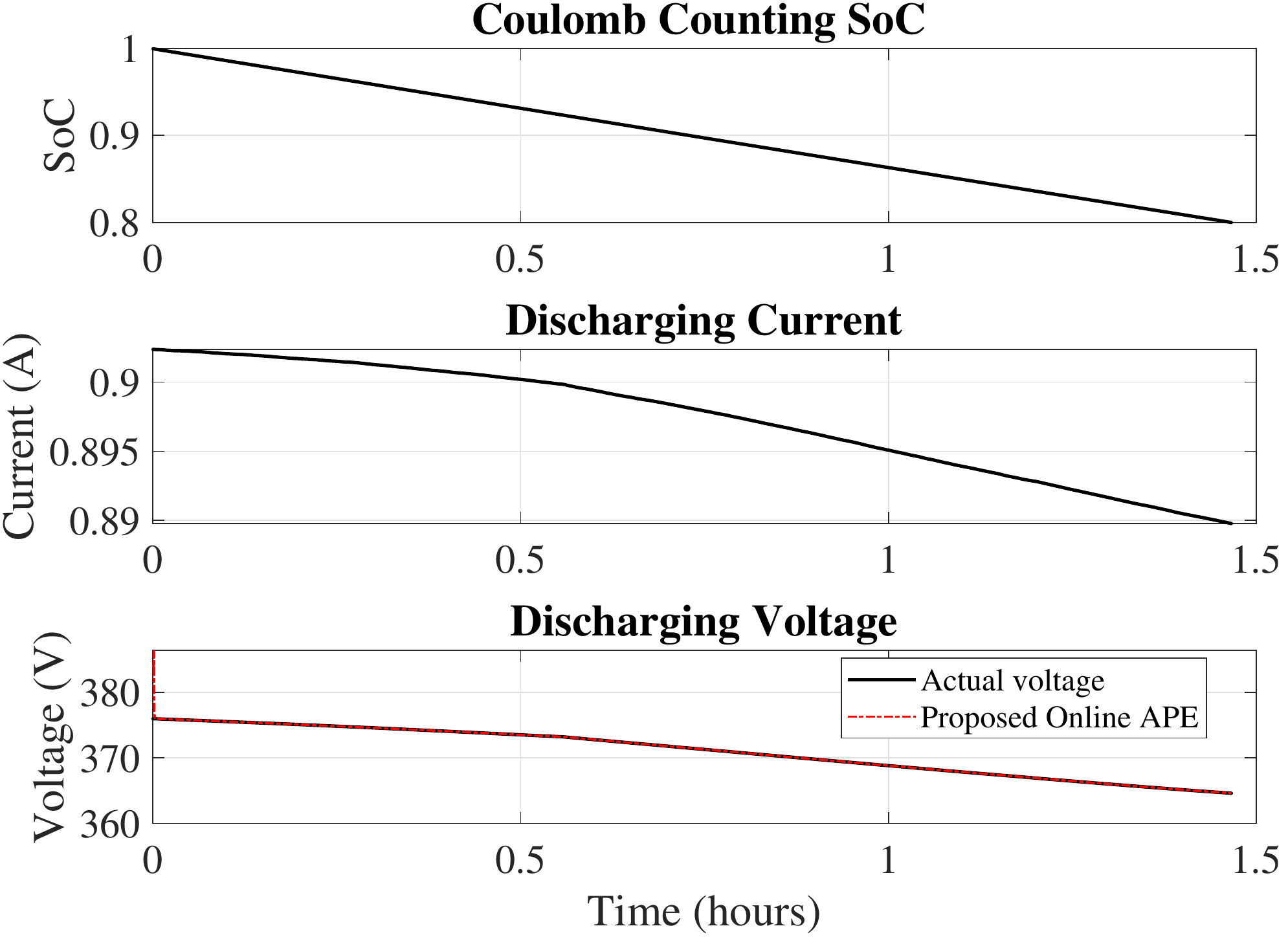}
	\caption{Lithium-Polymer battery bank discharging SoC, current, and voltage profiles during adaptation process.}
	\label{figpf13}
\end{figure}

In this section, the proposed online APE strategy is implemented and validated on a 400 V, 6.6 Ah Lithium-Polymer battery bank which is powering an indirect field-oriented induction motor driven electric vehicle (EV) traction system. The validation of real-time estimated parameters against the offline experimentations shows the suitability of the proposed online APE strategy for real-time parameters estimation of a Li-ion battery either at pack level or bank level. The picture of a complete prototype EV traction testbench is shown in Figure \ref{figsetup} \cite{usman2019}. The real-time adaptive parameters estimation of a 400 V, 6.6 Ah Li-ion battery bank is performed by running algorithm \ref{algo2} with all the required conditions described in section \ref{sec4}. The Li-ion battery bank powers an indirect field-orientation induction motor based EV traction system. The no-load operation of an induction motor in EV traction system draws around 0.2 amperes current and, thus, satisfies one of the essential conditions, i.e. the low discharge current requirement, of UAS based parameters estimation method. The estimated parameters at no-load operation of an induction motor in EV traction system are presented in Table \ref{tab400}. Note that in Table \ref{tab400}, certain values related to parameters $\widehat{r}_3$ and $\widehat{r}_{21}$ are shown by dashes. This is because $\widehat{r}_3$ and $\widehat{r}_{21}$ disappear from the observer equations used in the proposed online APE strategy. However, these parameters are calculated in real-time using equations (\ref{ls1})-(\ref{ls2}). The battery parameters estimated at no-load condition can be employed for SoC and SoH estimation, open circuit voltage and series resistance estimation, and fault detection in a battery management system during any loading condition of EV traction system.

\begin{figure*}[!t]
	\centering
	\includegraphics[scale=0.5]{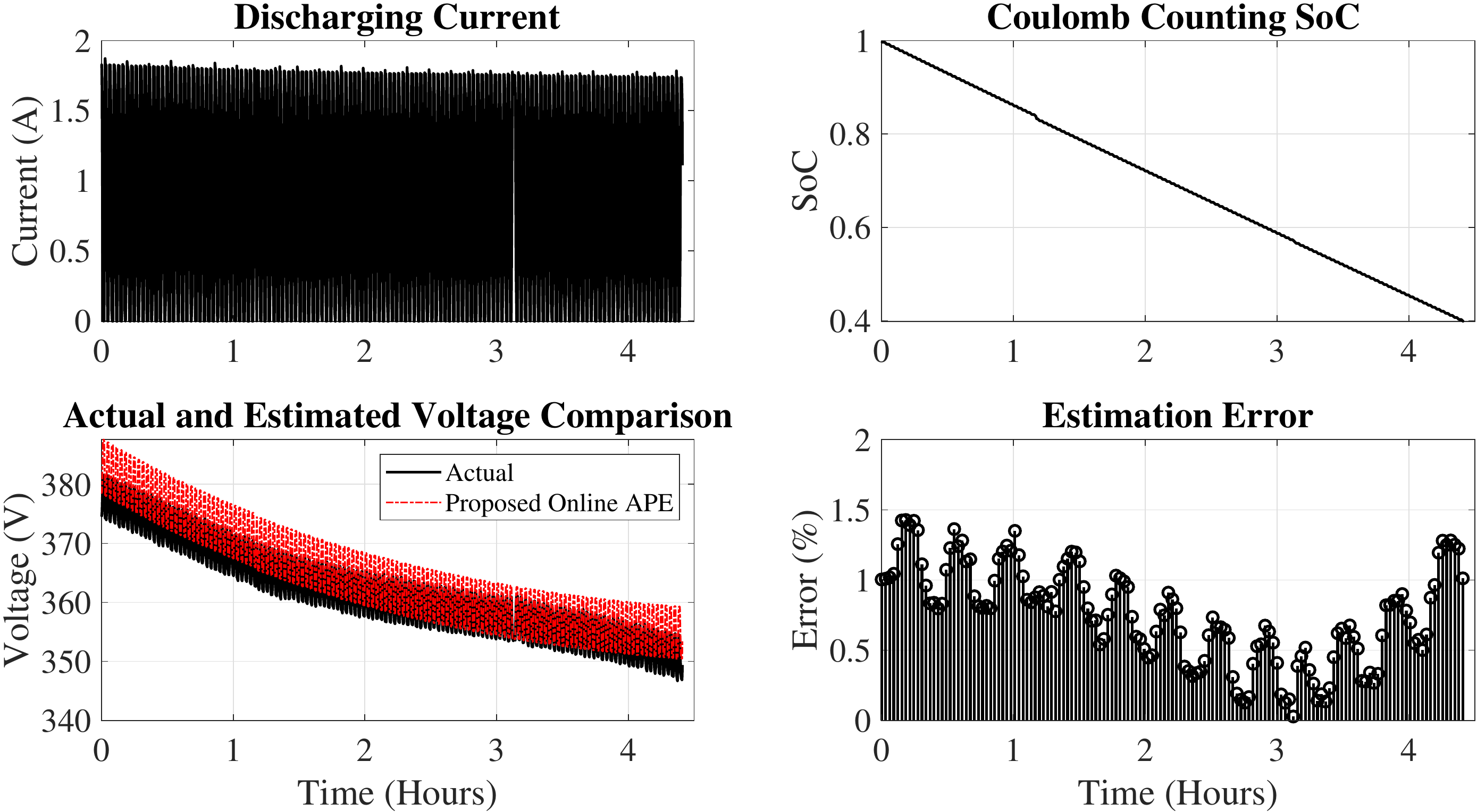}
	\caption{Terminal voltage estimation and absolute error $|e(t)|$ comparison for resistive load of 230 ohms, 1000 W, with 1 minutes ON and 1 minute OFF times.}
	\label{figpf14}
\end{figure*}

\begin{figure}[!t]
	\centering
	\includegraphics[scale=0.4]{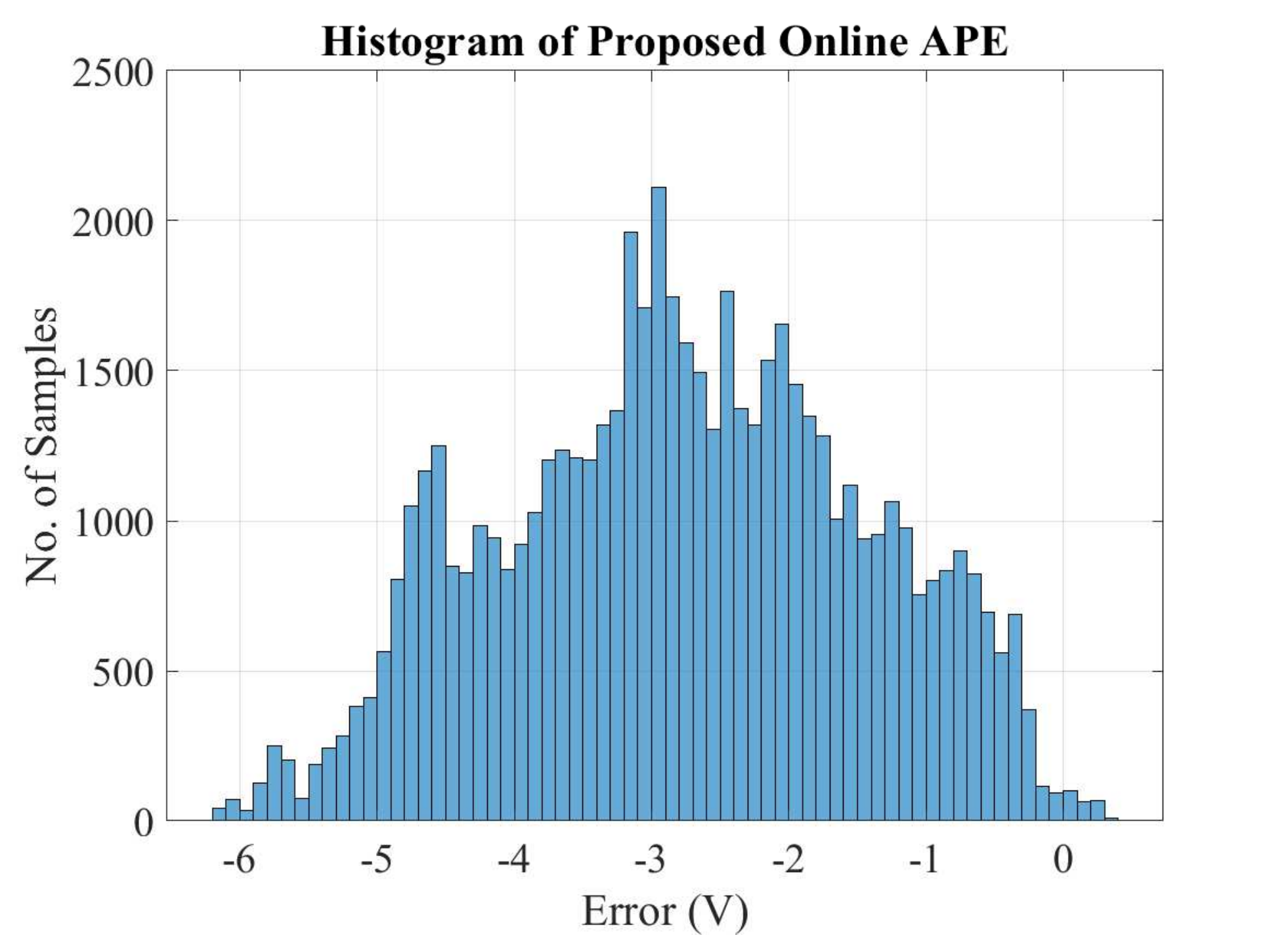}
	\caption{Histogram of terminal voltage estimation error for proposed online APE under Figure \ref{figpf14} battery bank discharge profile.}
	\label{figpf15}
\end{figure}

\begin{figure}[!t]
	\centering
	\includegraphics[scale=0.4]{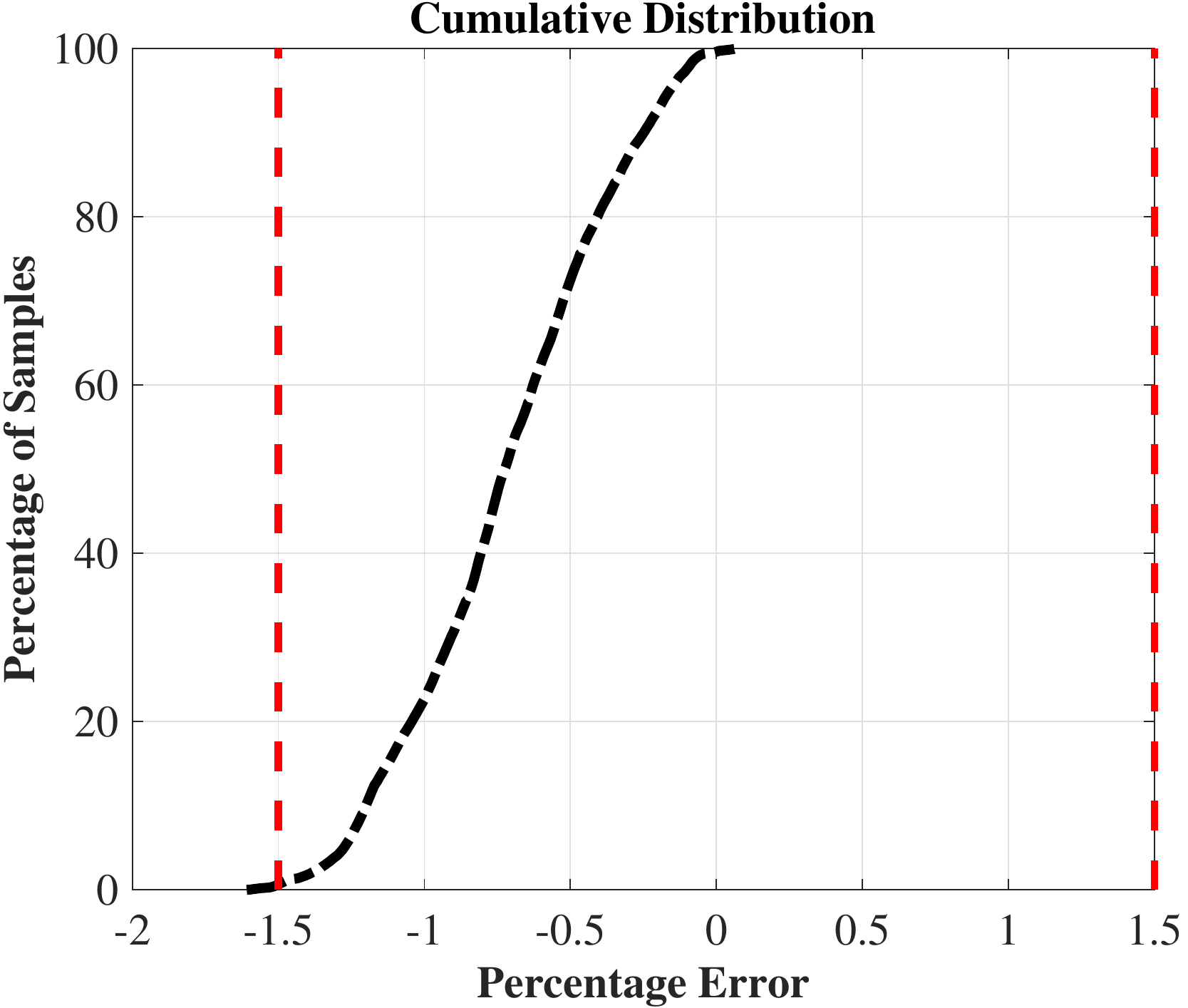}
	\caption{Cumulative distribution of terminal voltage estimation error for proposed online APE under Figure \ref{figpf14} battery bank discharge profile.}
	\label{figpf16}
\end{figure}

\subsection{Accuracy assessment of estimated parameters via battery bank discharging test}

\quad The effectiveness of the proposed online APE strategy is further quantified by comparing estimated parameters with the ones obtained through offline experimentation. For that purpose, the 400 V, 6.6 Ah Li-ion battery bank is discharged through a 384 ohms, 600 W resistive load. The battery bank discharge current and voltage profiles along with the estimated terminal voltage during the adaptation process are shown in Figure \ref{figpf13}. The detailed procedure of the proposed online APE strategy has been described in Section \ref{sec3}, and the results of estimated battery bank parameters are given in Table \ref{tab400}. The real-time estimated parameters of a 400 V, 6.6 Ah Li-ion battery bank model are quantified against the parameters obtained through offline mode. The estimation error in Table \ref{tab400} shows the accuracy of real-time parameters. 
The accuracy of offline estimated parameters is assessed by analyzing the estimated terminal voltage against an offline and fast periodic discharge  profile through a resistive load rated at 230 ohms, 1000 W. The time period of discharging profile is two minutes with 50\% duty cycle. The measured and estimated terminal voltage along with the estimation error are illustrated in Figure \ref{figpf14}. The terminal voltage estimation error in Figure \ref{figpf14} is around 1\% which proves the effectiveness of the proposed online APE strategy.

The statistical analysis of terminal voltage estimation error is also performed. Note that the total number of samples collected in the estimation error array during the discharging test are 73,529.  The mean, median, mode, and standard deviation analysis of the error array for the proposed online APE strategy are provided in Table \ref{tab10}. Moreover, the histogram and cumulative distribution graphs of terminal voltage estimation error are shown Figure \ref{figpf15} and Figure \ref{figpf16}, respectively. The red vertical lines in Figure \ref{figpf16} indicate the $\pm$ 1.5\% terminal voltage estimation error i.e. $\pm$ 6 V. The statistical analysis of terminal voltage estimation error shows the effectiveness of the proposed APE strategy for real-time parameters estimation of EV traction system.

\begin{table}[!t]
	\centering
	\caption{Terminal voltage estimation error statistics under Figure \ref{figpf14} battery bank discharge profile.}
	\begin{adjustbox}{width=1\linewidth}
		\begin{tabular}{c c c c c} 
			\hline\hline
			\thead{Parameters Estimation \\ Methods}  & \thead{Mean of\\ error (V)} & \thead{Median of\\ error (V)} & \thead{Mode of\\ error (V)} & \thead{Standard deviation\\ of error (V)}  \\
			\hline \\    
			Proposed online APE & -2.7754 & -2.7828 & -6.1766
			& 1.3199  \\   
			\hline
		\end{tabular}
		\label{tab10}
	\end{adjustbox}
\end{table}

%%%%%%%%%%%%%%%%

\section{Conclusion} \label{sec8}

An online UAS-based effective method for estimating Li-ion battery model parameters has been presented in this paper. The applicability of the developed method has been rigorously verified at the battery cell, pack and bank levels. In contrast to the reference offline UAS-based Li-ion battery parameters estimation;  the proposed technique does not require prior offline experimentation for open circuit voltage estimation, and also eliminates post-processing for series resistance estimation. Numerical simulations are performed on a 4.1 V, 270 mAh Li-ion battery model to quantify the accuracy of estimated parameters by comparing them against well-known results obtained experimentally by Chen and Mora. Mathematical proofs are provided to support
the proposed online APE strategy. Moreover, the results of the online APE strategy are experimentally compared with the reference offline APE technique on a 22.2 V, 6.6 Ah Li-ion battery test setup. The proposed strategy is further validated by performing a comprehensive statistical analysis of the terminal voltage estimation error for sixteen different discharging and sixteen constant charging protocols. It can be inferred from the results that the proposed online APE strategy produces similar results when compared with the existing offline APE strategy, yet minimizing the experimental effort and time required for the parameters estimation process. Furthermore, the proposed online APE strategy is implemented for real time, online parameters estimation of a 400 V, 6.6 Ah Li-ion battery bank; powering an indirect field-oriented induction motor driven EV traction system. The real time results are validated against an offline and fast periodic discharging battery bank voltage profile. The terminal voltage estimation error is around 1\%, which proves the accuracy of the proposed online APE strategy for real time battery bank parameters estimation of an EV traction system.

\appendix
\section{{Convergence speed of the proposed UAS-based method}}

{By definition of Nussbaum type switching function in equation} (\ref{pf25})-(\ref{pf26}), {the adaptive high-gain of the proposed UAS-based method ensures quick convergence of the estimation error. Therefore, the computational time of the proposed UAS-based strategy is not greatly affected by dynamic conditions, or by any type of driving cycle, including DST or UDDS. The parameter $k(t)$, adaptive gain $N(k(t))$, control input $u(t)$, and voltage estimation error $e(t)$ are shown below in Figure} \ref{fig_convergence} {during a 4.1 V Li-ion battery model parameters estimation process. The adaptive gain $N(k(t))$ settles to a steady state value in less than 150 samples, which implies $k(t) \to k_\infty$  by definition of Nussbaum function from equation} (\ref{pf25})-(\ref{pf26}). {From equation} (\ref{pf22}), {this further implies $\dot{k}(t) \to 0$, or $e(t) \to 0$ as $t \to t_c$, where $t_c$ denotes the convergence time. Since, the sampling time of the proposed algorithm is set to 0.01 seconds, which indicates $t_c = 0.01 \times 150 = 1.5$ seconds. Therefore, irrespective of any driving cycle/dynamic condition, the proposed strategy does not need to run for the entire driving cycle track, rather it is run for a few seconds and enables self-update of battery parameters in run-time for battery management systems (BMS) and real-time electric vehicle (EV) applications.}

{In Figure} \ref{fig_paramconv}, {the convergence of all battery parameters $\hat{r}_1,\hat{r}_2,\cdots,\hat{r}_{21}$ is shown during the adaptive estimation process for a 4.1 V Li-ion battery. Note that the parameters $\hat{r}_1,\hat{r}_2,\cdots,\hat{r}_{21}$ are normalized in Figure}  \ref{fig_paramconv} {for clearly observing convergence. Many of the traces of 21 battery parameters overlap in Figure} \ref{fig_paramconv}, {and show convergence in under 0.25 seconds, but it can also be clearly seen that all the parameters  $\hat{r}_1,\hat{r}_2,\cdots,\hat{r}_{21}$ achieve convergence in less than 150 samples or 1.5 seconds.}
\begin{figure}[!t]
	\centering
	\includegraphics[scale=0.26]{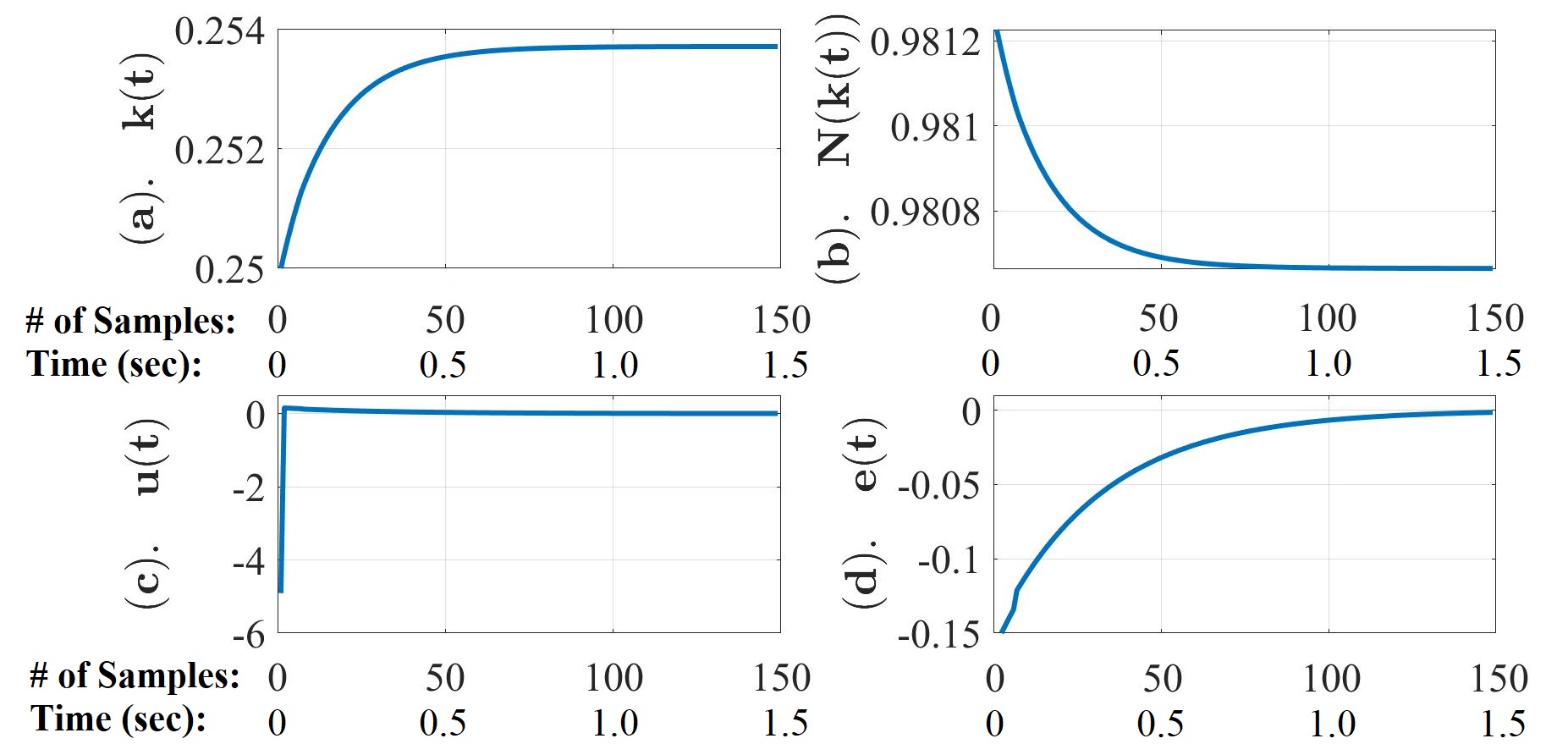}
	\caption{{(a) Parameter $k(t)$, (b) adaptive gain $N(k(t))$, (c) control input $u(t)$, and (d) voltage estimation error $e(t)$ during a 4.1 V Li-ion battery model parameters estimation process.}}
	\label{fig_convergence}
\end{figure}

\begin{figure}[!t]
	\centering
	\includegraphics[scale=0.22]{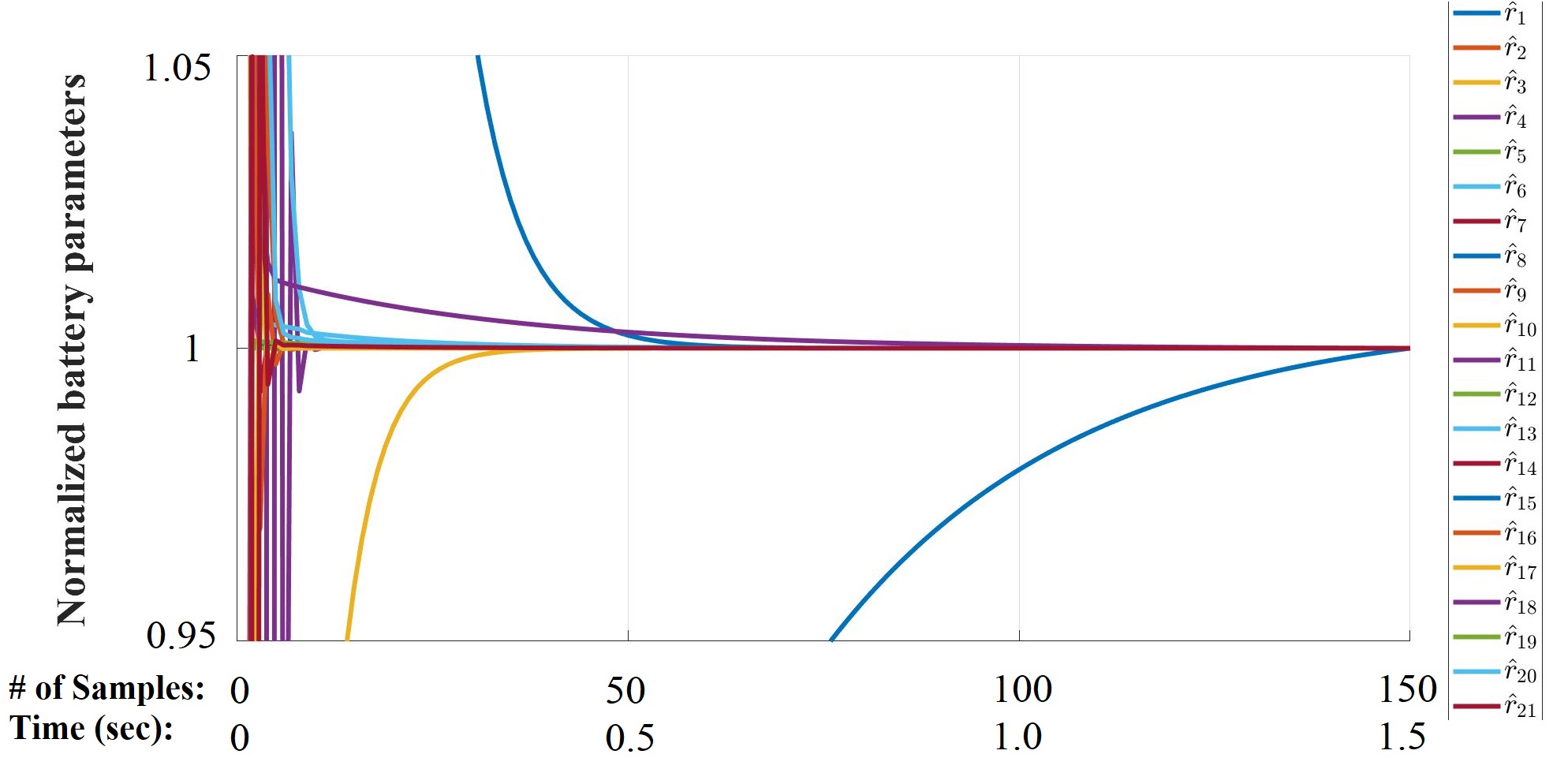}
	\caption{{Parameters $\hat{r}_1,\hat{r}_2,\cdots,\hat{r}_{21}$ convergence during the adaptive estimation process for a 4.1 V Li-ion battery.}}
	\label{fig_paramconv}
\end{figure}

\begin{table}[!t]
	\centering
	\caption{{Computation time comparison of the proposed UAS-based scheme with TRO-based Lease-Squares method, optimization-based methods, and two-stage adaptive scheme \& optimization-based methods.}}
	\begin{adjustbox}{width=1\linewidth}
		\begin{tabular}{c c} 
			\hline\hline
			\thead{Methods}  & \thead{Computation\\Time (seconds)}   \\
			\hline    
			Particle Swarm Optimization (PSO) & 34,200 \\
			Fmincon optimization & 1,500 \\
			Hybrid (PSO-fmincon) optimization &	39,780 \\
			Two-stage: Adaptive scheme \& PSO &	5,640 \\
			Two-stage: Adaptive scheme \& fmincon & 1,300 \\
			Two-stage: Adaptive scheme \& Hybrid optimization & 10,620 \\
			TRO-based Least Squares method & 46 \\
			\textbf{Proposed UAS-based approach} & \textbf{1.5} \\			  
			\hline
		\end{tabular}
		\label{tab_time}
	\end{adjustbox}
\end{table}

{Recently, a Trust Region Optimization (TRO) based Least-Squares method has been introduced in} \cite{new3} {to address the high computation time and slow convergence issues of the conventional Least-Squares method for battery parameters estimation. The work in} \cite{Access} {also presents the computation time of TRO-based Least-Squares method under various experiments, where the lowest computation time reported is 46 seconds. Emphasizing that the conventional Least-Squares method has significantly higher computation time with poor convergence compared to TRO-based Least-Squares method. Further, in our previous work} \cite{Access}, {we significantly reduced the computation time of optimization-based methods by employing an adaptive strategy to fine-tune the search space interval required by optimization method. In Table} \ref{tab_time} {, we compare the computation/execution time of the proposed UAS-based scheme with TRO-based Least-Squares method, optimization-based methods, and two-stage adaptive scheme \& optimization-based methods.}

{It is worth noting that the computation time of TRO-based Least-Squares method (an improved version of conventional Least-Squares method) it almost 30 times more than the proposed UAS-based approach for battery parameters estimation. The lower computation time shows the best suitability of the proposed UAS-based approach for real-time battery parameters estimation of an electric vehicle. The proposed strategy is run for a few seconds and enables self-update of battery parameters in run-time for battery management systems (BMS) and real-time electric vehicle (EV) applications.} 

\bibliographystyle{ieeetr}
\bibliography{usmanieeetranforarxiv_2022}

\begin{IEEEbiography}[{\includegraphics[width=1in,height=1.25in,clip,keepaspectratio]{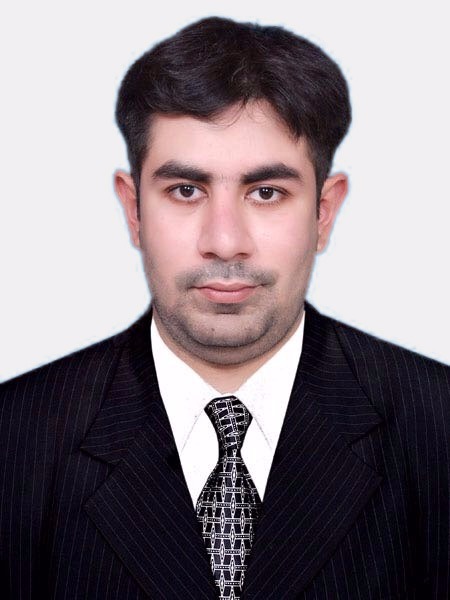}}]{Hafiz M. Usman} received his B.Sc. in Electrical Engineering from the University of Engineering and Technology, Lahore, Pakistan, in 2016, and M.Sc. in Electrical Engineering from the American University of Sharjah (AUS), United Arab Emirates (UAE), in 2019. He worked as a research assistant at the American University of Sharjah from 2017 to 2019. Currently, he is pursuing Ph.D. in Power and Energy Systems at the University of Waterloo, ON, Canada. His research interests include Li-ion batteries, power electronics and electric drives, control systems, renewable energy and power distribution systems.
\end{IEEEbiography}

\begin{IEEEbiography}[{\includegraphics[width=1in,height=1.25in,clip,keepaspectratio]{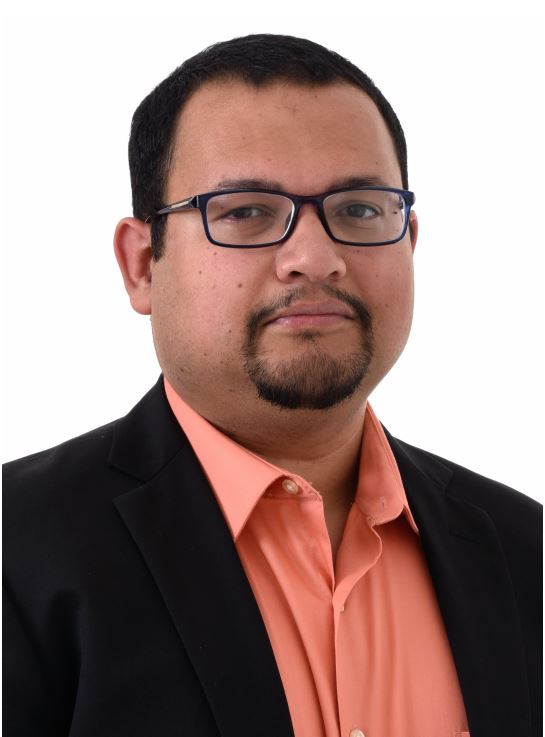}}]{Shayok Mukhopadhyay} received the B.E. degree in electrical engineering from the College of Engineering Pune, Savitribai Phule Pune University (formerly known as the University of Pune), India, in 2006, the M.Sc. degree in electrical engineering from Utah State University, Logan, UT, USA, in 2009, and the Ph.D. degree in electrical engineering from the Georgia Institute of Technology, Atlanta, GA, USA, in 2014. He has been with the Department of Electrical Engineering, American University of Sharjah, United Arab Emirates (UAE), since 2014, where he is currently an Associate Professor. His research interests include control systems, nonlinear systems, computational methods, battery modeling and failure detection, and robotic path planning. He received the Award for the best presentation in the Nonlinear Systems III Session from the American Control Conference 2014. He was a part of five-person team that received the national category of the AI and Robotics for Good Award for developing an in-pipe inspection robot at UAE in 2017. 
\end{IEEEbiography}

\begin{IEEEbiography}[{\includegraphics[width=1in,height=1.25in,clip,keepaspectratio]{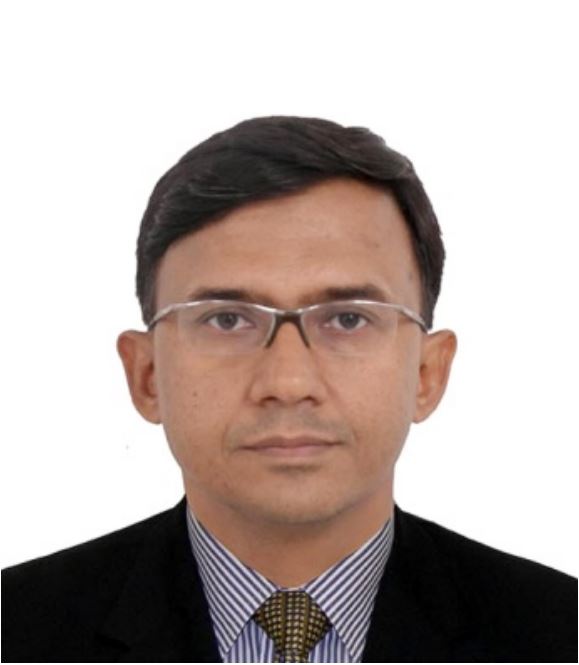}}]{Habibur Rehman} received the B.Sc. degree in electrical engineering from the University of Engineering and Technology at Lahore, Lahore, Pakistan, in 1990, and the M.S. and Ph.D. degrees in electrical engineering from The Ohio State University, Columbus, OH, USA, in 1995 and 2001, respectively. He has a wide experience in the areas of power electronics and motor drives in both industry and academia. From 1998 to 1999, he was a Design Engineer with Ecostar Electric Drive Systems and Ford Research Laboratory, where he was a member of the Electric, Hybrid, and Fuel Cell Vehicle Development Programs. From 2001 to 2006, he was with the Department of Electrical Engineering, United Arab Emirates (UAE) University, Al Ain, UAE, as an Assistant Professor. In 2006, he joined the Department of Electrical Engineering, American University of Sharjah, where he is currently working as a Professor. His primary research interests are in the areas of power electronics and their application to power systems, adjustable-speed drives, and alternative energy vehicles. He was a recipient of the Best Teacher Award (2002-2003) from the College of Engineering, UAE University.
\end{IEEEbiography}

\end{document}